\documentclass[preprint,12pt]{elsarticle}

\usepackage{amssymb}
\usepackage{amsmath}

\usepackage{tabularx}
\usepackage{algorithm}
\usepackage{booktabs}
\usepackage{algpseudocode}
\usepackage{subcaption}

\usepackage{xcolor}
\usepackage{booktabs}
\usepackage{multirow}
\usepackage{siunitx}
\usepackage{graphicx} 

\begin{document}

\begin{frontmatter}
	
	
	
	\title{MAPPO-LCR: Multi-Agent Proximal Policy Optimization with Local Cooperation Reward in Spatial Public Goods Games}
	
	
	\author[1,2]{Zhaoqilin Yang}
	
	
	\ead{zqlyang@gzu.edu.cn}
	
	\tnotetext[1]{https://github.com/geek12138/MAPPO-LCR}
	
	\affiliation[1]{organization={State Key Laboratory of Public Big Data, College of Computer Science and Technology},
		addressline={Guizhou University}, 
		city={Guiyang},
		postcode={550025}, 
		state={Guizhou},
		country={China}}

	\affiliation[2]{organization={Institute of Cryptography and Data Security},
		addressline={Guizhou University}, 
		city={Guiyang},
		postcode={550025}, 
		state={Guizhou},
		country={China}}
		
		\author[3,2]{Axin Xiang}

		\ead{axxiang@gzu.edu.cn}
		
		\author[3,2]{Kedi Yang}
		
		\ead{kdyang@gzu.edu.cn}
		
		\affiliation[3]{organization={State Key Laboratory of Public Big Data, College of Big Data and Information Engineering},
			addressline={Guizhou University}, 
			city={Guiyang},
			postcode={550025}, 
			state={Guizhou},
			country={China}}
			
		\author[1]{Tianjun Liu}
		\ead{tjliu_cumt@126.com}
		
		\author[3,2]{Youliang Tian\corref{cor1}}
		\ead{yltian@gzu.edu.cn}
		\cortext[cor1]{Corresponding author}
		
	\begin{abstract}
		
		Spatial public goods games model collective dilemmas where individual payoffs depend on population-level strategy configurations. Most existing studies rely on evolutionary update rules or value-based reinforcement learning methods. These approaches struggle to represent payoff coupling and non-stationarity in large interacting populations. This work introduces Multi-Agent Proximal Policy Optimization (MAPPO) into spatial public goods games for the first time. In these games, individual returns are intrinsically coupled through overlapping group interactions. Proximal Policy Optimization (PPO) treats agents as independent learners and ignores this coupling during value estimation. MAPPO addresses this limitation through a centralized critic that evaluates joint strategy configurations. To study neighborhood-level cooperation signals under this framework, we propose MAPPO with Local Cooperation Reward, termed MAPPO-LCR. The local cooperation reward aligns policy updates with surrounding cooperative density without altering the original game structure. MAPPO-LCR preserves decentralized execution while enabling population-level value estimation during training. Extensive simulations demonstrate stable cooperation emergence and reliable convergence across enhancement factors. Statistical analyses further confirm the learning advantage of MAPPO over PPO in spatial public goods games.
		
	\end{abstract}
	
	
	
	\begin{keyword}
		Spatial public goods games \sep Deep reinforcement learning \sep Multi-agent proximal policy optimization \sep Local cooperation reward
		
		
	\end{keyword}
	
\end{frontmatter}

	
	
	\section{Introduction}
	\label{sec1}
	
	Cooperation is a fundamental concept in the study of collective dynamics within structured populations \cite{dawes_1988_anomalies,perc_2016_phase,perc_2017_statistical}.
	Energy-sharing microgrids provide a practical illustration, where households coordinate local decisions to maintain stable distributed power systems.
	These systems reveal how localized cooperative actions generate collective benefits that exceed individual contributions.
	Such patterns also highlight the persistent tension between personal incentives and community welfare observed in many social dilemmas \cite{pennisi_2005_did,kennedy_2005_don}.
	Spatial public goods games (SPGG) provide a formal framework for studying cooperation in structured populations \cite{nowak_1993_spatial,macy_2002_learning}.
	By explicitly defining payoff rules for strategic interactions, SPGG captures the tension between individual gain and collective welfare.

	Insights from evolutionary game theory further reveal how interaction topology and population structure influence the stability of cooperation across multiple settings \cite{nowak_1992_evolutionary,hauert_2005_game,szabo_2007_evolutionary}.
	Prior studies outline several pathways that help curb free-riding, including incentive-based designs \cite{chen_2015_first,dos_2015_evolution}, reputation-mediated strategies \cite{tang_2024_cooperative,shen_2022_high}, punishment systems \cite{helbing_2010_punish,chen_2015_competition}, and exclusion rules \cite{liu_2017_competitions,szolnoki_2017_alliance}.
	A tangible real-world example arises in community microgrids, where residents share locally generated energy.
	Stable operation depends on households consistently contributing energy rather than exploiting shared resources.
	Empirical observations indicate that localized cooperative signals can enhance overall performance even without centralized enforcement.
	These observations highlight the importance of localized coordination mechanisms, forming the conceptual foundation for the Local Cooperation Reward (LCR) in our framework.
	
	Reinforcement learning (RL) has reshaped the study of strategic adaptation in social dilemmas.
	Instead of instantaneous imitation rules, it adopts a sequential decision-making paradigm based on state evaluation, action selection, and reward feedback \cite{sutton_1998_reinforcement,izquierdo_2007_transient}.
	Unlike classical evolutionary approaches based on Fermi updates and replicator dynamics \cite{szabo_1998_evolutionary,schuster_1983_replicator}, RL offers a more flexible modeling framework.
	It captures how agents gradually refine their behaviors over time through accumulated experience.
	This perspective aligns more closely with real socio-economic systems, where individuals adjust decisions in response to both immediate incentives and long-term expectations \cite{jia_2021_local,song_2022_reinforcement}.
	Among value-based RL algorithms, Q-learning has shown a notable ability to sustain cooperation even when agents face strong temptations to defect \cite{watkins_1992_q,shi_2022_analysis}.
	Its temporal-difference updates adapt well across heterogeneous spatial structures \cite{szolnoki_2009_topology,szolnoki_2010_impact}.
	Recent extensions further incorporate periodic strategy adjustment \cite{yan_2024_periodic} and hybrid mechanisms that combine Q-learning with Fermi dynamics to reinforce cooperative tendencies \cite{shen_2024_learning}.
	Although these developments broaden the applicability of Q-learning, its representational capacity remains limited.
	This constraint hinders performance in high-dimensional multi-agent environments, where agents must infer complex dependencies and coordinate under non-stationary dynamics.
	These challenges motivate a transition toward deep reinforcement learning (DRL) frameworks that can approximate more expressive policies.
	Such frameworks are well suited for managing large-scale state–action spaces and pave the way for policy-gradient methods in multi-agent evolutionary settings.

	DRL has emerged as a key analytical framework for studying cooperation in high-dimensional strategic environments.
	Early work demonstrated that neural networks can effectively approximate value functions, as shown in deep Q-learning \cite{Mnih_2015_nature}.
	This capability supports applying RL to large-scale public goods environments, where cooperation can arise under multiple learning mechanisms \cite{Tamura_2024_RSOC}. 
	Actor–critic methods advanced this direction, with A3C \cite{Mnih_2016_ICML} improving stability through asynchronous updates and revealing how social preferences influence multi-agent behavior \citep{hughes_2018_NeurIPS}. 
	Additional efforts explore mean-based pursuit strategies \cite{Wang_2025_ESWA} and knowledge-driven cooperation on complex networks \cite{Chen_2024_ESWA}.
	Proximal Policy Optimization (PPO) \cite{John_2017_arxiv} further strengthened policy-gradient methods by constraining updates with clipped objectives, a design that stabilizes training in complex decision spaces \cite{sun_2024_intuitionistic,yu_2022_surprising}. 
	PPO and its extensions have been applied successfully to SPGG \cite{YANG_2025_116762,YANG_2025_116928}, where curriculum strategies and group-oriented objectives improve cooperative outcomes. 
	Despite these advances, existing DRL methods still face challenges in representing fine-grained interaction patterns and stabilizing cooperation in spatially distributed systems. 
	These gaps motivate the development of new multi-agent RL frameworks tailored to localized coordination and complex evolutionary dynamics.

	 We introduce a framework based on Multi-Agent Proximal Policy Optimization (MAPPO) \cite{Yu_2022_nips} for analyzing cooperation dynamics in SPGG. 
	 To the best of our knowledge, this is the first study to adapt MAPPO to the SPGG setting. 
	 The framework integrates decentralized policy learning with a centralized value function, allowing agents to update individual strategies while jointly capturing population-level payoff structures. 
	 This design improves learning stability in highly non-stationary strategic environments.
	 Classical evolutionary update rules often fail to recover cooperation from full-defection initial states.
	 They may also exhibit oscillatory behavior under strong free-riding incentives.
	 Moreover, the MAPPO architecture provides richer behavioral representations, enabling agents to form more adaptive long-term strategies than tabular RL or deterministic evolutionary rules.
	 
	 We propose Multi-Agent Proximal Policy Optimization with Local Cooperation Reward (MAPPO-LCR), a multi-agent DRL framework for modeling cooperation in SPGG.
	 To the best of our knowledge, this work constitutes the first application of Multi-Agent Proximal Policy Optimization (MAPPO) \cite{Yu_2022_nips} to the SPGG setting.
	 Compared with PPO, agents learn independently under non-stationary interactions.
	 MAPPO employs a centralized critic that explicitly captures the coupled payoff structure arising from collective participation.
	 This learning paradigm better aligns with SPGG dynamics, in which individual returns depend strongly on group-level strategy configurations.
	 Within this framework, we further examine how MAPPO responds to explicit neighborhood-level cooperation signals.
	 We introduce a LCR that encodes the cooperative density of an agent’s focal interaction group.
	 Rather than enforcing cooperation directly, this signal provides structured local feedback that shapes policy updates during learning.
	 As a result, MAPPO-LCR enables a controlled analysis of how advanced multi-agent DRL internalizes spatial cooperation cues.
	 It further reveals how these cues are translated into macroscopic cooperation patterns in SPGG.
	 
	 Our contributions are threefold.
	 \begin{itemize}
	 	\item To the best of our knowledge, this work is the first to apply MAPPO to SPGG. Compared with PPO, MAPPO provides a learning framework that better reflects the coupled payoff structure inherent in collective interaction settings.
	 	\item We introduce a LCR mechanism specifically for MAPPO, supplying neighborhood-level cooperative feedback without introducing additional sensitive hyperparameters.
	 	\item We conduct controlled experiments to examine how MAPPO combined with LCR influences cooperation emergence in SPGG.
	 	The results show that localized cooperative feedback can reshape long-term strategic configurations.
	 \end{itemize}

	\section{Model}
	\label{sec:model}
	
	We formulate the SPGG on a periodic $L\times L$ grid, where each location is occupied by a single agent.
	Spatial interactions follow a von Neumann neighborhood, so each individual is linked to the four orthogonally adjacent sites.
	Because a focal agent also appears in the neighborhoods of its neighbors, it becomes a member of five partially overlapping game groups.
	Each agent selects one of two actions: cooperation (C) or defection (D).
	Within a local group $g$, a cooperator contributes one unit resource to the collective pool, while a defector contributes nothing.
	Let
	\begin{equation}
		N_C^g = \sum_{j \in g} \mathbb{I}(s_j^g = C)
	\end{equation}
	denote the number of cooperators in group $g$,
	where $\mathbb{I}(\cdot)$ is an indicator function that equals $1$ if the condition holds and $0$ otherwise.
	The accumulated contribution is multiplied by the enhancement factor $r>1$ and equally divided among the $k+1=5$ group members.
	Thus, the payoff assigned to agent $i$ from group $g$ is
	
	\begin{equation}
	\Pi_i^g=
	\begin{cases}
		\dfrac{rN_C^g}{5}-1, & s_i^g=C,\\[6pt]
		\dfrac{rN_C^g}{5}, & s_i^g=D.
	\end{cases}
	\end{equation}
	
	Since every agent appears in five different groups, its total payoff results from summing the earnings from all associated groups:
	
	\begin{equation}
	\Pi_i=\sum_{g\in\mathcal{G}_i}\Pi_i^g ,
\end{equation}
	where $\mathcal{G}_i$ refers to the set of groups in which agent i participates.
	These definitions specify the strategic environment on which subsequent learning or evolutionary mechanisms operate.
	They capture the essential tension between individual temptation and collective benefit that characterizes SPGG dynamics.
	
	\subsection{MAPPO}
	
	MAPPO is a Centralized-Training and Decentralized-Execution (CTDE) framework that extends the original PPO algorithm to cooperative multi-agent systems.
	Let $\pi_\theta(a_t^i \mid s_t^i)$ denote the decentralized policy of agent i with parameters $\theta$. Each agent selects an action $a_t^i$ based solely on its local observation $s_t^i$.
	To stabilize learning in non-stationary multi-agent environments, MAPPO employs a centralized value function $V_\phi(\mathbf{S}_t)$ parametrized by $\phi$, where $\mathbf{S}_t \in \{0,1\}^{L \times L}$ denotes the global state at time $t$. This centralized critic provides a consistent learning signal by evaluating joint configurations rather than independent local states.
	
	The policy is optimized through the clipped surrogate objective:
	\begin{equation}
		L^{\text{CLIP}}(\theta)=
		-\mathbb{E}_t
		\left[
		\min\left(
		r_t(\theta)\hat A_t\ ,\
		\mathrm{clip}\big(r_t(\theta),1-\epsilon,1+\epsilon\big)\hat A_t
		\right)
		\right],
	\end{equation}
	where
	$r_t(\theta)={\pi_\theta(a_t^i\mid s_t^i)}/{\pi_{\theta_{\text{old}}}(a_t^i\mid s_t^i)}$ is the policy ratio,
	$\epsilon$ is the clipping threshold,
	and $\hat A_t$ denotes the advantage estimate obtained from the centralized critic.
	
	The centralized critic is optimized by minimizing the value regression loss:
	\begin{equation}
	L^{\text{VF}}(\phi)=
	\mathbb{E}_t\!\left[\,\big(V_\phi(\mathbf{S}_t)-R_t\big)^2\right],
	\end{equation}
	where $R_t$ is the empirical return computed through temporal recursion.
	
	Entropy regularization promotes exploration through:
	\begin{equation}
	L^{\text{ENT}}(\theta)=
	\mathbb{E}_t\left[-\sum_a
	\pi_\theta(a\mid s_t)\log \pi_\theta(a\mid s_t) \right].
	\end{equation}
	
	The overall optimization objective of MAPPO combines the policy loss,
	the centralized value loss, and the entropy regularization term:
	\begin{equation}
		L(\theta,\phi)
		=
		L^{\text{CLIP}}(\theta)
		+
		\delta\,L^{\text{VF}}(\phi)
		-
		\rho\,L^{\text{ENT}}(\theta),
	\end{equation}
	where $\delta$ controls the relative weight of the value function loss and $\rho$ denotes the entropy coefficient.
	
	MAPPO thus couples decentralized actors with a shared centralized critic, allowing agents to learn coordinated behaviors while acting independently during execution.
	
	\subsection{MAPPO-LCR}
	
	To investigate how MAPPO responds to spatial cooperation cues, we introduce a local cooperation reward into SPGG.
	MAPPO captures coupled payoffs effectively, but its response to neighborhood-level cooperation remains underexplored.
	We therefore study how repeated exposure to local cooperation density shapes policy learning dynamics.
	
	For an agent $i$, the local cooperation indicator measures the cooperative
	density within its focal interaction group and is defined as
	\begin{equation}
		r_t^{\text{LCR}}(i)=\frac{N_t^i}{4},
	\end{equation}
	where $N_t^i$ denotes the number of cooperators in the focal group of agent $i$,
	including itself and its four von Neumann neighbors.
	The normalization by four provides a bounded local feedback signal that reflects
	the surrounding cooperative intensity while preserving relative payoff scales.
	Integrating this term with the original SPGG payoff yields the augmented reward
	\begin{equation}
	r_t(i)=r_t^{\text{SPGG}}(i)+\zeta{r}_t^{\text{LCR}}(i),
	\end{equation}
	with $\zeta$ controlling the strength of local feedback.
	
	The MAPPO-LCR framework follows the CTDE paradigm of MAPPO.
	A centralized critic evaluates the global state, while decentralized actors condition decisions on local features.
	The LCR is introduced as an auxiliary feedback term without modifying the MAPPO architecture.
	This design reshapes policy gradients while preserving the original learning structure.
	The formulation enables principled analysis of how neighborhood cooperation information influences policy learning.
	Architectural design is decoupled from reward modulation to isolate learning effects.
	MAPPO-LCR supports systematic study of convergence behavior, stability under different enhancement factors, and sensitivity to spatial initialization.
	The framework characterizes how local cooperation signals translate into population-level cooperation patterns in SPGG.
	
	\subsection{Actor-Critic Network Architecture}

\begin{figure}[h]
		\centering
		\includegraphics[width=0.9\linewidth]{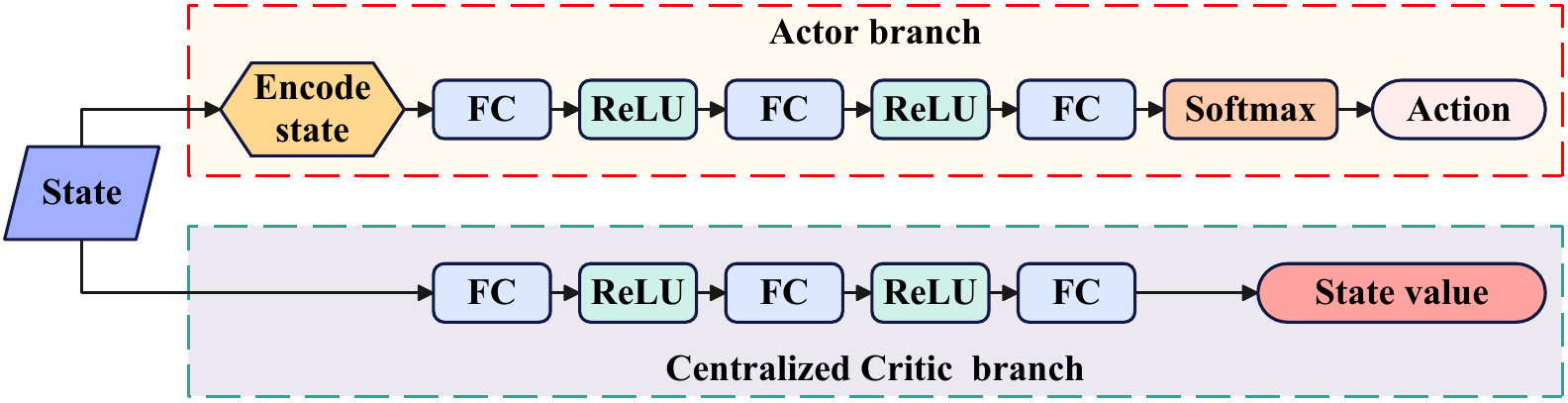}
		\caption{The Actor--Critic network of MAPPO and MAPPO-LCR.The network is divided into an Actor branch and a Centralized Critic branch. The Actor branch operates on locally encoded features for decentralized action selection.
			The Centralized Critic branch evaluates the global strategy matrix during training.}
		\label{fig:mappo-lcr}
	\end{figure}

	At time $t$, the global configuration of the game is given by  $S_t \in \{0,1\}^{L\times L}$.  
	The actor and the centralized critic operate on two different representations of this matrix.  
	The full architecture is illustrated in Figure~\ref{fig:mappo-lcr}.
	
	Each agent $i$ receives a locally encoded feature vector $s_t^i$ produced by the encode state module.
	The vector is defined as $s_t^i = (x_t^i, n_t^i, g_t)$.
	Here, $x_t^i$ denotes the agent’s current strategy at time $t$.
	The term $n_t^i$ counts the number of cooperators in the focal group of agent $i$,
	including the agent itself and its four von Neumann neighbors.
	The scalar $g_t$ represents the global cooperation ratio.
	This definition aligns with the SPGG interaction structure, where payoffs are
	determined by contributions within the same five-member focal group.
	The actor computes a categorical policy using a three-layer feed-forward transformation:
	\begin{equation}
		\pi_\theta(a_t^i \mid s_t^i)
		= \mathrm{Softmax}\big(\mathrm{FC}(\mathrm{ReLU}(\mathrm{FC}(\mathrm{ReLU}(\mathrm{FC}(s_t^i)))))\big)
	\end{equation}
	where $\mathrm{FC}(\cdot)$ denotes a fully connected linear mapping,  
	$\mathrm{ReLU}(\cdot)$ is the ReLU activation function \cite{Glorot_2011_relu},  
	and $\mathrm{Softmax}(\cdot)$ converts the output into a probability distribution.  
	Because the actor parameters are shared among all agents, the joint policy factorizes as
	\begin{equation}
		\pi_\theta(a_t \mid S_t)
		= \prod_{i=1}^{L^2} \pi_\theta(a_t^i \mid s_t^i).
	\end{equation}
	
	The centralized critic follows a separate pathway and operates directly on the unencoded global matrix.  
	After vectorization as
	\begin{equation}
		z_t = \mathrm{vec}(S_t),
	\end{equation}
	the critic evaluates the global state using a three-layer transformation:
	\begin{equation}
		V_{\mathrm{cent}}(S_t)
		= \mathrm{FC}\!\left(\mathrm{ReLU}\!\left(
		\mathrm{FC}\!\left(\mathrm{ReLU}(\mathrm{FC}(z_t))\right)\right)\right).
	\end{equation}
	In this expression, $\mathrm{vec}(\cdot)$ flattens the $L\times L$ matrix into a vector.  
	The critic is used only during training under the CTDE paradigm,  
	while execution relies exclusively on the actor’s local features.
	
	The complete optimization procedure of MAPPO-LCR for SPGG is summarized in Algorithm~\ref{alg:mappo_lcr}.
	
	\begin{algorithm}[H]
		\caption{MAPPO-LCR Optimization Procedure for SPGG}
		\label{alg:mappo_lcr}
		\begin{algorithmic}[1]

			\State Initialize environment state $S_0$
			
			\For{each training epoch $t=1$ to $T$}
			
			\State Encode local features $s_t^i$ from $S_t$
			\State Sample actions $a_t^i \sim \pi_\theta(a_t^i \mid s_t^i)$
			\State Execute actions and obtain next state $S_{t+1}$
			\State Compute individual reward and LCR-augmented reward
			
			\State Evaluate values using centralized critic $V_{\mathrm{cent}}(S_t)$
			\State Compute advantages $A_t$ using GAE
			
			\For{each PPO update iteration}
			\State Compute policy ratio $r_t$
			\State Compute clipped loss $L_{\mathrm{clip}}$
			\State Compute value loss $L_{\mathrm{value}}$
			\State Compute entropy bonus $L_{\mathrm{ent}}$
			\State Update parameters:
			\[
			\theta \leftarrow \theta - \alpha \nabla_\theta
			\left( L_{\mathrm{clip}} + \delta L_{\mathrm{value}} - \rho L_{\mathrm{ent}} \right)
			\]
			\EndFor

			\State $S_t \leftarrow S_{t+1}$
			
			\EndFor
			
		\end{algorithmic}
	\end{algorithm}

	\section{Experimental results}
	\label{sec:exp}
	\subsection{Experimental setup}
	
	Experiments were conducted on a $200\times200$ lattice with von Neumann neighborhoods.
	The clipping coefficient, GAE parameter, entropy weight, value-loss weight, and discount factor all follow the configuration used in TUC-PPO \citep{YANG_2025_116928}.
	Accordingly, both MAPPO and MAPPO-LCR adopt the Adam optimizer \cite{Diederik_2015_ICLR} with a learning rate of $1\times10^{-3}$, clip $\epsilon=0.2$, GAE $\lambda=0.95$, entropy weight $\rho=0.001$, value-loss weight $\delta=0.5$, and discount factor $\gamma=0.99$.
	The total training horizon is 1000 iterations. Cooperators and defectors are denoted as C and D in the figures.
	
	\subsection{Ablation study of MAPPO}
	
\begin{figure}[htbp]
		\centering
		\includegraphics[width=0.65\linewidth]{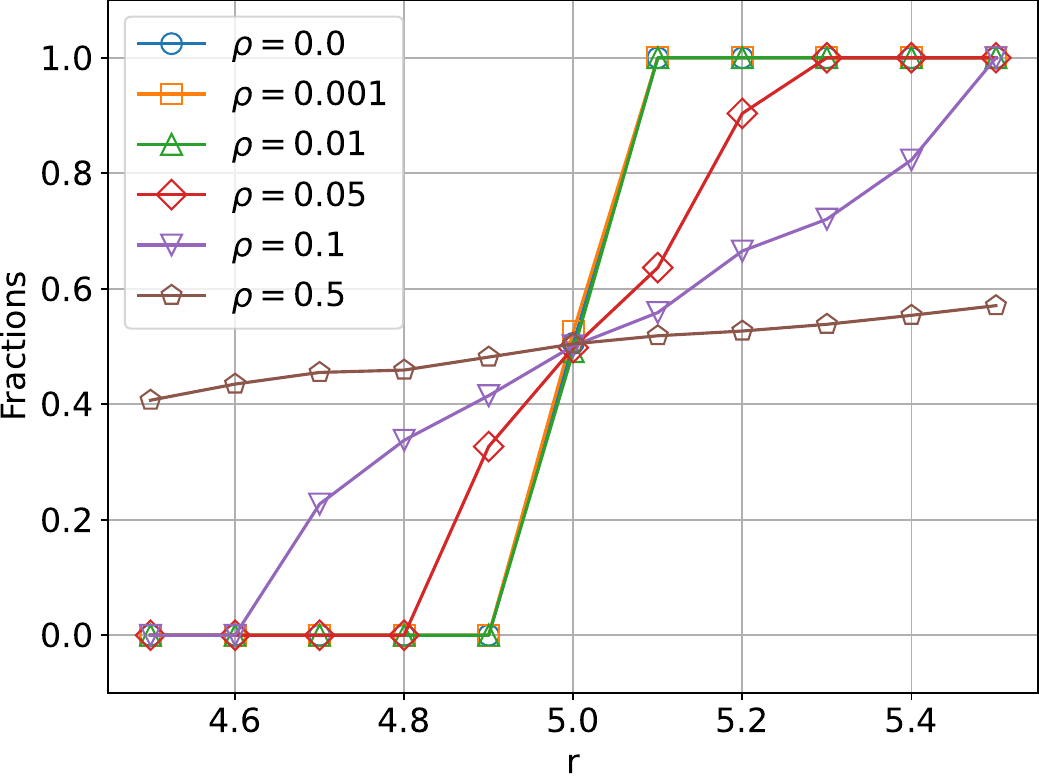}
		\caption{Cooperation fraction under different entropy coefficients $\rho$.}
		\label{fig:rho_ablation}
	\end{figure}

	For MAPPO, only the entropy coefficient $\rho$ produces clear behavioral differences,  so we report ablation results solely on $\rho$. 
	The initial state contains pure defection in the upper region and pure cooperation in the lower region.
	Figure~\ref{fig:rho_ablation} shows cooperation levels under $\rho\in\{0.0, 0.001, 0.01, 0.05, 0.1, 0.5\}$.
	The horizontal axis is the enhancement factor $r$, and the vertical axis is the
	fraction of cooperators.
	All MAPPO settings remain fixed except for $\rho$.
	Cooperation emerges for $r<5$ once $\rho>0.01$.
	Larger $\rho$ shifts this emergence to even smaller $r$, because stronger entropy regularization increases the chance of sampling cooperative actions during early learning.
	These samples give the centralized critic additional cooperative trajectories, helping the policy escape the defect-dominated basin in the low-$r$ region.
	For $r>5$, the effect reverses.
	High $\rho$ slows convergence to full cooperation because persistent exploration keeps the policy stochastic even when cooperation is optimal.
	This stochasticity prevents fast alignment to a uniform cooperative state.
	Entropy regularization thus acts as a mutation force.
	Higher $\rho$ helps the system leave the defection attractor when $r$ is small, but weakens the cooperative attractor when $r$ is large.
	We adopt $\rho=0.001$, which preserves useful exploration while allowing stable convergence in cooperative regimes.
	
	\subsection{Ablation study of MAPPO-LCR}
	
\begin{figure}[htbp]
		\centering
		\includegraphics[width=0.7\linewidth]{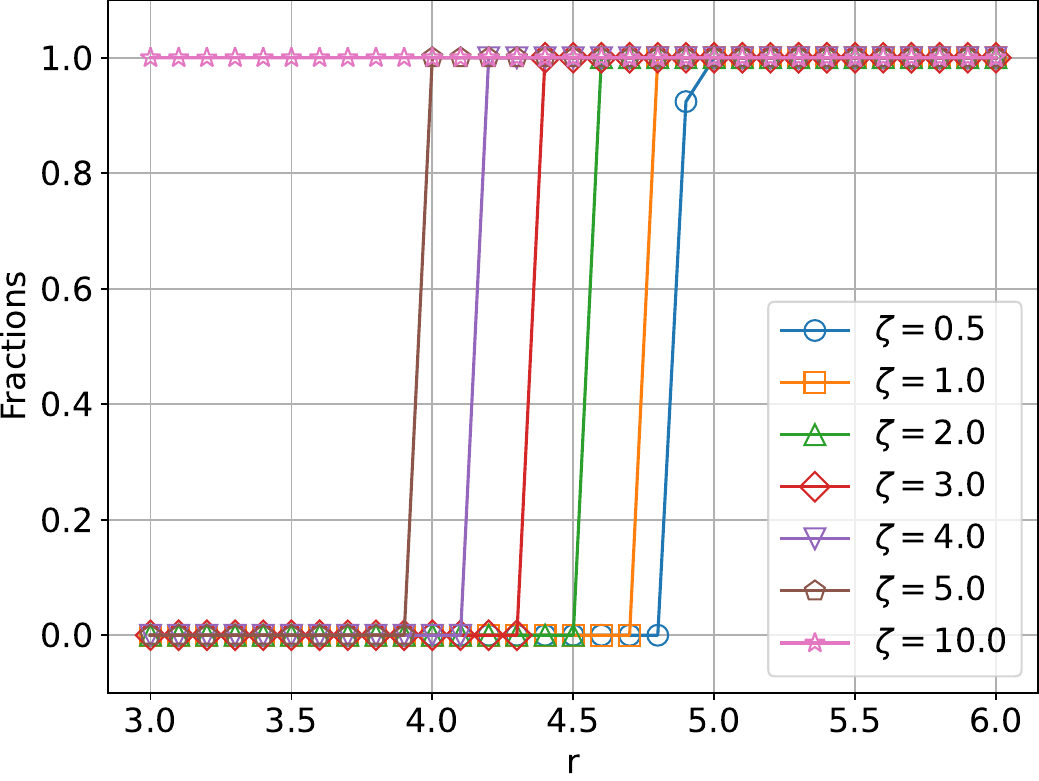}
		\caption{Effect of $\zeta$ on cooperation dynamics in MAPPO-LCR.}
		\label{fig:exp_zeta}
	\end{figure}

	For MAPPO-LCR, we examine how different values of the local cooperation weight $\zeta$ influence cooperative behavior.
	The grid is initialized with defectors in the upper region and cooperators in the lower region.
	Figure~\ref{fig:exp_zeta} presents the cooperation fraction under $\zeta\in{0.5,1.0,2.0,3.0,4.0,5.0,10.0}$ across different $r$.
	LCR increases the influence of neighborhood cooperation within MAPPO.
	A larger $\zeta$ strengthens the gradient weight of local cooperative feedback.
	This adjustment pushes policy updates toward regions with stronger cooperation.
	As a result, the required $r$ for stable cooperation becomes smaller.
	The curves in Figure~\ref{fig:exp_zeta} clearly reflect this trend.
	Excessively large $\zeta$ decreases the comparability of results across $r$.
	The shaping term becomes dominant and weakens the meaning of SPGG payoffs.
	Such extreme settings provide limited interpretability for game dynamics.
	We use $\zeta=3.0$ in later experiments.
	This choice keeps the SPGG structure meaningful while supporting full cooperation for $r\ge4.4$.
	The setting offers a stable balance between shaping strength and interpretability.
	
	\subsection{Statistical analysis of MAPPO-LCR}
	\label{exp:compare_stat}
	
\begin{figure*}[htbp!]
		\begin{minipage}{\linewidth}
			\centering
			\includegraphics[width=0.55\linewidth]{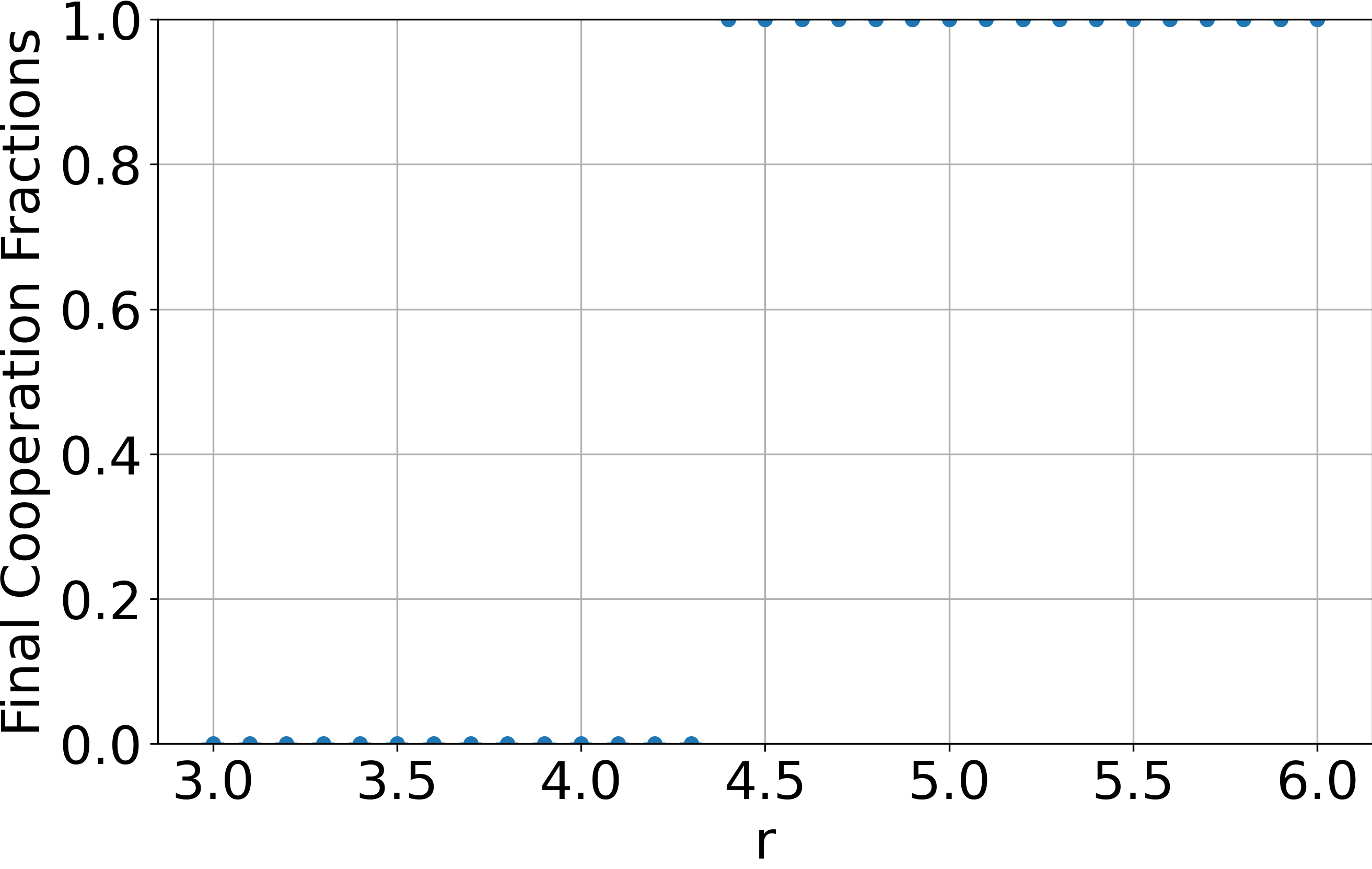}\\
			\vspace{-2mm}
			\caption*{\footnotesize (a) MAPPO-LCR}
		\end{minipage}
		\\[3mm]
		\begin{minipage}{\linewidth}
			\centering
			\includegraphics[width=0.55\linewidth]{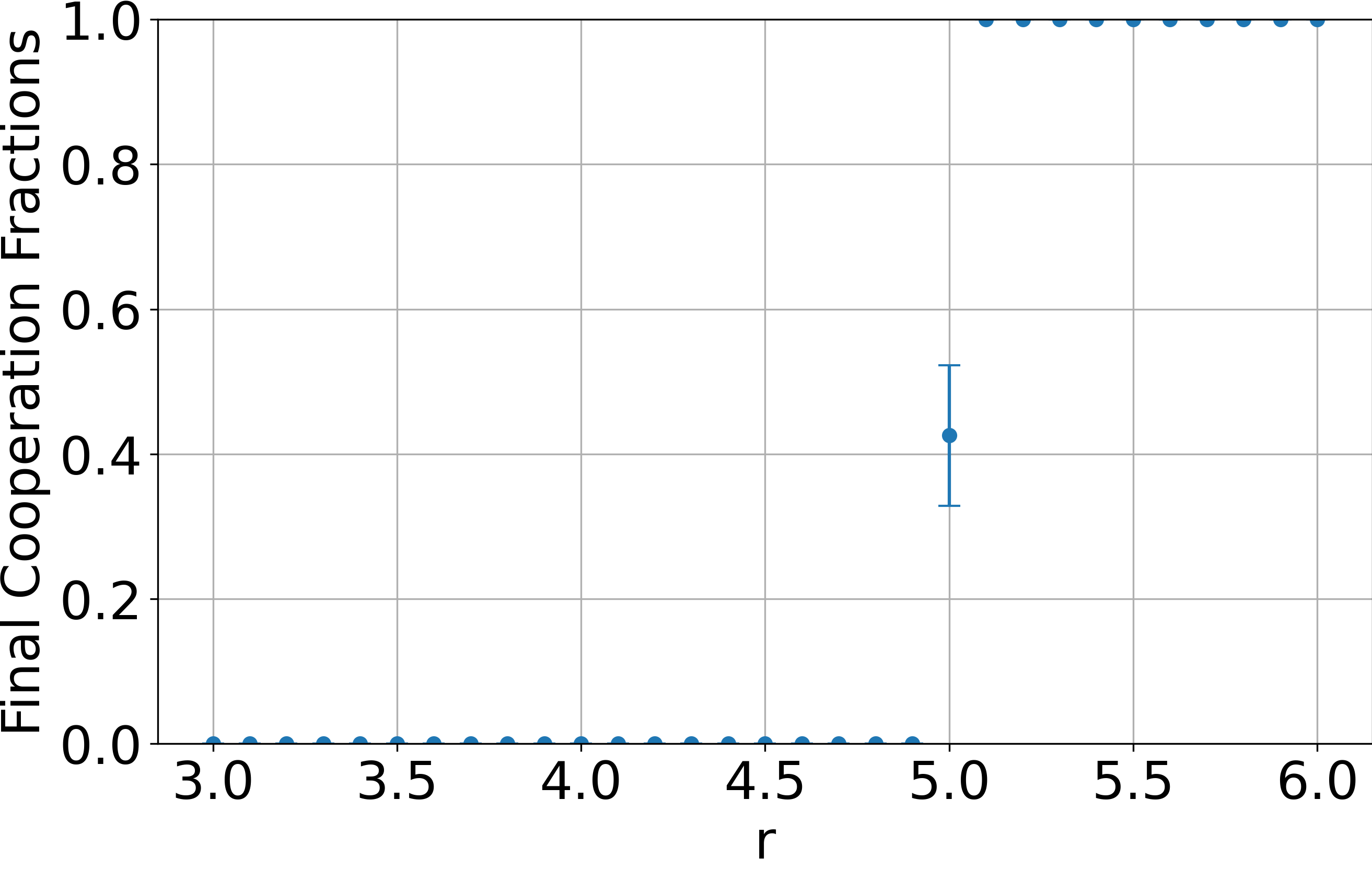}\\
			\vspace{-2mm}
			\caption*{\footnotesize (b) MAPPO}
		\end{minipage}
		\\[3mm]
		\begin{minipage}{\linewidth}
			\centering
			\includegraphics[width=0.55\linewidth]{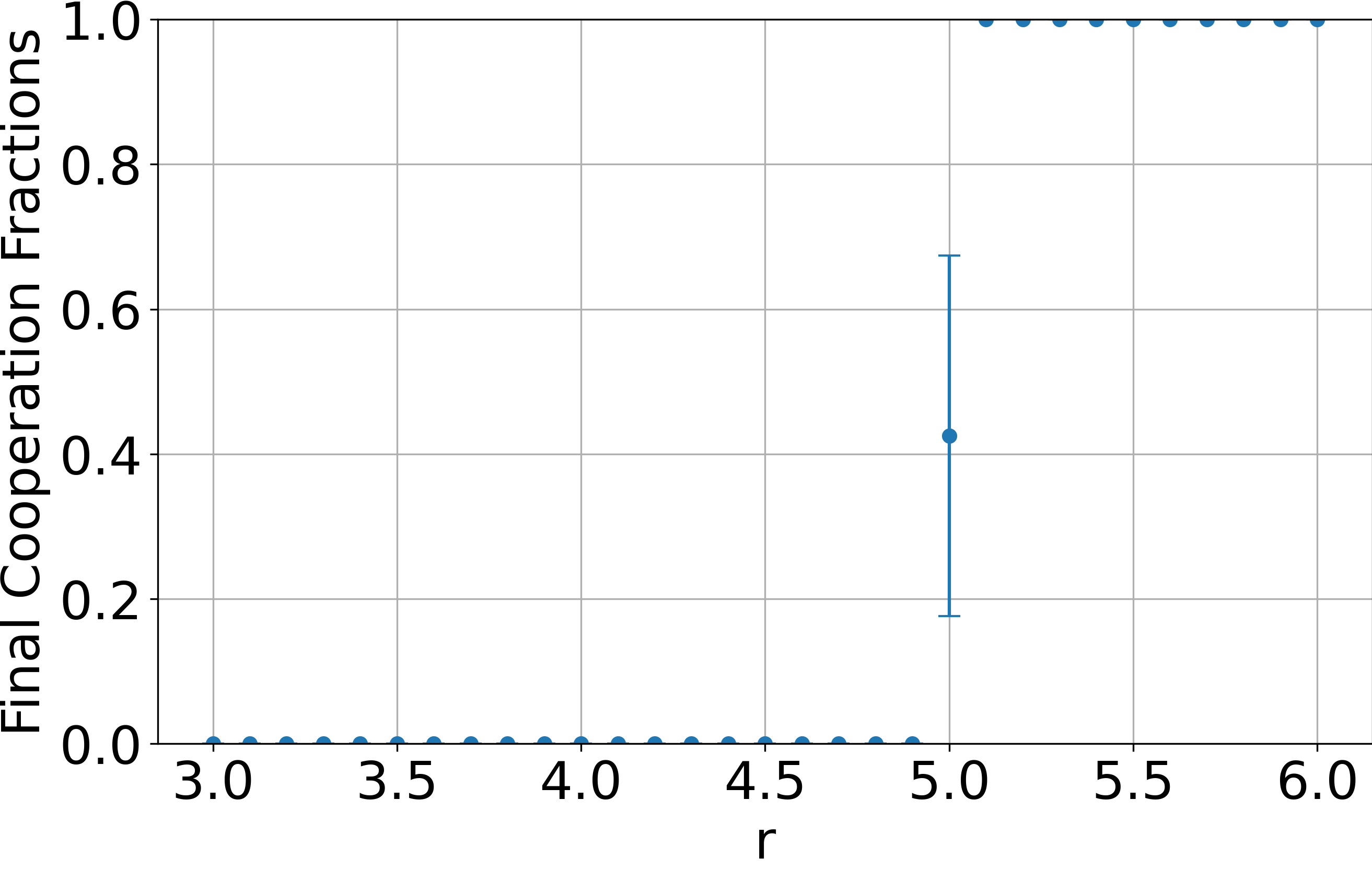}\\
			\vspace{-2mm}
			\caption*{\footnotesize (c) PPO}
		\end{minipage}
		\caption{Statistical comparison of cooperation fractions based on 50 independent runs.
			Error bars show the mean and standard deviation across runs
			for (a) MAPPO-LCR, (b) MAPPO, and (c) PPO.
			MAPPO-LCR exhibits a sharp and deterministic transition at $r=4.4$,
			while MAPPO and PPO display delayed and more variable transitions.}
		\label{fig:MAPPO-LCR_r_stat_err}
	\end{figure*}

	To assess statistical robustness, MAPPO-LCR, MAPPO, and PPO are each evaluated over 50 independent runs for every enhancement factor $r \in [3.0, 6.0]$.
	All three methods are compared using error bars, violin plots, and $95\%$ confidence intervals.
	The error bar results in Fig.~\ref{fig:MAPPO-LCR_r_stat_err} show that MAPPO-LCR exhibits highly stable and near-deterministic outcomes.
	When $r < 4.4$, all MAPPO-LCR runs converge to full defection.
	When $r \ge 4.4$, all MAPPO-LCR runs converge to full cooperation.
	The variance therefore collapses to zero on both sides of the transition.
	MAPPO exhibits a delayed transition compared with MAPPO-LCR.
	For $r < 5.0$, all MAPPO runs converge to full defection.
	For $r > 5.0$, all MAPPO runs converge to full cooperation.
	At the critical point $r = 5.0$, MAPPO shows pronounced variability across independent runs.
	The mean cooperation level is $42.5\%$ with a standard deviation of $9.7\%$.
	This reflects unstable convergence near the transition regime.
	Across the same enhancement factors, MAPPO shows consistently smaller standard deviations than PPO.
	PPO exhibits substantially larger variability, indicating weaker learning stability under non-stationary multi-agent interactions.
	This contrast highlights the advantage of MAPPO’s centralized critic over independently learned PPO policies.
	
\begin{figure*}[htbp!]
		\centering
		\begin{minipage}{\linewidth}
			\centering
			\includegraphics[width=0.8\linewidth]{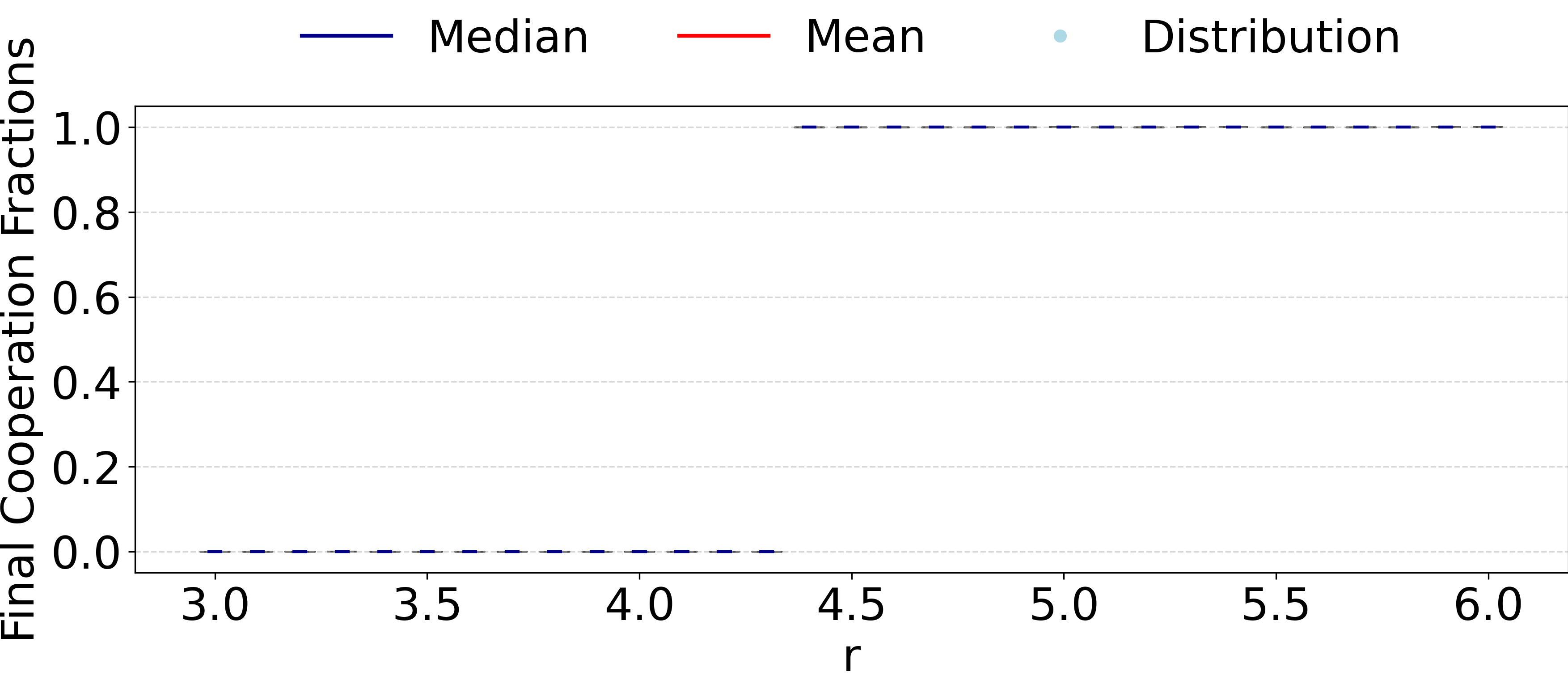}\\
			\vspace{-2mm}
			\caption*{\footnotesize (a) MAPPO-LCR}
		\end{minipage}
		\\[3mm]
		\begin{minipage}{\linewidth}
			\centering
			\includegraphics[width=0.8\linewidth]{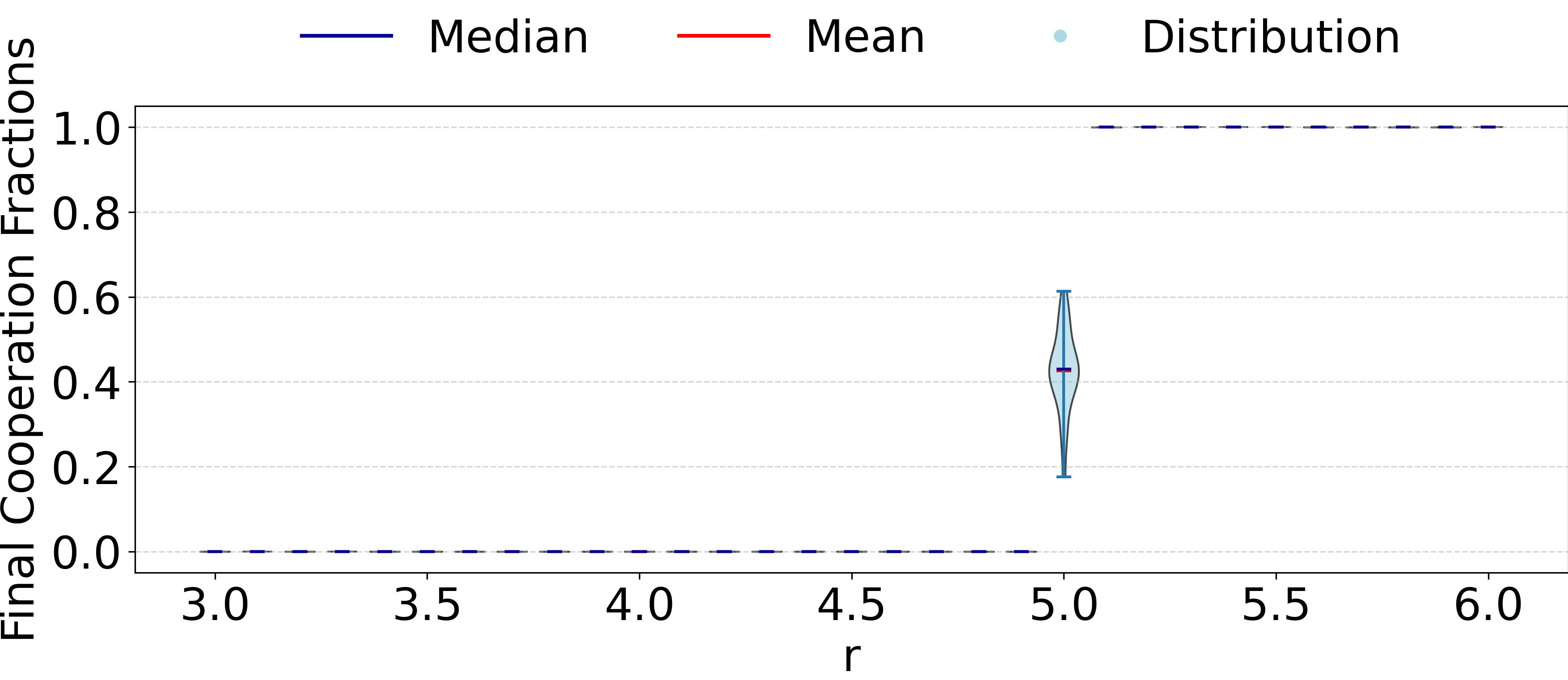}\\
			\vspace{-2mm}
			\caption*{\footnotesize (b) MAPPO}
		\end{minipage}	
		\\[3mm]
		\begin{minipage}{\linewidth}
			\centering
			\includegraphics[width=0.8\linewidth]{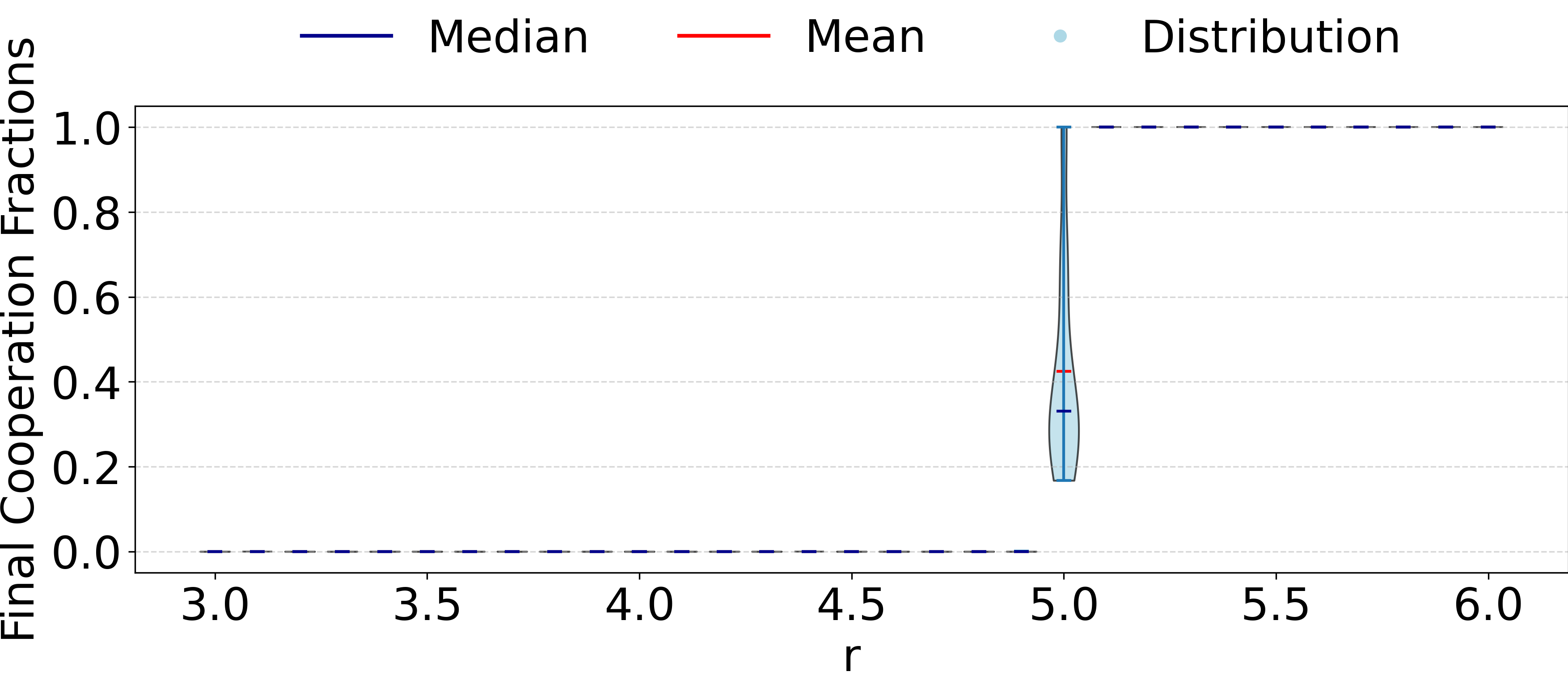}\\
			\vspace{-2mm}
			\caption*{\footnotesize (c) PPO}
		\end{minipage}	
		\caption{Distribution of final cooperation fractions across 50 independent runs for (a) MAPPO-LCR, (b) MAPPO, and (c) PPO.
			MAPPO-LCR yields concentrated distributions at $0\%$ or $100\%$ cooperation, indicating stable convergence.
			MAPPO and PPO exhibit broader distributions near their transition regions, reflecting higher sensitivity to exploration noise.}
		\label{fig:MAPPO-LCR_r_stat_vio}
	\end{figure*}

	The violin plots in Figure~\ref{fig:MAPPO-LCR_r_stat_vio} further corroborate these observations.
	MAPPO-LCR produces narrow and sharply concentrated distributions collapsing to either $0\%$ or $100\%$ cooperation.
	This pattern indicates strong attraction toward deterministic equilibria.
	MAPPO exhibits broader distributions near $r = 5.0$, reflecting increased sensitivity to exploration noise.
	PPO shows even wider distributions with pronounced asymmetry.
	The characteristic top-narrow and bottom-wide shape indicates amplified stochasticity under independent learning.
	Compared with PPO, MAPPO achieves more stable convergence by modeling payoff coupling through a centralized critic.
	
	The $95\%$ confidence intervals in Table~\ref{tab:CI_comparison} provide additional quantitative evidence.
	For MAPPO-LCR, confidence intervals become undefined (NaN) whenever all trials reach identical final states.
	This behavior confirms perfect consistency across independent runs.
	For MAPPO, non-zero confidence intervals appear at $r = 5.0$, indicating incomplete convergence after 1000 training iterations.
	PPO exhibits wider confidence intervals at the same enhancement factor.
	This reflects stronger sensitivity to exploration noise and higher convergence uncertainty under independent learning.
	
	\begin{table}[h]
		\centering
		\footnotesize
		\caption{$95\%$ confidence intervals comparison for cooperation fractions}
		\label{tab:CI_comparison}
		\resizebox{\textwidth}{!}{ 
			\begin{tabular}{@{}c*{7}{S[table-format=1.2]@{\,--\,}S[table-format=1.2]}@{}}
				\toprule
				r & \multicolumn{2}{c}{3.5} & \multicolumn{2}{c}{3.6} & \multicolumn{2}{c}{3.7} & \multicolumn{2}{c}{3.8} & \multicolumn{2}{c}{3.9} & \multicolumn{2}{c}{4.0} & \multicolumn{2}{c}{4.1} \\
				\cmidrule(lr){2-3} \cmidrule(lr){4-5} \cmidrule(lr){6-7} \cmidrule(lr){8-9} \cmidrule(lr){10-11} \cmidrule(lr){12-13} \cmidrule(lr){14-15}
				MAPPO-LCR &  0.00 & 0.00  &  0.00 & 0.00  &  0.00 & 0.00  &  0.00 & 0.00  &  0.00 & 0.00  &  0.00 & 0.00  &  0.00 & 0.00  \\
				MAPPO & 0.00 & 0.00 &  0.00 & 0.00  &  0.00 & 0.00 &  0.00 & 0.00  &  0.00 & 0.00  &  0.00 & 0.00  &  0.00 & 0.00    \\
				PPO & 0.00 & 0.00 &  0.00 & 0.00  &  0.00 & 0.00 &  0.00 & 0.00  &  0.00 & 0.00  &  0.00 & 0.00  &  0.00 & 0.00    \\
				\midrule
				
				r & \multicolumn{2}{c}{4.2} & \multicolumn{2}{c}{4.3} & \multicolumn{2}{c}{4.4} & \multicolumn{2}{c}{4.5} & \multicolumn{2}{c}{4.6} & \multicolumn{2}{c}{4.7} & \multicolumn{2}{c}{4.8} \\
				\cmidrule(lr){2-3} \cmidrule(lr){4-5} \cmidrule(lr){6-7} \cmidrule(lr){8-9} \cmidrule(lr){10-11} \cmidrule(lr){12-13} \cmidrule(lr){14-15}
				MAPPO-LCR  &  0.00 & 0.00  &  0.00 & 0.00  &  1.00  & 1.00 &  1.00  & 1.00  & 1.00  & 1.00 &  1.00  & 1.00  &  1.00  & 1.00   \\
				MAPPO  &  0.00 & 0.00  &  0.00 & 0.00  &  0.00 & 0.00  &  0.00 & 0.00  &  0.00 & 0.00 &  0.00 & 0.00  &  0.00 & 0.00  \\
				PPO & 0.00 & 0.00 &  0.00 & 0.00  &  NaN & NaN &  0.00 & 0.00  &  0.00 & 0.00  &  0.00 & 0.00  &  0.00 & 0.00    \\
				\midrule
				
				r & \multicolumn{2}{c}{4.9} & \multicolumn{2}{c}{5.0} & \multicolumn{2}{c}{5.1} & \multicolumn{2}{c}{5.2} & \multicolumn{2}{c}{5.3} & \multicolumn{2}{c}{5.4} & \multicolumn{2}{c}{5.5} \\
				\cmidrule(lr){2-3} \cmidrule(lr){4-5} \cmidrule(lr){6-7} \cmidrule(lr){8-9} \cmidrule(lr){10-11} \cmidrule(lr){12-13} \cmidrule(lr){14-15}
				MAPPO-LCR  &  1.00  & 1.00  &  NaN &  NaN  & 1.00 & 1.00 &  1.00 & 1.00  &  1.00 & 1.00   &  NaN &  NaN  &  1.00 & 1.00    \\
				MAPPO  &  0.00 & 0.00  & 0.40 & 0.45 & 1.00 & 1.00 & NaN & NaN &  NaN & NaN  &  NaN & NaN  &  NaN & NaN   \\
				PPO & 0.00 & 0.00 &  0.35 & 0.50  &  NaN &  NaN  &  NaN & NaN &  NaN & NaN  &  NaN & NaN  &  NaN & NaN   \\
				\bottomrule
			\end{tabular}
		}
	\end{table}

	MAPPO and PPO exhibit cooperation transitions near $r = 5.0$, consistent with theoretical expectations in spatial public goods games. 
	Statistical results reveal pronounced stability differences between the two learning frameworks. 
	Under identical enhancement factors $r$, MAPPO consistently shows smaller standard deviations in error bar analysis. 
	This indicates more reliable convergence across independent experimental runs. 
	PPO exhibits substantially larger standard deviations under the same conditions. 
	This reflects higher sensitivity to exploration noise and multi-agent non-stationarity. 
	Violin plots further clarify these differences near the critical region $r = 5.0$. 
	MAPPO produces more compact cooperation distributions around the transition point. 
	PPO displays wider distributions with pronounced lower tails. 
	This pattern indicates unstable convergence behavior. 
	Although both methods share the same critical threshold, MAPPO demonstrates superior statistical stability. 
	MAPPO-LCR further shifts the cooperation transition toward smaller $r$ values. 
	Local cooperation feedback reshapes the reward landscape and stabilizes learning dynamics.

	\subsection{Comparative analysis of algorithms}
	\label{exp:Compare}
	
\begin{figure*}[htbp!]
		\begin{minipage}{\linewidth}
			\begin{minipage}{0.24\linewidth}
				\centering
				\includegraphics[width=\linewidth]{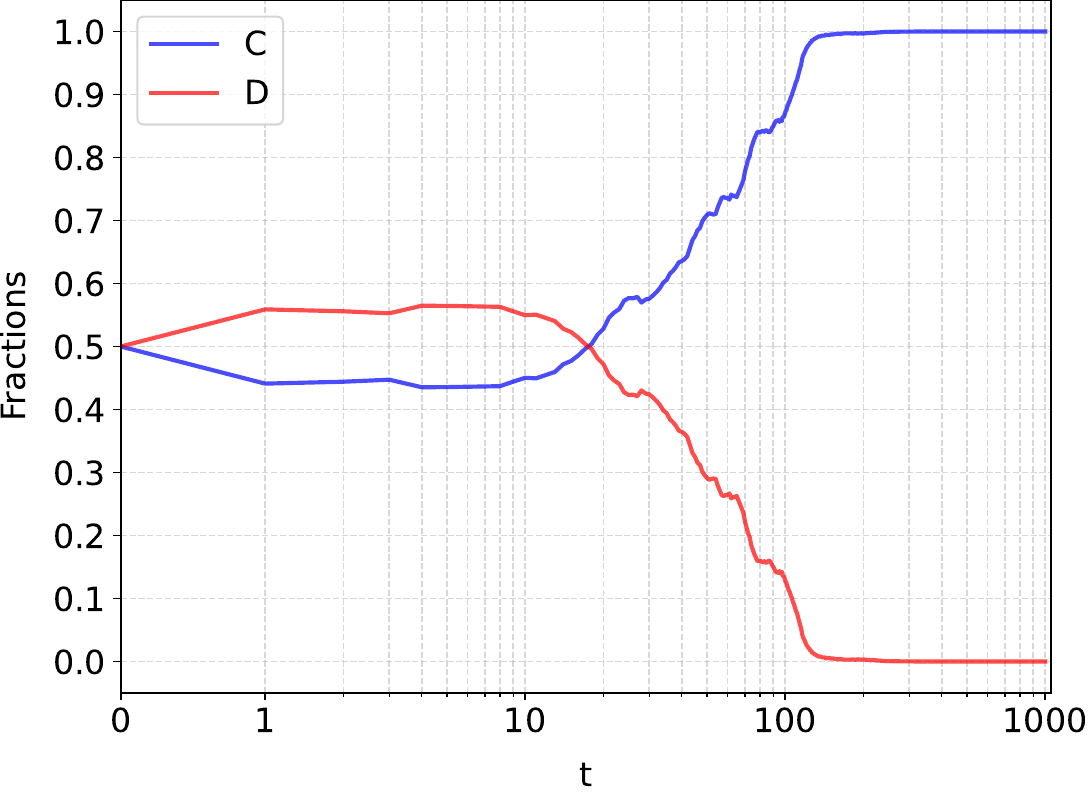}\\
			\end{minipage}
			\begin{minipage}{0.14\linewidth}
				\centering
				\fbox{\includegraphics[width=\linewidth]{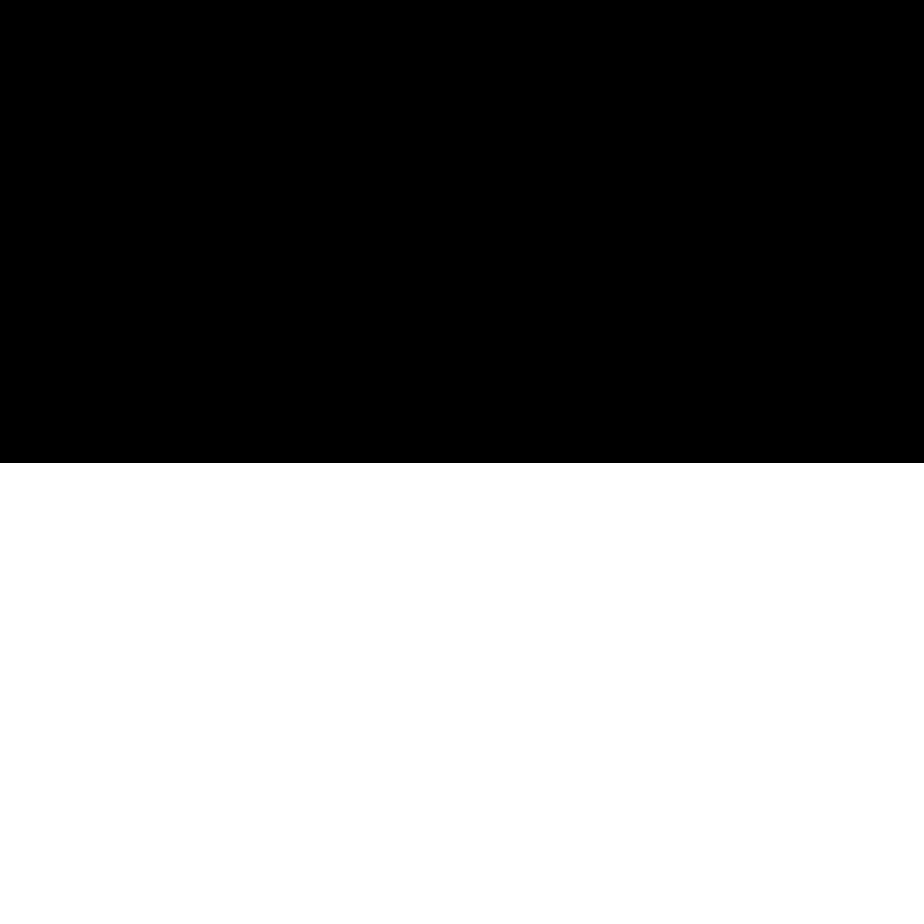}}\\
				\vspace{-2mm}
				{\footnotesize t=0}
			\end{minipage}
			\begin{minipage}{0.14\linewidth}
				\centering
				\fbox{\includegraphics[width=\linewidth]{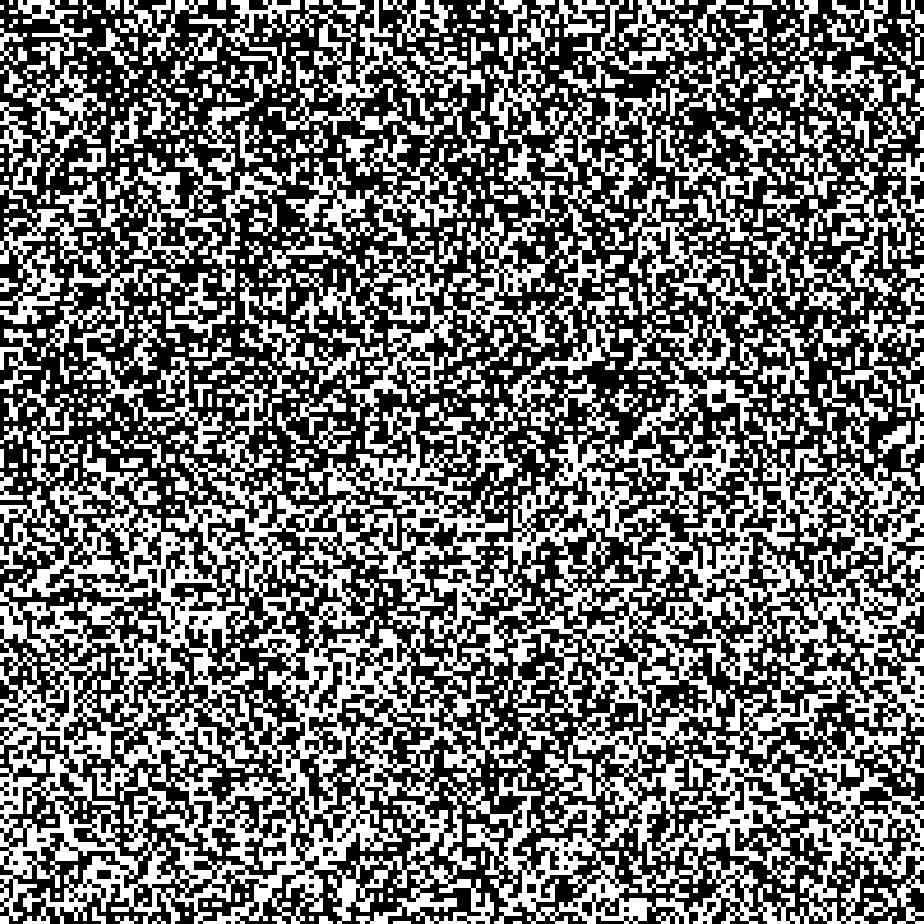}}\\
				\vspace{-2mm}
				{\footnotesize t=1}
			\end{minipage}
			\begin{minipage}{0.14\linewidth}
				\centering
				\fbox{\includegraphics[width=\linewidth]{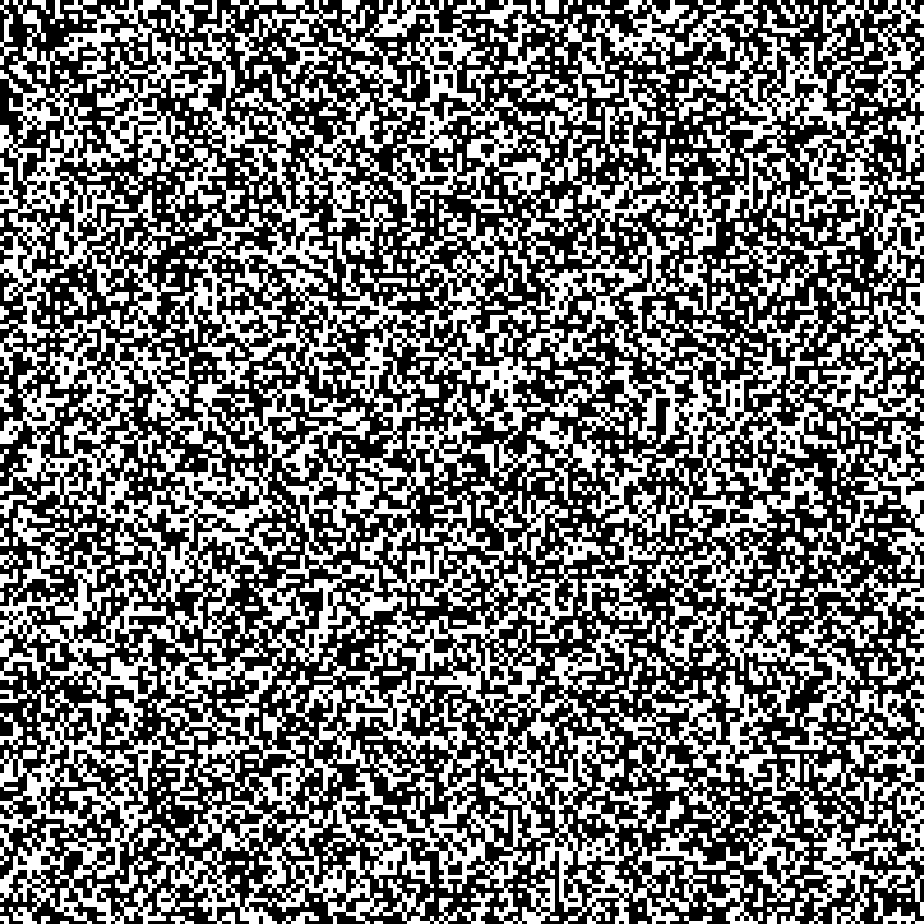}}\\
				\vspace{-2mm}
				{\footnotesize t=10}
			\end{minipage}
			\begin{minipage}{0.14\linewidth}
				\centering
				\fbox{\includegraphics[width=\linewidth]{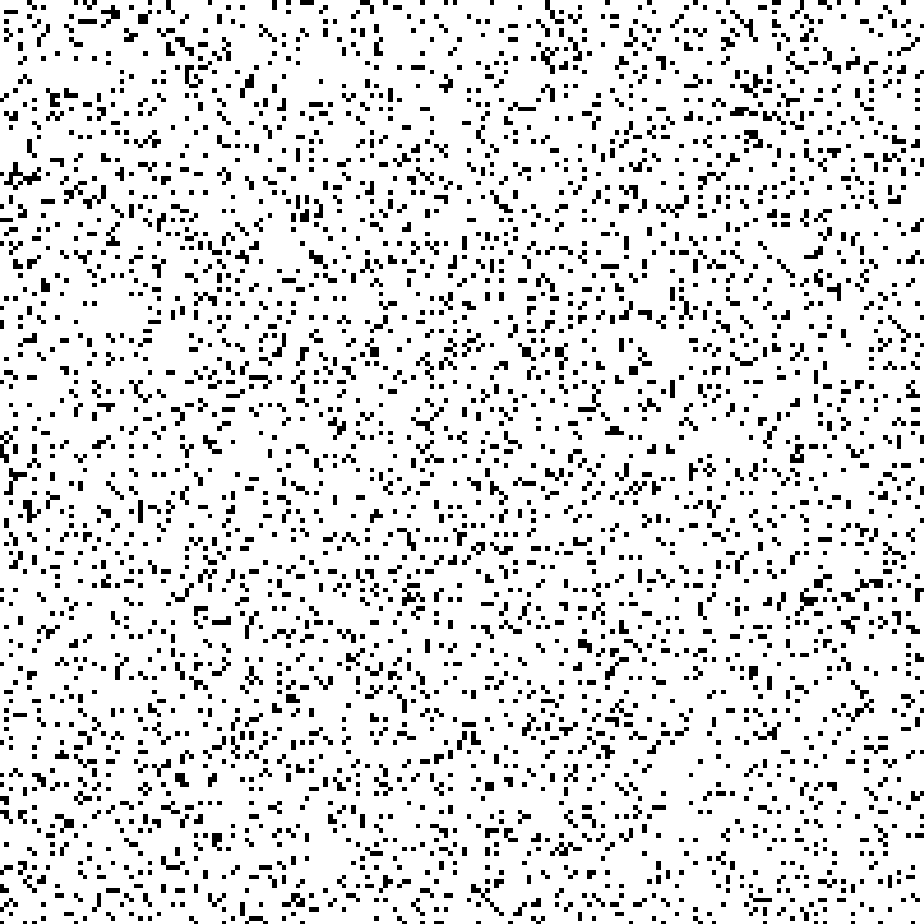}}\\
				\vspace{-2mm}
				{\footnotesize t=100}
			\end{minipage}
			\begin{minipage}{0.14\linewidth}
				\centering
				\fbox{\includegraphics[width=\linewidth]{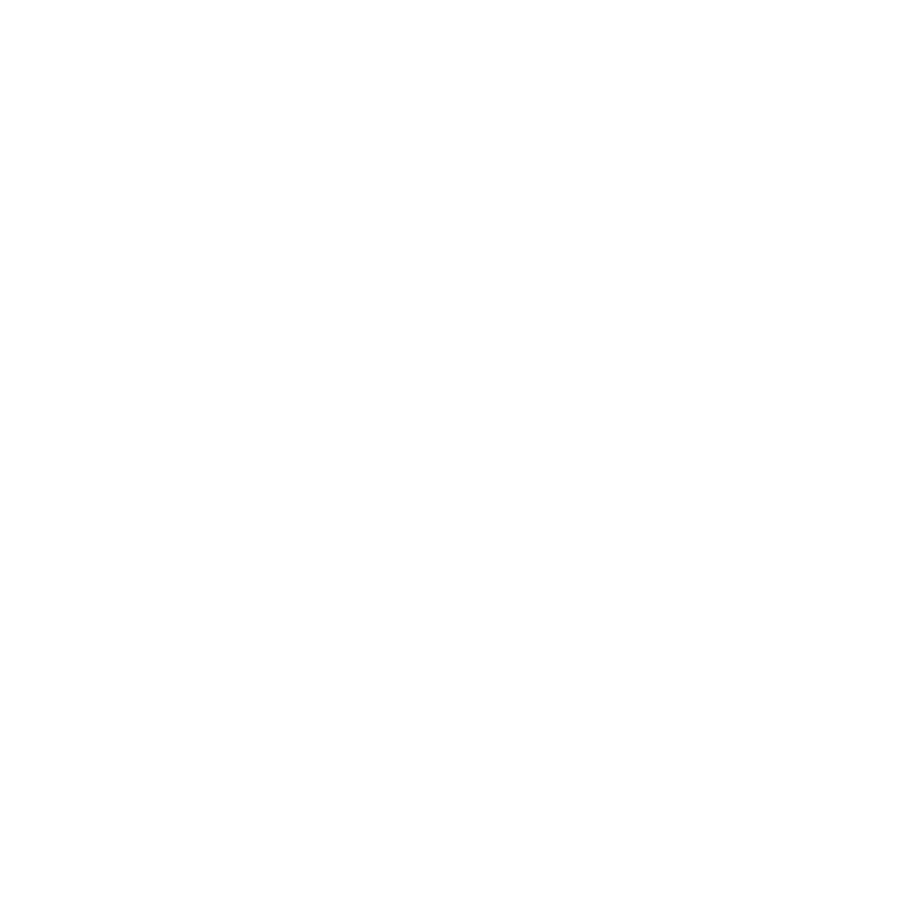}}\\
				\vspace{-2mm}
				{\footnotesize t=1000}
			\end{minipage}
			\vspace{-3mm}
			\caption*{\footnotesize (a) MAPPO-LCR}
		\end{minipage}
		\\[2mm]
		\begin{minipage}{\linewidth}
			\begin{minipage}{0.24\linewidth}
				\centering
				\includegraphics[width=\linewidth]{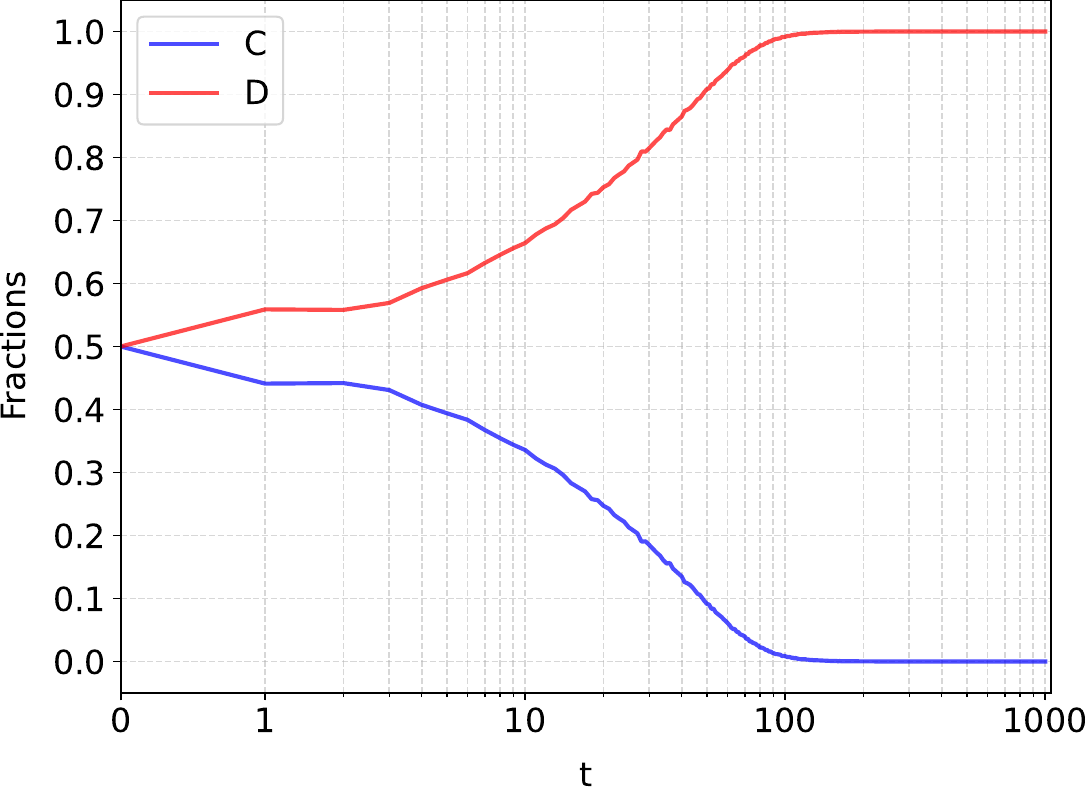}\\
			\end{minipage}
			\begin{minipage}{0.14\linewidth}
				\centering
				\fbox{\includegraphics[width=\linewidth]{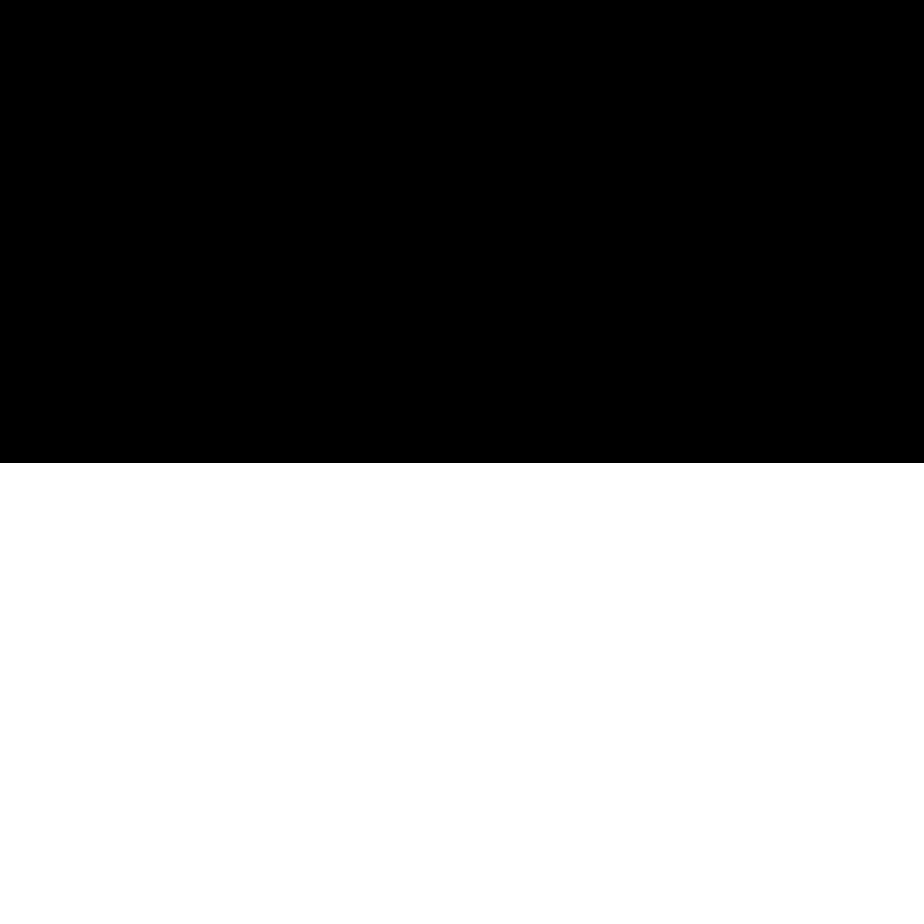}}\\
				\vspace{-2mm}
				{\footnotesize t=0}
			\end{minipage}
			\begin{minipage}{0.14\linewidth}
				\centering
				\fbox{\includegraphics[width=\linewidth]{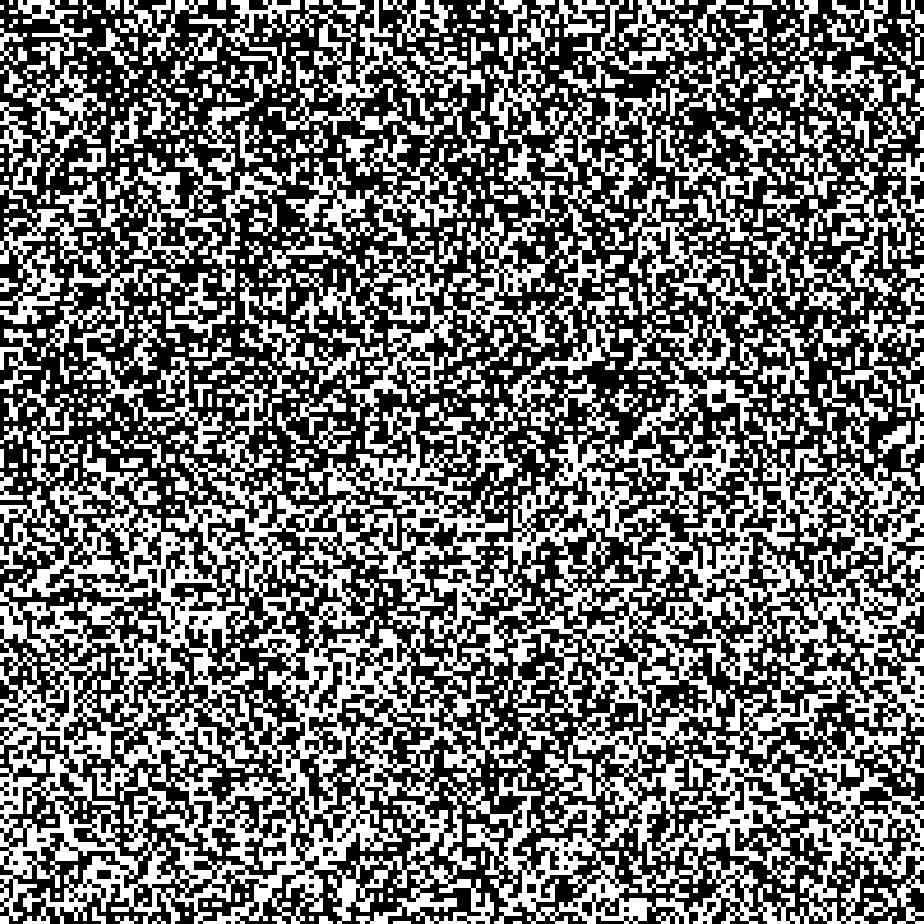}}\\
				\vspace{-2mm}
				{\footnotesize t=1}
			\end{minipage}
			\begin{minipage}{0.14\linewidth}
				\centering
				\fbox{\includegraphics[width=\linewidth]{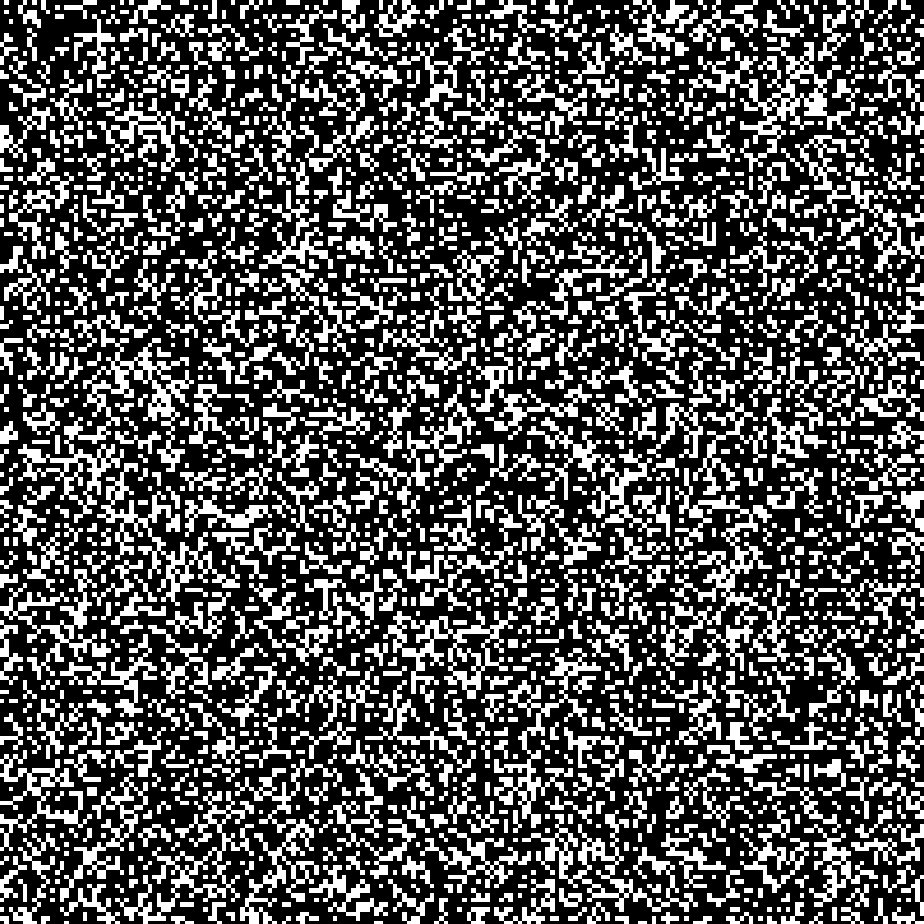}}\\
				\vspace{-2mm}
				{\footnotesize t=10}
			\end{minipage}
			\begin{minipage}{0.14\linewidth}
				\centering
				\fbox{\includegraphics[width=\linewidth]{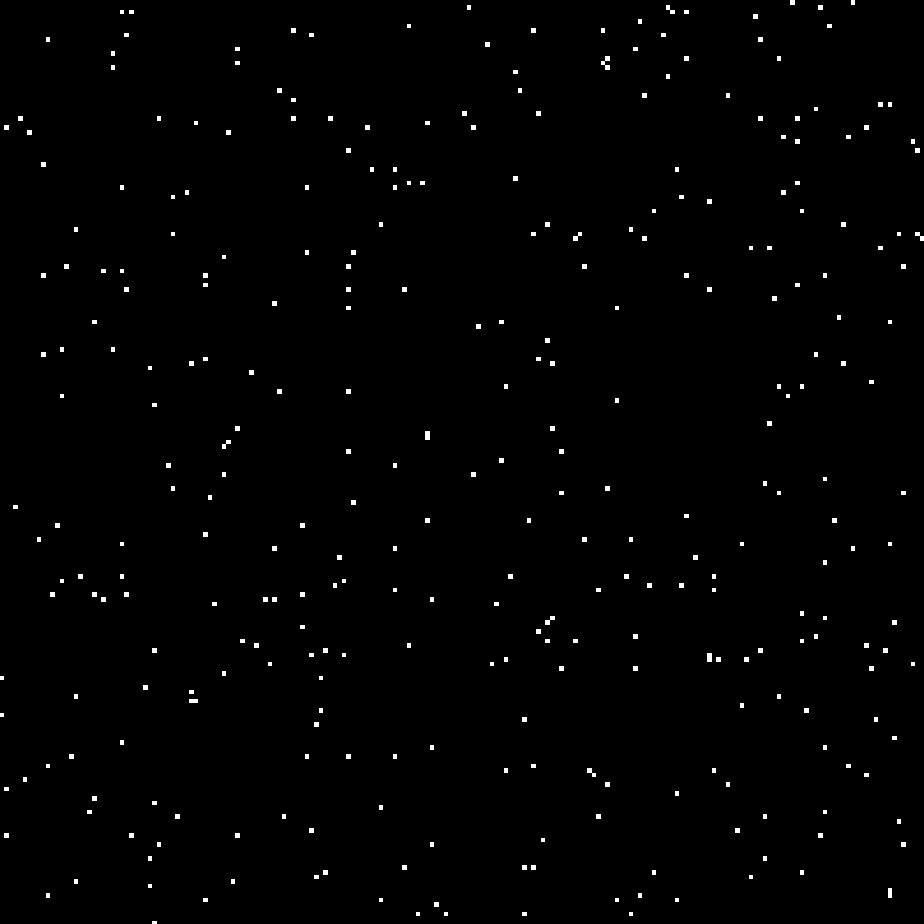}}\\
				\vspace{-2mm}
				{\footnotesize t=100}
			\end{minipage}
			\begin{minipage}{0.14\linewidth}
				\centering
				\fbox{\includegraphics[width=\linewidth]{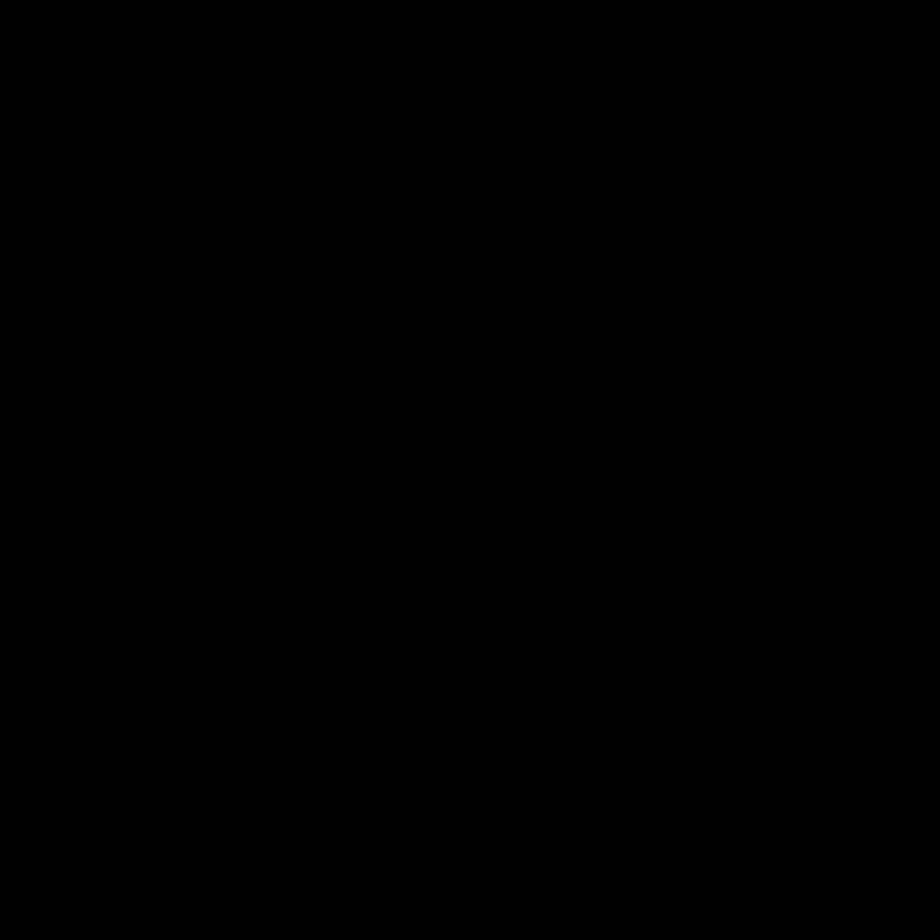}}\\
				\vspace{-2mm}
				{\footnotesize t=1000}
			\end{minipage}
			\vspace{-3mm}
			\caption*{\footnotesize (b) MAPPO}
		\end{minipage}
		\\[2mm]
		\begin{minipage}{\linewidth}
			\begin{minipage}{0.24\linewidth}
				\centering
				\includegraphics[width=\linewidth]{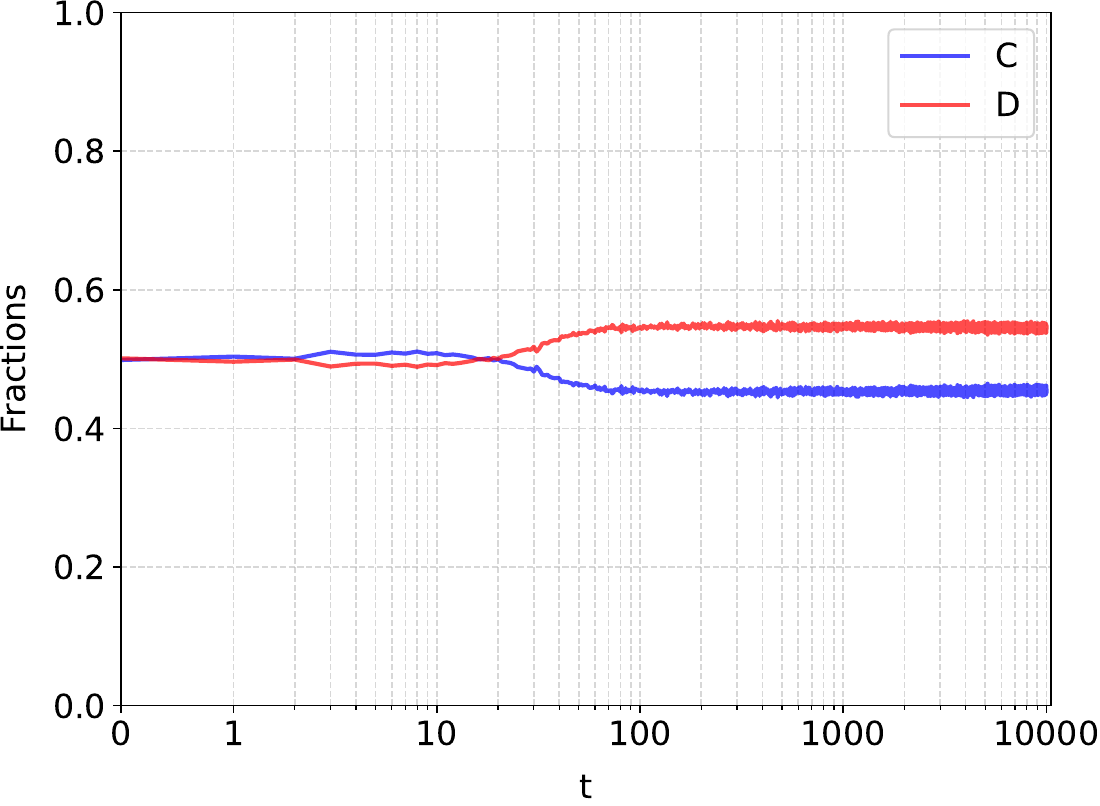}\\
			\end{minipage}
			\begin{minipage}{0.14\linewidth}
				\centering
				\fbox{\includegraphics[width=\linewidth]{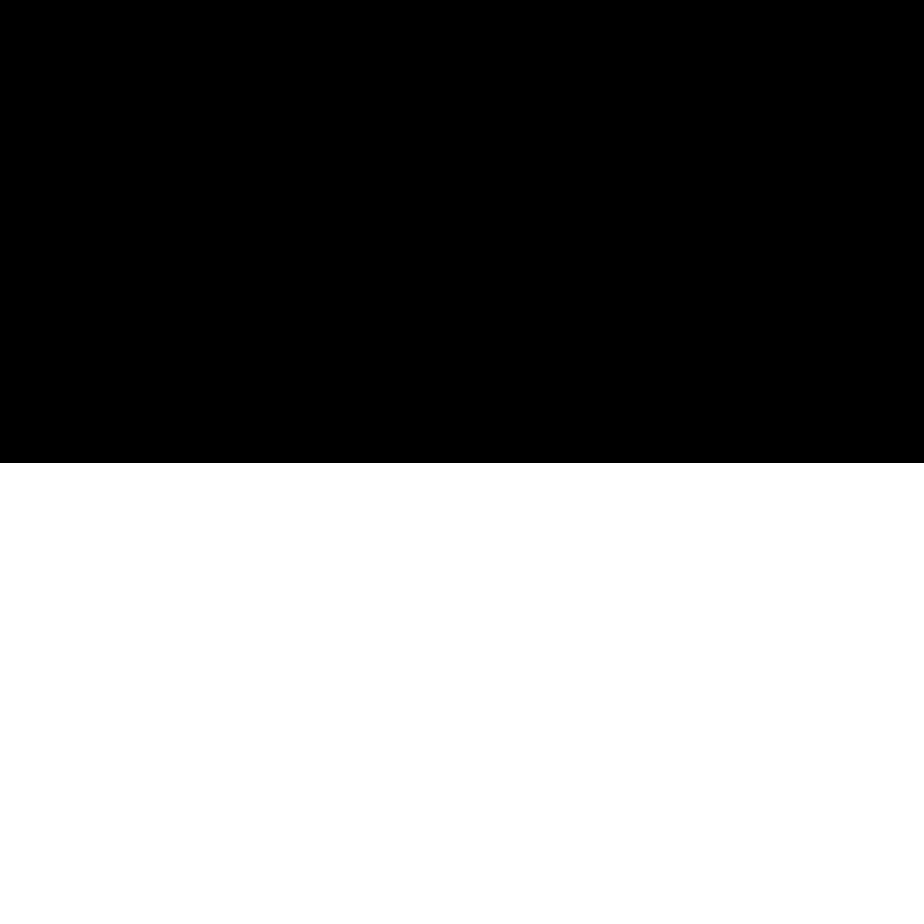}}\\
				\vspace{-2mm}
				{\footnotesize t=0}
			\end{minipage}
			\begin{minipage}{0.14\linewidth}
				\centering
				\fbox{\includegraphics[width=\linewidth]{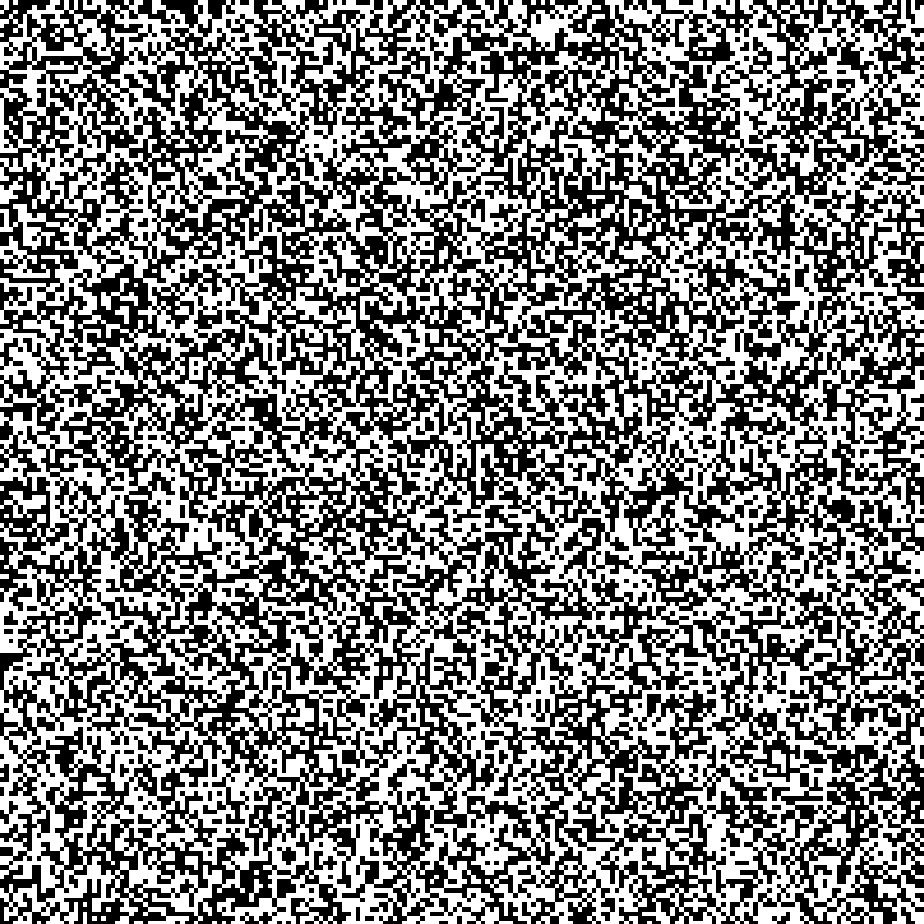}}\\
				\vspace{-2mm}
				{\footnotesize t=10}
			\end{minipage}
			\begin{minipage}{0.14\linewidth}
				\centering
				\fbox{\includegraphics[width=\linewidth]{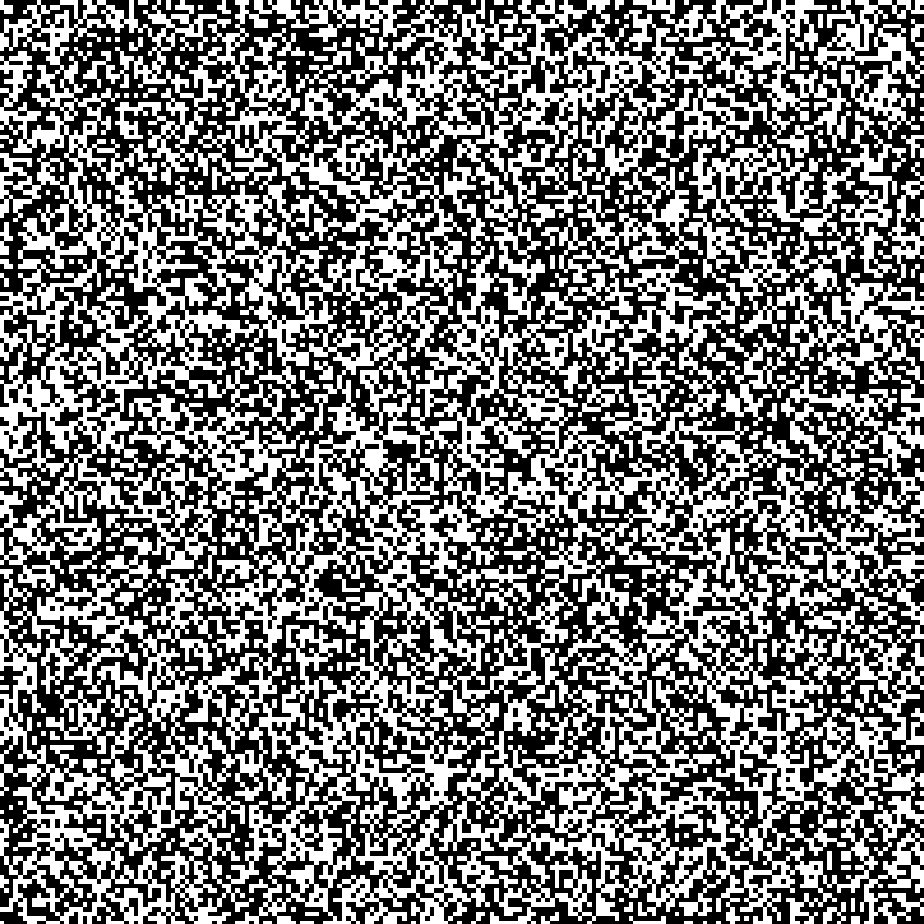}}\\
				\vspace{-2mm}
				{\footnotesize t=100}
			\end{minipage}
			\begin{minipage}{0.14\linewidth}
				\centering
				\fbox{\includegraphics[width=\linewidth]{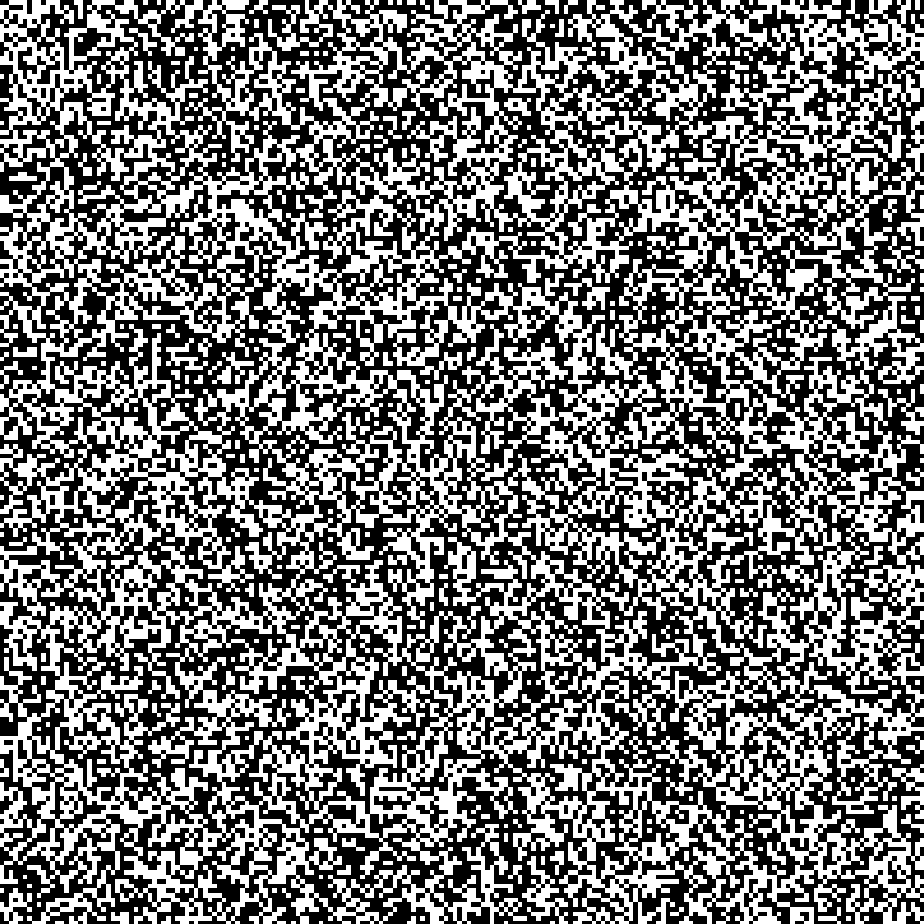}}\\
				\vspace{-2mm}
				{\footnotesize t=1000}
			\end{minipage}
			\begin{minipage}{0.14\linewidth}
				\centering
				\fbox{\includegraphics[width=\linewidth]{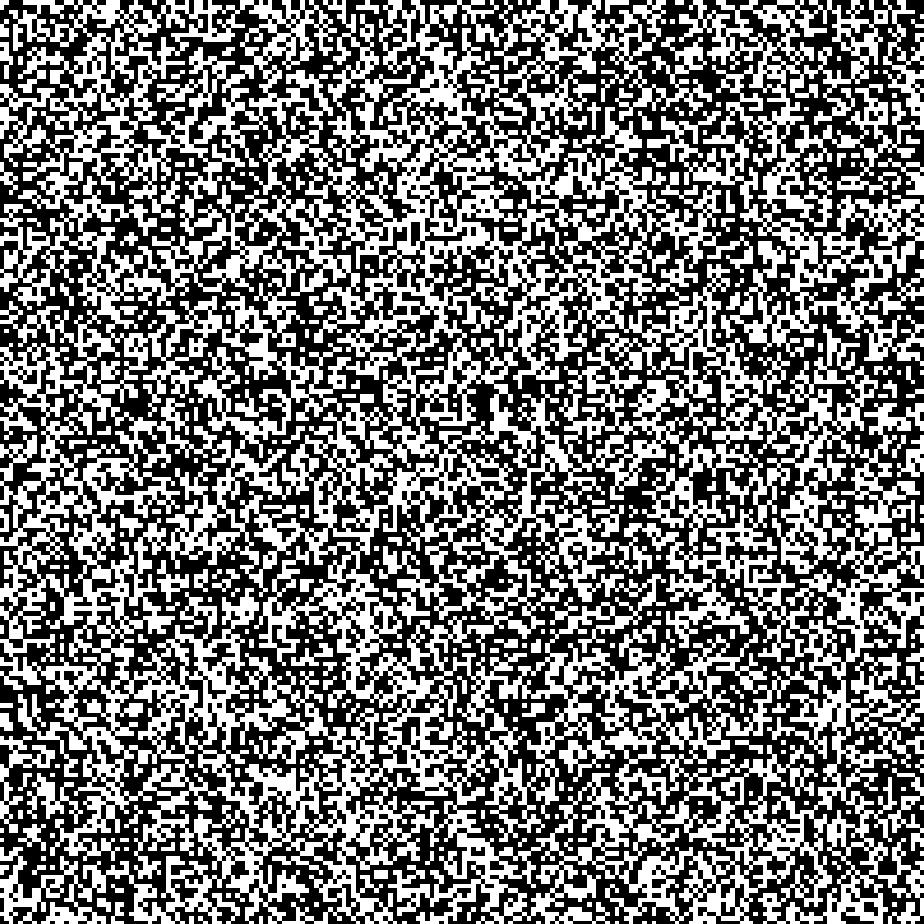}}\\
				\vspace{-2mm}
				{\footnotesize t=10000}
			\end{minipage}
			\vspace{-3mm}
			\caption*{\footnotesize (c) Q-learning}
		\end{minipage}	
		\\[2mm]
		\begin{minipage}{\linewidth}
			\begin{minipage}{0.24\linewidth}
				\centering
				\includegraphics[width=\linewidth]{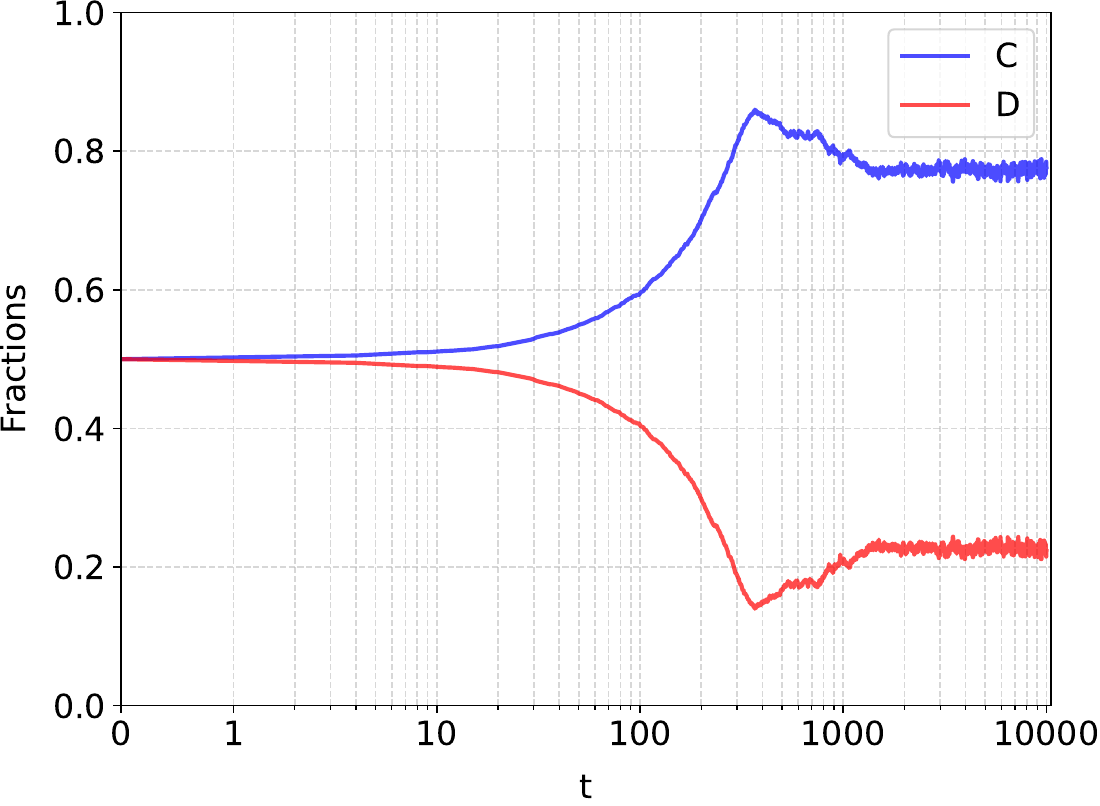}\\
			\end{minipage}
			\begin{minipage}{0.14\linewidth}
				\centering
				\fbox{\includegraphics[width=\linewidth]{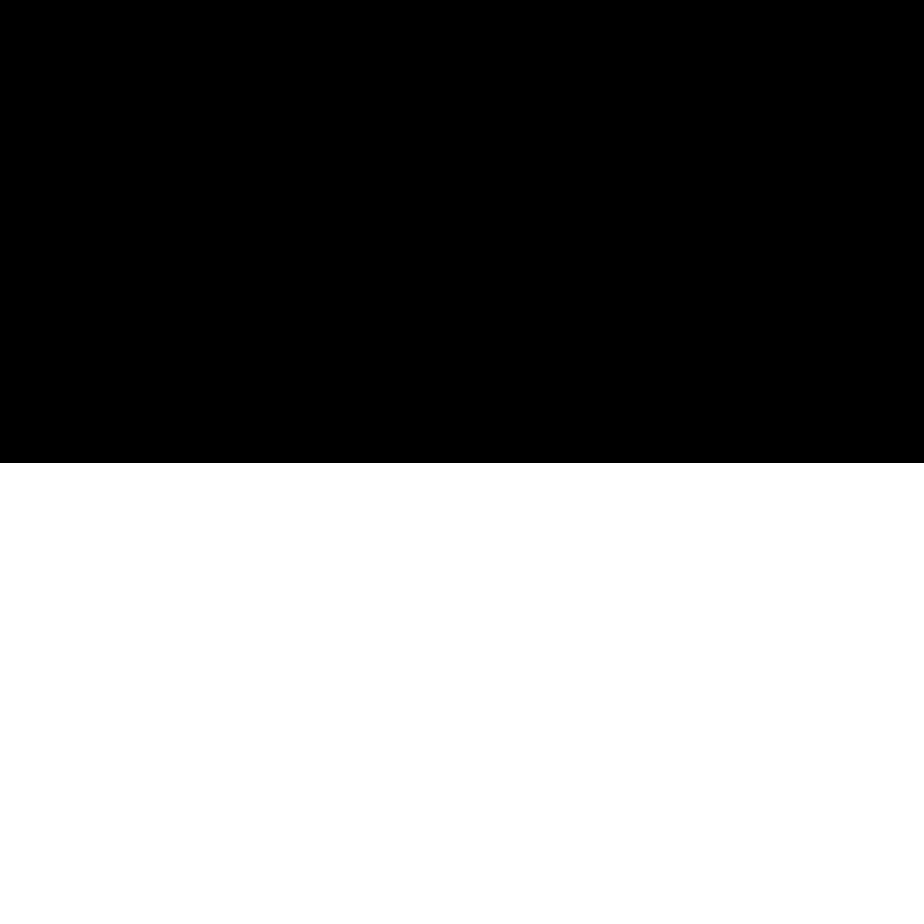}}\\
				\vspace{-2mm}
				{\footnotesize t=0}
			\end{minipage}
			\begin{minipage}{0.14\linewidth}
				\centering
				\fbox{\includegraphics[width=\linewidth]{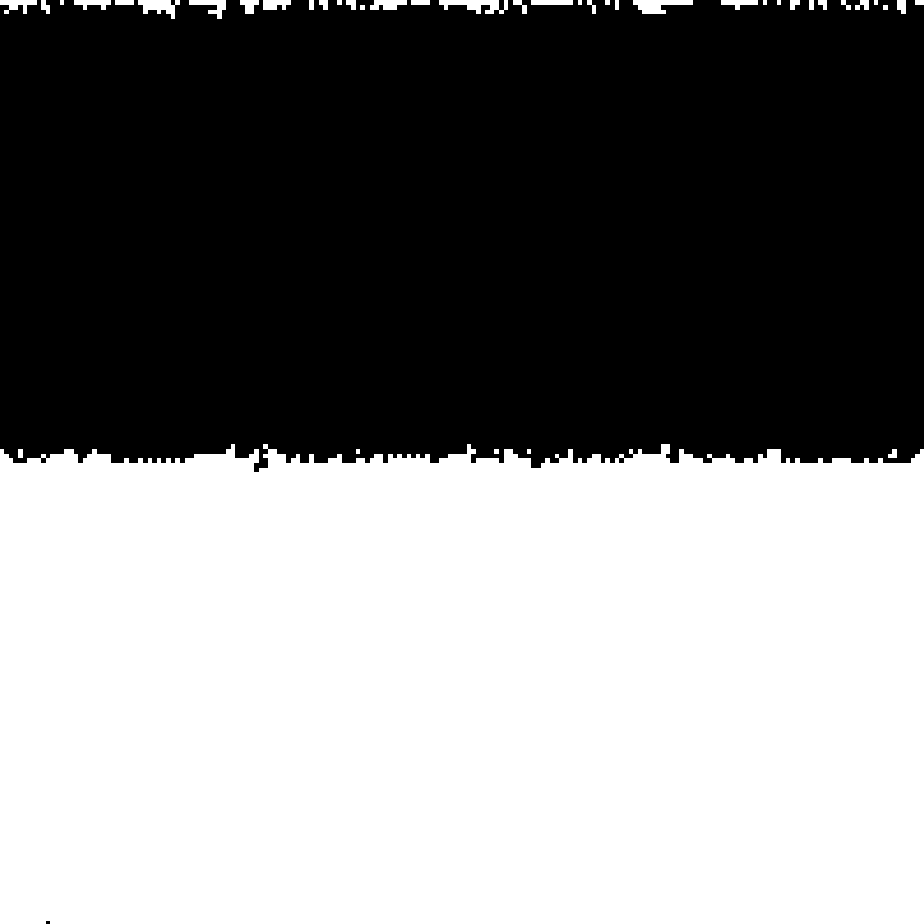}}\\
				\vspace{-2mm}
				{\footnotesize t=10}
			\end{minipage}
			\begin{minipage}{0.14\linewidth}
				\centering
				\fbox{\includegraphics[width=\linewidth]{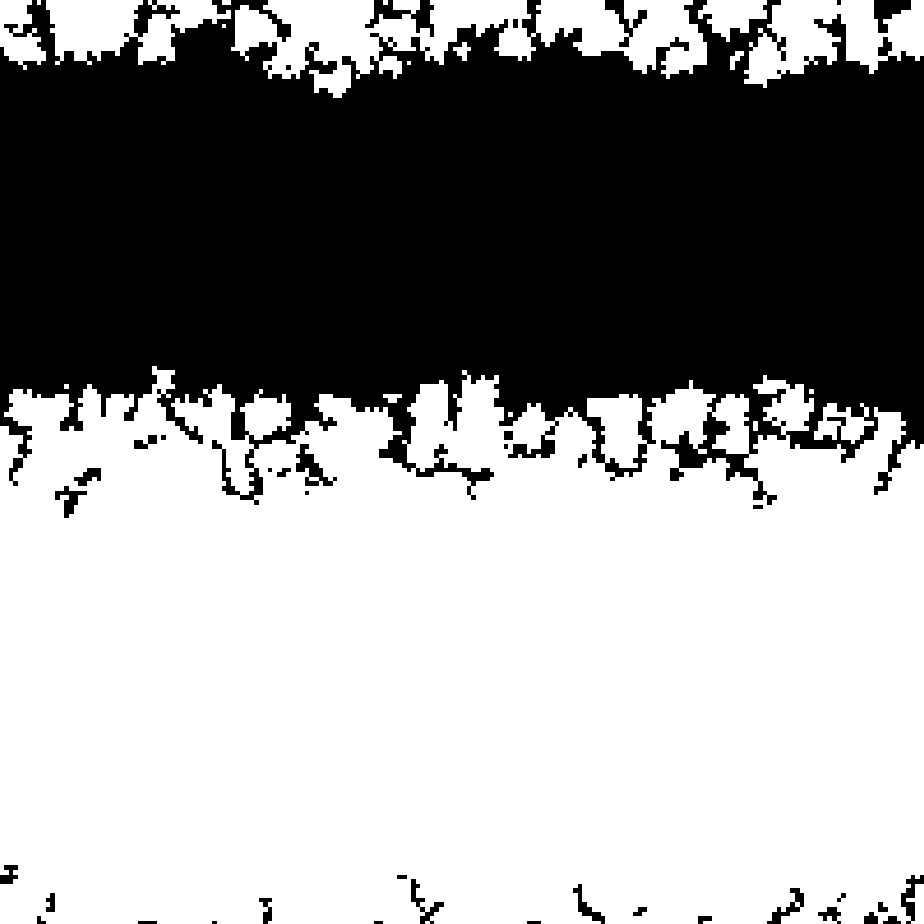}}\\
				\vspace{-2mm}
				{\footnotesize t=100}
			\end{minipage}
			\begin{minipage}{0.14\linewidth}
				\centering
				\fbox{\includegraphics[width=\linewidth]{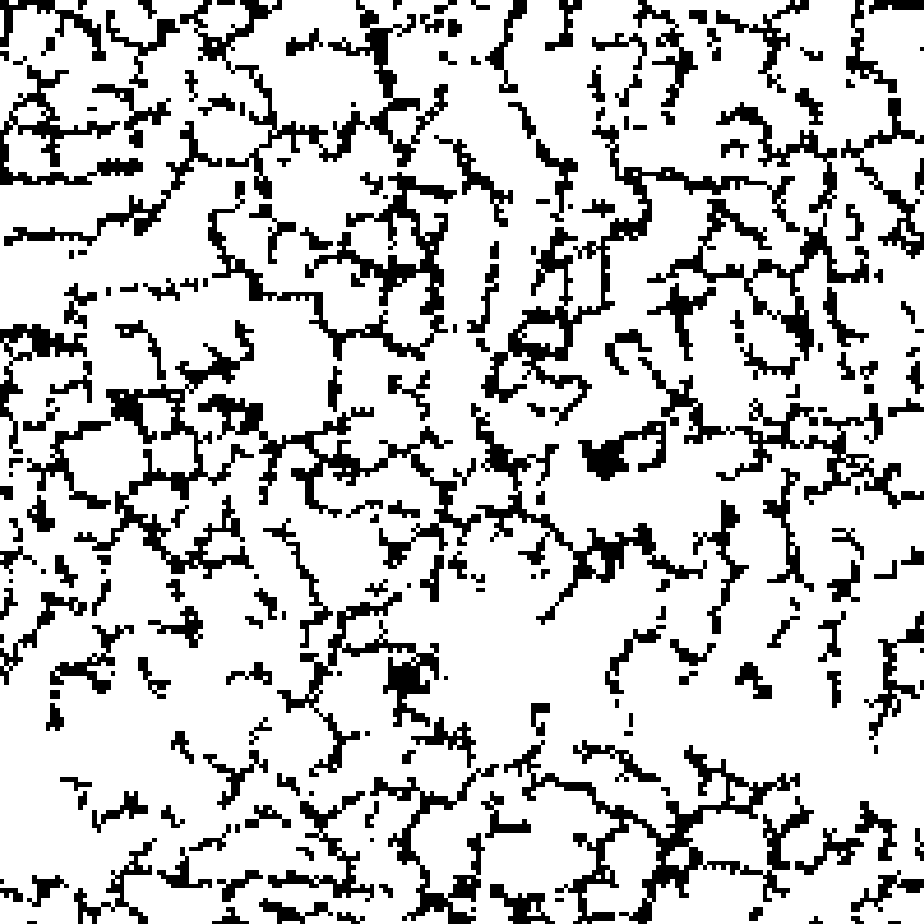}}\\
				\vspace{-2mm}
				{\footnotesize t=1000}
			\end{minipage}
			\begin{minipage}{0.14\linewidth}
				\centering
				\fbox{\includegraphics[width=\linewidth]{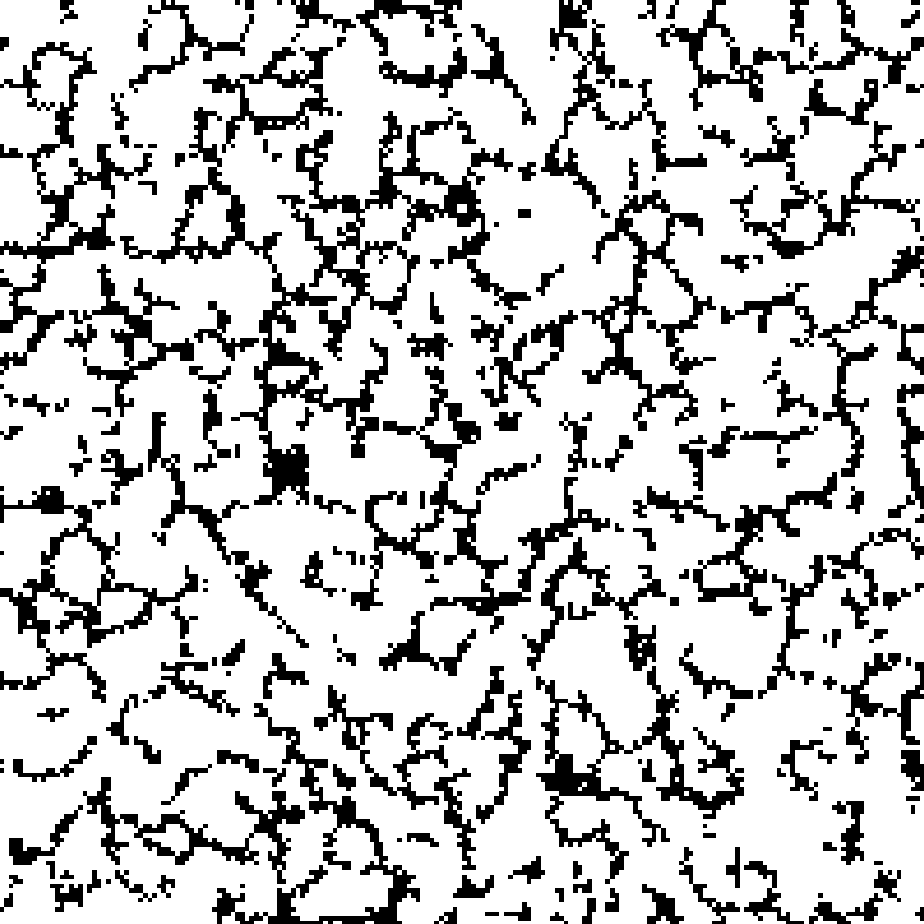}}\\
				\vspace{-2mm}
				{\footnotesize t=10000}
			\end{minipage}
			\vspace{-3mm}
			\caption*{\footnotesize (d) Fermi update rule}
		\end{minipage}	
		\caption{Comparative dynamics of four update rules at $r=4.4$. Rows (a)--(d) correspond to MAPPO-LCR, MAPPO, Q-learning, and the Fermi update rule, respectively. In each row, the leftmost panel shows the temporal evolution of cooperation and defection fractions over time $t$ (blue for cooperators $C$, red for defectors $D$). The remaining five panels display strategy snapshots at representative iterations. White cells indicate cooperators, and black cells indicate defectors, highlighting the different spatial organization patterns induced by each algorithm.}
		\label{fig:exp_comp_alg}
	\end{figure*}

	We compare four representative update rules at an enhancement factor of $r=4.4$, including MAPPO-LCR, MAPPO, Q-learning, and the Fermi update rule.
	All methods adopt the same deterministic initialization, where defectors occupy the upper half and cooperators occupy the lower half. 
	Figure~\ref{fig:exp_comp_alg} summarizes their macroscopic dynamics and microscopic spatial patterns.
	For each algorithm, the leftmost panel shows the temporal evolution of cooperation and defection fractions. 
	The horizontal axis is time step $t$, and the vertical axis is the fraction of cooperators $C$ and defectors $D$. 
	Blue curves indicate $C$, while red curves indicate $D$. 
	The remaining five panels show strategy snapshots at representative iterations. White sites denote cooperators, and black sites denote defectors.
	
	As shown in Figure~\ref{fig:exp_comp_alg}.
	Under MAPPO-LCR, the cooperation fraction rises in a clear step-like manner after about 11 iterations. 
	The system reaches the all-cooperator state before $t=200$. This behavior follows from the shaped reward. 
	LCR amplifies gradients that align local policies with cooperative neighborhoods. 
	Once small cooperative regions appear, MAPPO-LCR quickly reinforces them through centralized value guidance and clipped policy updates.
	In contrast, MAPPO without LCR exhibits a smooth decline in cooperation. After roughly 100 iterations, all agents become defectors. 
	At $r=4.4$, the intrinsic public goods payoff does not fully offset the temptation to defect for the MAPPO critic. 
	The centralized value function therefore estimates higher long-term returns for defection, and gradient updates drive the population into the full-defection attractor.
	Q-learning produces an intermediate outcome. 
	The final fraction of defectors is slightly higher than the fraction of cooperators and remains stable over time. 
	This reflects the more local, value-based learning process. 
	At $r=4.4$, individual Q-updates favor neither pure cooperation nor pure defection. 
	The population settles near a mixed configuration where exploitation and cooperation coexist.
	The Fermi update rule shows a distinct pattern. 
	Cooperators remain more numerous than defectors across the entire run. 
	After cooperators and defectors become well mixed, the cooperation level decreases slightly and then stabilizes around $76\%$. 
	Spatial snapshots reveal that defector regions form elongated, filament-like clusters. 
	This is consistent with spatial reciprocity theory. 
	Clustered cooperators protect themselves along interior sites, while defectors survive mainly on thin interfaces.
	Overall, MAPPO-LCR demonstrates the strongest tendency toward full cooperation at $r=4.4$. 
	MAPPO collapses to defection, Q-learning stabilizes in a mixed state, and the Fermi rule maintains a high but incomplete cooperation plateau.

	\subsection{Algorithm performance evaluation under varying enhancement factors $r$}
	\label{exp_r}
	
\begin{figure*}[h]
		\begin{minipage}{0.48\linewidth}
			\centering
			\includegraphics[width=\linewidth]{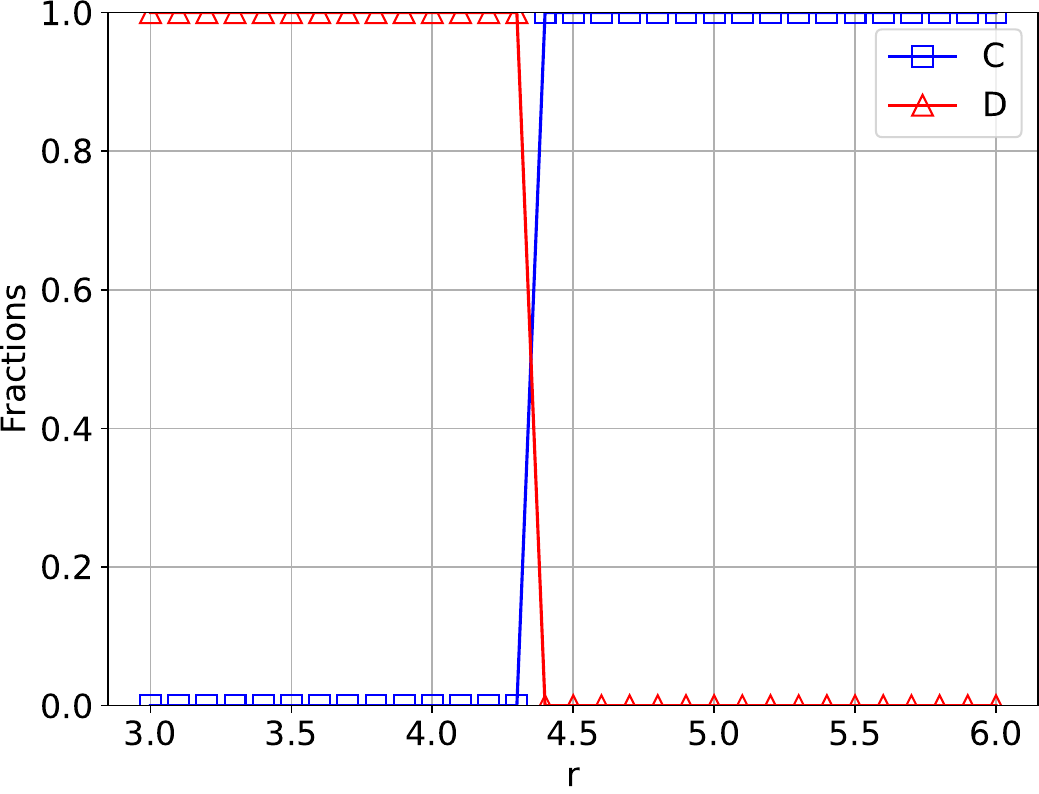}\\
			\vspace{-2mm}
			\caption*{\footnotesize (a) MAPPO-LCR}
		\end{minipage}
		\begin{minipage}{0.48\linewidth}
			\centering
			\includegraphics[width=\linewidth]{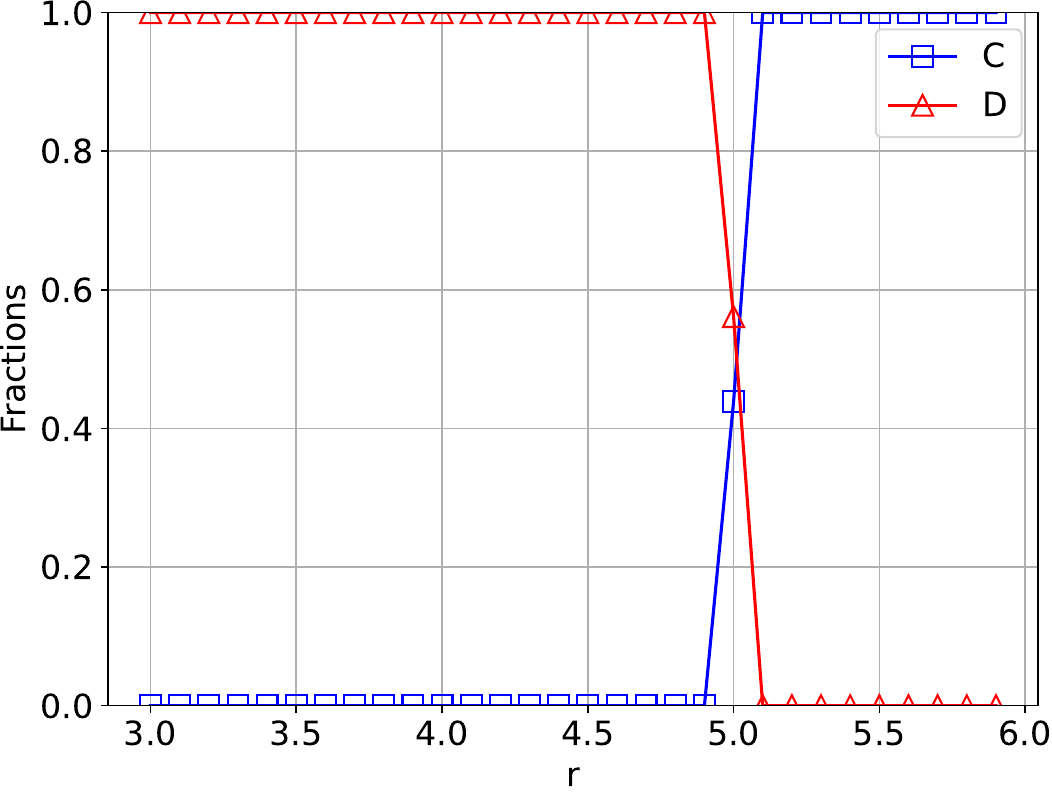}\\
			\vspace{-2mm}
			\caption*{\footnotesize (b) MAPPO}
		\end{minipage}
		\\
		[3mm]
		\begin{minipage}{0.48\linewidth}
			\centering
			\includegraphics[width=\linewidth]{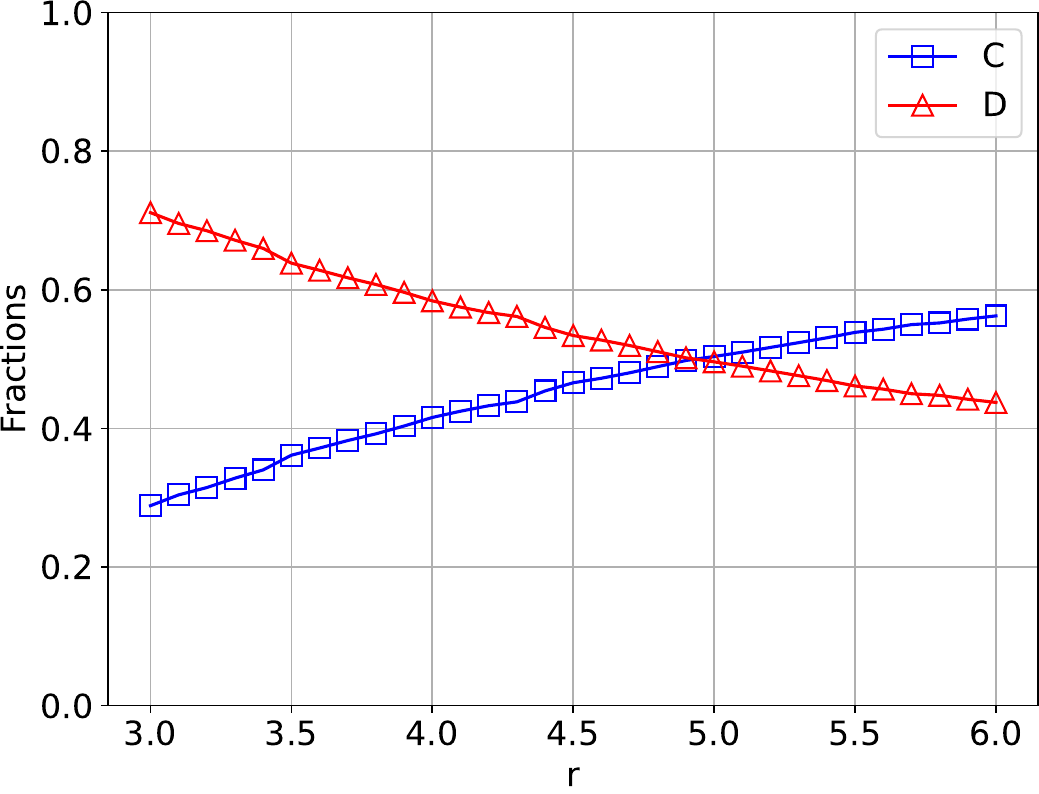}\\
			\vspace{-2mm}
			\caption*{\footnotesize (c) Q-learing}
		\end{minipage}	
		\begin{minipage}{0.48\linewidth}
			\centering
			\includegraphics[width=\linewidth]{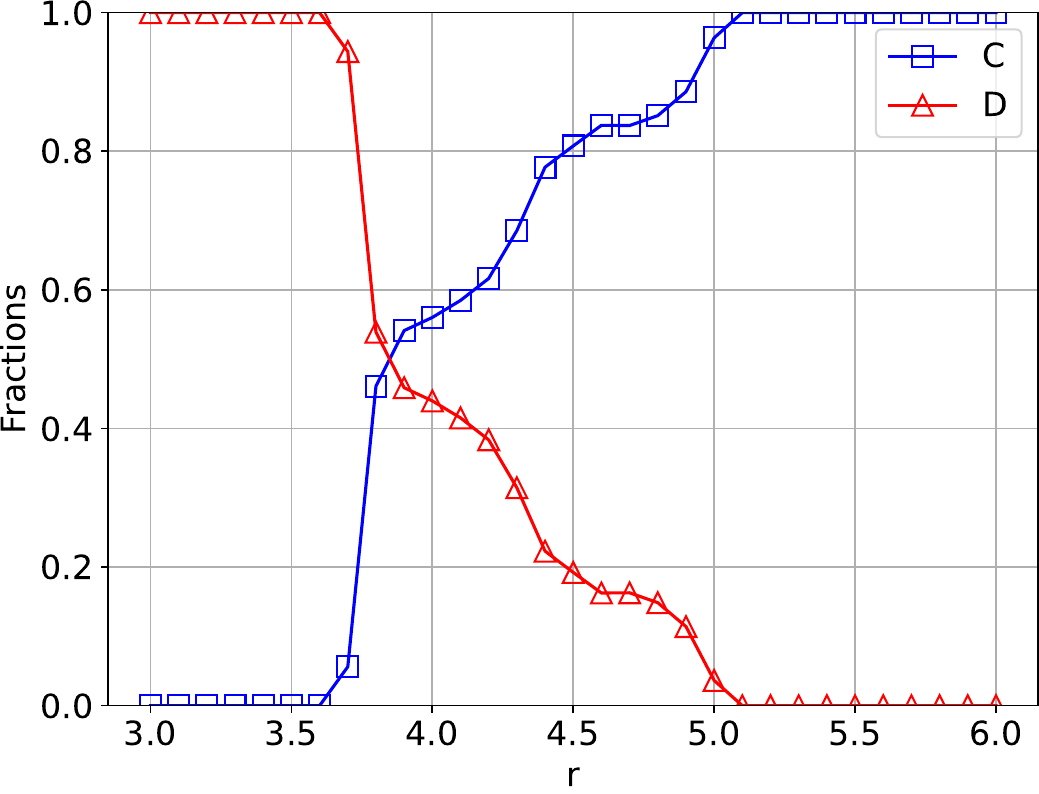}\\
			\vspace{-2mm}
			\caption*{\footnotesize (d) Fermi update rule}
		\end{minipage}	
		\caption{Final cooperation and defection fractions across enhancement factor
			$r$ for MAPPO-LCR, MAPPO, Q-learning, and the Fermi update rule. Blue squares
			indicate cooperators and red triangles indicate defectors.}
		\label{fig:exp_r}
	\end{figure*}

	Figure~\ref{fig:exp_r} presents the final cooperation and defection fractions of four algorithms under different values of the enhancement factor $r$. 
	Each subfigure reports the fractions of cooperators and defectors, where the horizontal axis represents $r$ and the vertical axis represents the fractions.
	Blue squares denote cooperators and red triangles denote defectors.
	
	As shown in Figure~\ref{fig:exp_r}.
	MAPPO-LCR reaches full cooperation when $r\ge4.4$, and full defection appears when $r<4.4$. 
	This transition arises from the combination of centralized training and local shaping. 
	The centralized critic stabilizes gradient signals at the population level, while the LCR term increases the influence of local cooperative feedback. 
	The combined effect reduces the payoff requirement for cooperation to remain stable.
	MAPPO exhibits a different threshold. When $r<5.0$, the population converges to full defection. 
	When $r>5.0$, full cooperation emerges. At $r=5.0$, the two strategies remain in comparable proportions. The behavior reflects the fact that MAPPO relies solely on the centralized critic. 
	Without local shaping, agents require stronger cooperative payoffs before cooperative actions become self-reinforcing.
	Q-learning produces cooperative behavior even when $r$ is small. 
	The cooperation fraction increases with $r$, yet it does not exceed 60\% even at $r=6.0$. 
	The pattern follows from the tabular structure of Q-learning. 
	Agents depend on local rewards only, and global cooperation signals are weak. 
	As a result, partial cooperation appears early, while full cooperation remains difficult to maintain.
	The Fermi update rule displays cooperation at even smaller $r$. 
	Cooperation emerges when $r\ge3.7$, and full cooperation appears when $r\ge5.1$. 
	This trend results from the imitation mechanism. 
	Agents tend to adopt strategies with higher local payoffs, which enlarges local cooperative groups even when the global incentives are not strong.

	\subsection{MAPPO-LCR with half-and-half initialization}
	\label{exp_hh}

\begin{figure*}[htbp!]
		\begin{minipage}{0.47\linewidth}
			\begin{minipage}{\linewidth}
				\centering
				\includegraphics[width=\linewidth]{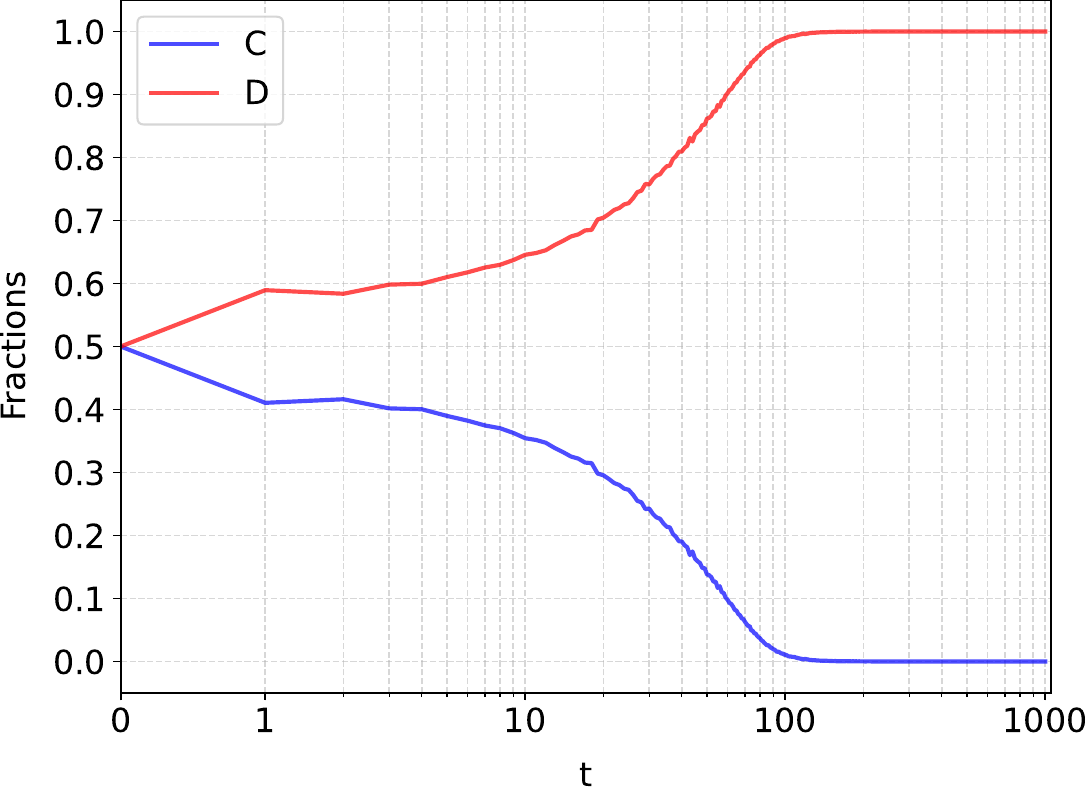}\\
			\end{minipage}
			\vspace{2mm}
			\\
			\begin{minipage}{0.188\linewidth}
				\centering
				\fbox{\includegraphics[width=\linewidth]{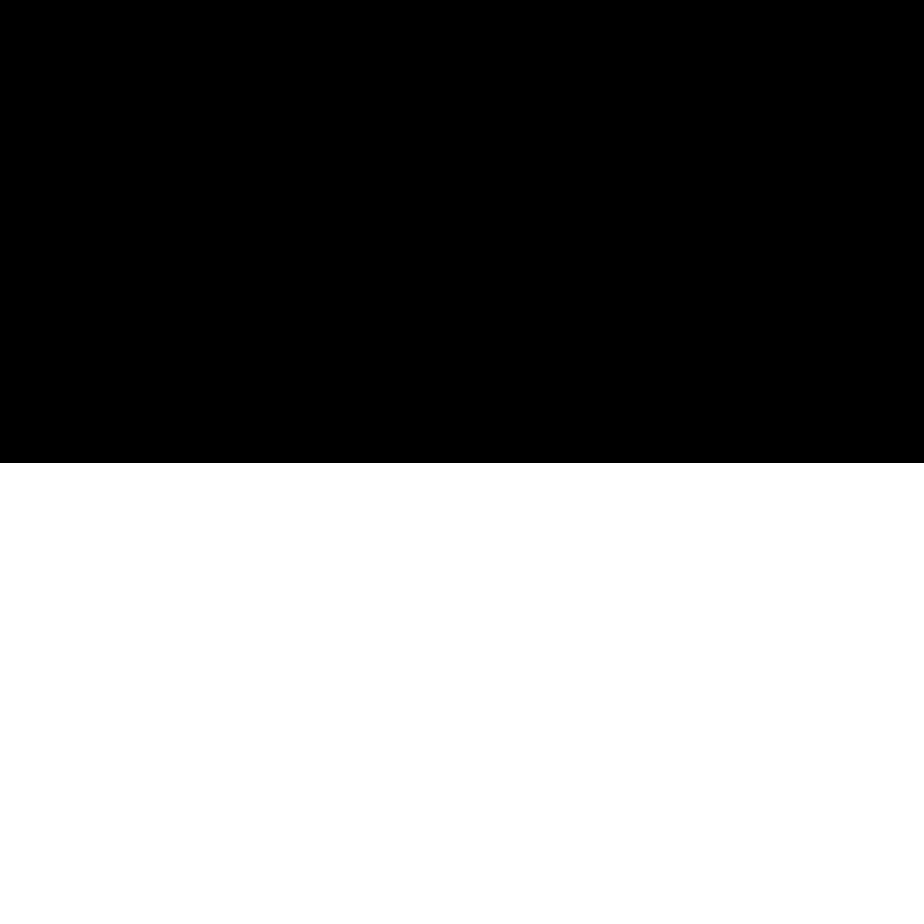}}\\
				\vspace{-2mm}
				{\footnotesize t=0}
			\end{minipage}
			\begin{minipage}{0.188\linewidth}
				\centering
				\fbox{\includegraphics[width=\linewidth]{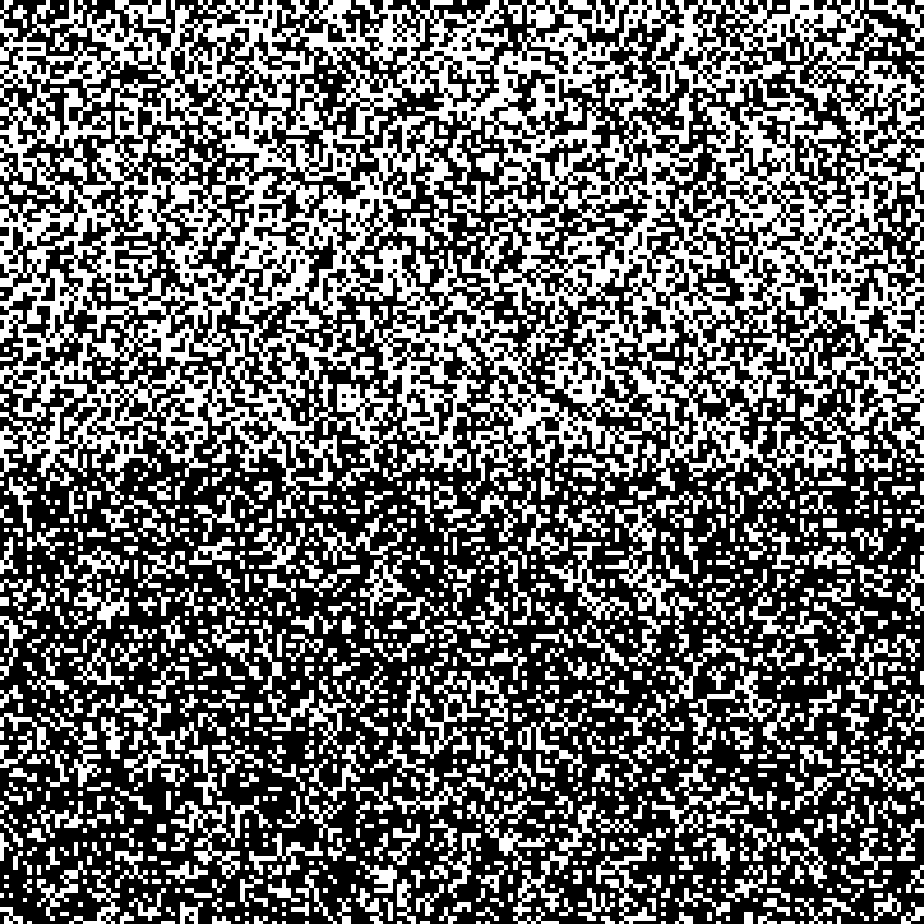}}\\
				\vspace{-2mm}
				{\footnotesize t=1}
			\end{minipage}
			\begin{minipage}{0.188\linewidth}
				\centering
				\fbox{\includegraphics[width=\linewidth]{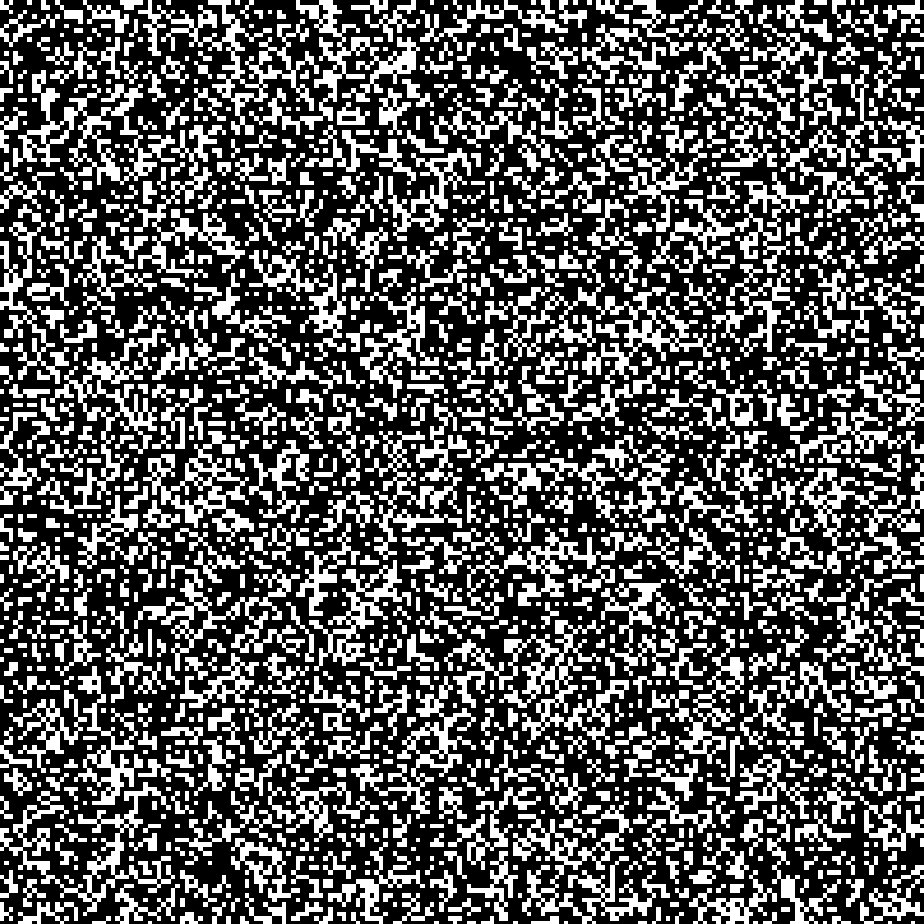}}\\
				\vspace{-2mm}
				{\footnotesize t=10}
			\end{minipage}
			\begin{minipage}{0.188\linewidth}
				\centering
				\fbox{\includegraphics[width=\linewidth]{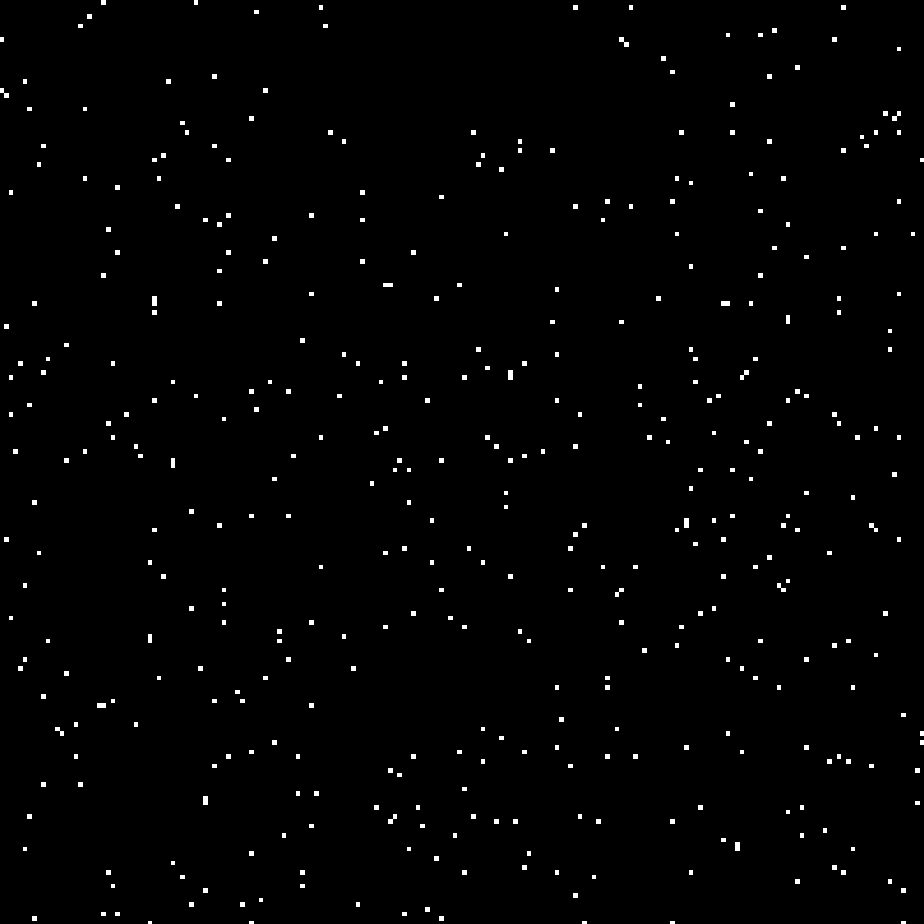}}\\
				\vspace{-2mm}
				{\footnotesize t=100}
			\end{minipage}
			\begin{minipage}{0.188\linewidth}
				\centering
				\fbox{\includegraphics[width=\linewidth]{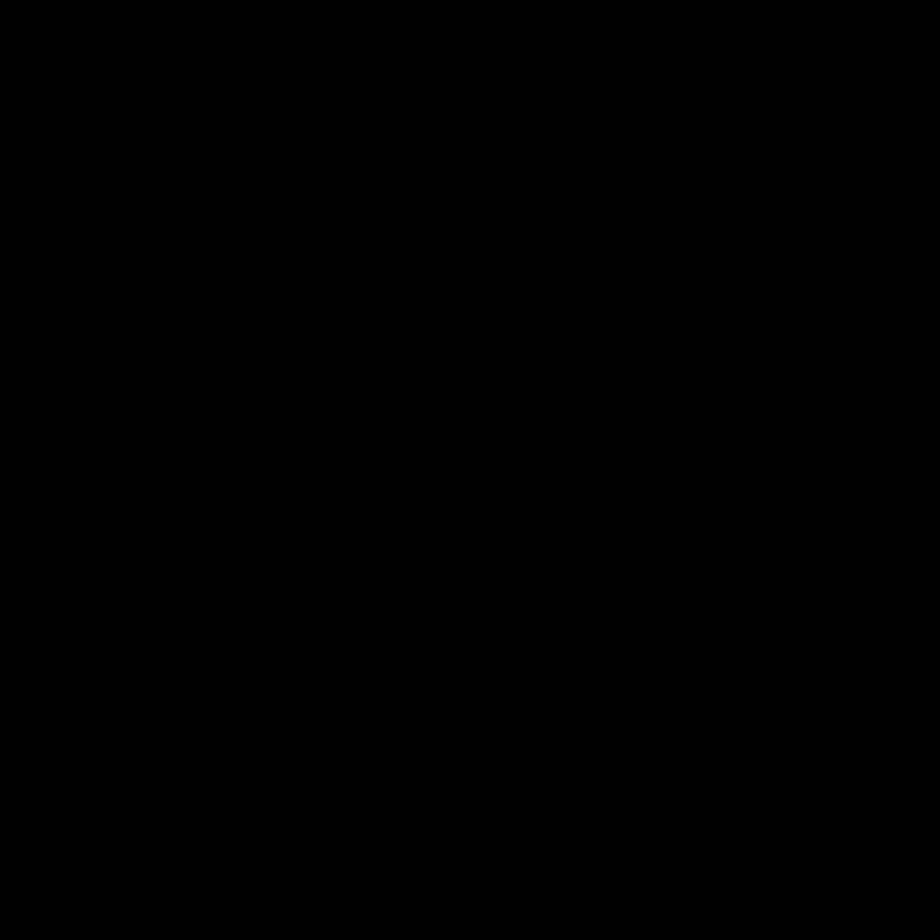}}\\
				\vspace{-2mm}
				{\footnotesize t=1000}
			\end{minipage}
			\vspace{-2mm}
			\caption*{\footnotesize (a) r=4.1}
		\end{minipage}
		\hfill
		\begin{minipage}{0.47\linewidth}
			\begin{minipage}{\linewidth}
				\centering
				\includegraphics[width=\linewidth]{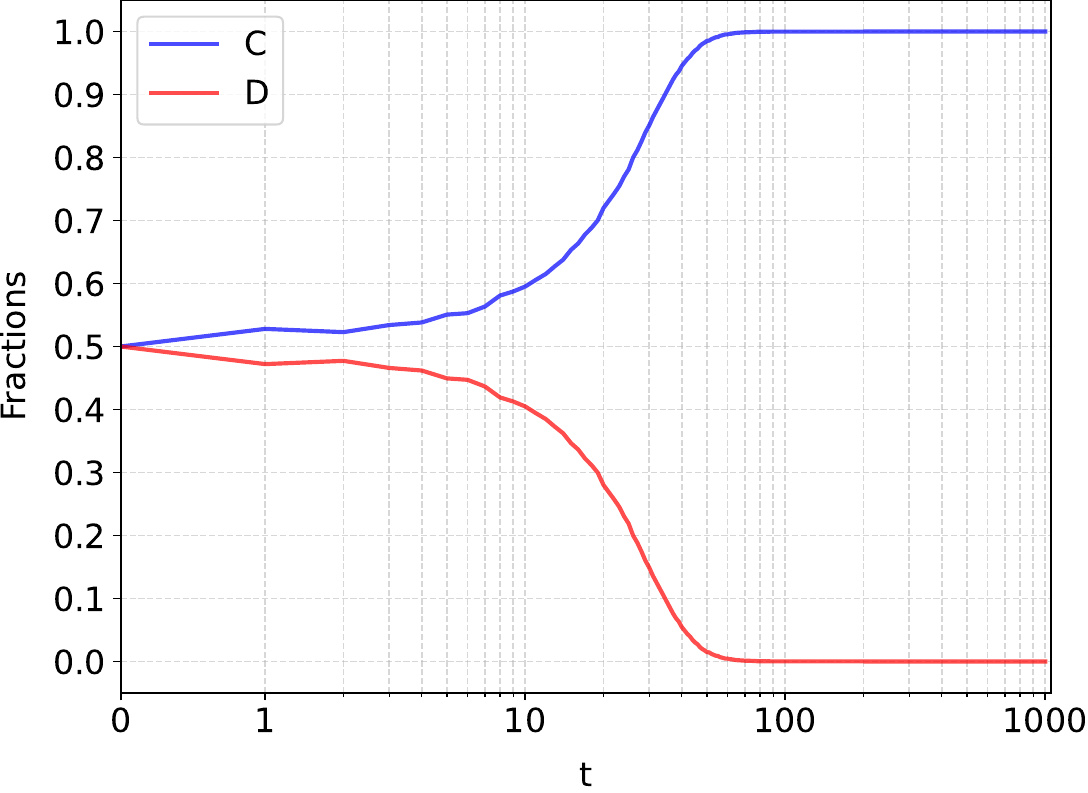}\\
			\end{minipage}
			\vspace{2mm}
			\\
			\begin{minipage}{0.188\linewidth}
				\centering
				\fbox{\includegraphics[width=\linewidth]{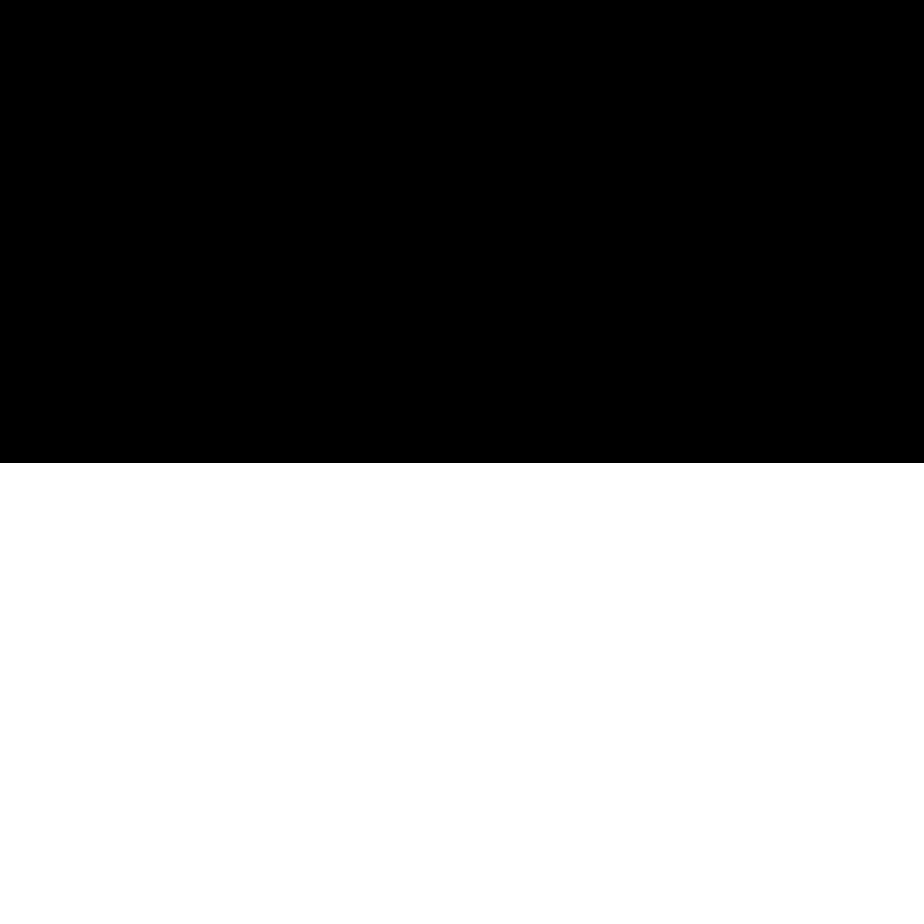}}\\
				\vspace{-2mm}
				{\footnotesize t=0}
			\end{minipage}
			\begin{minipage}{0.188\linewidth}
				\centering
				\fbox{\includegraphics[width=\linewidth]{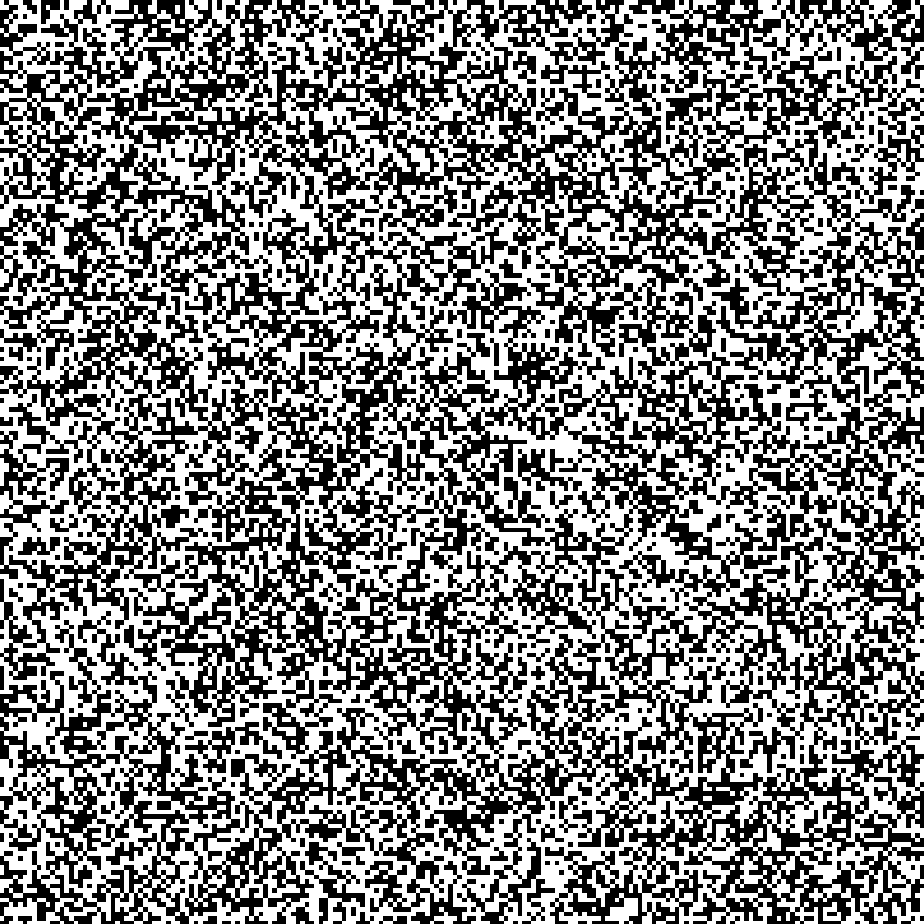}}\\
				\vspace{-2mm}
				{\footnotesize t=1}
			\end{minipage}
			\begin{minipage}{0.188\linewidth}
				\centering
				\fbox{\includegraphics[width=\linewidth]{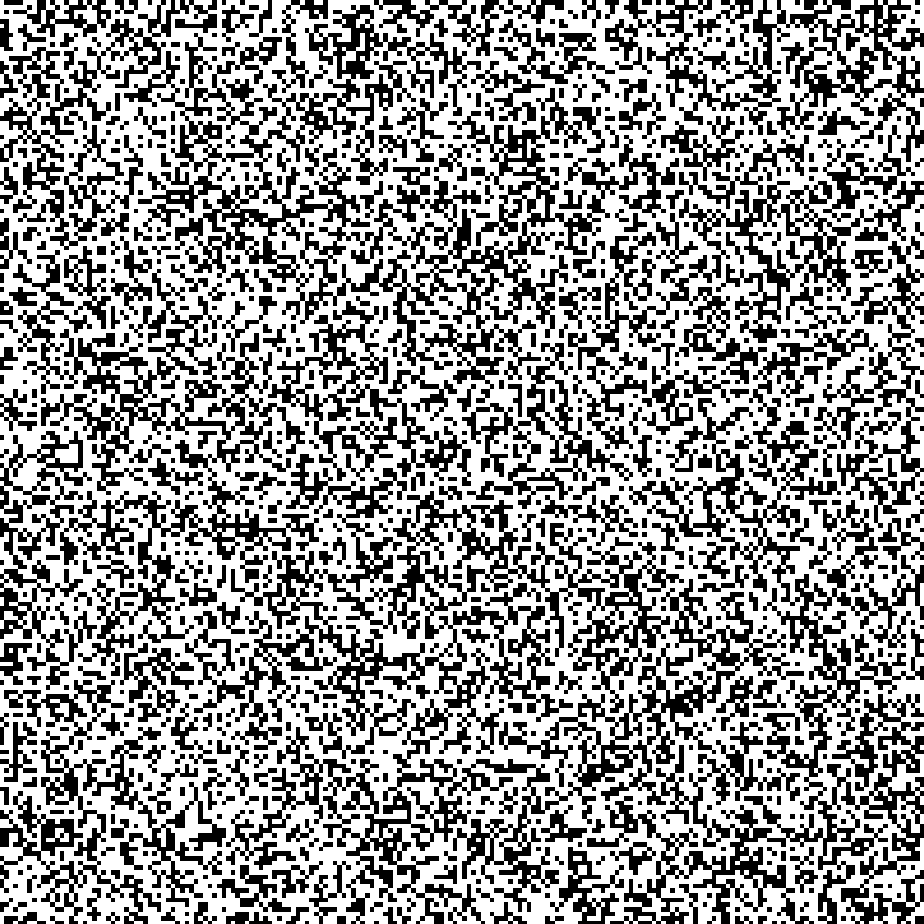}}\\
				\vspace{-2mm}
				{\footnotesize t=10}
			\end{minipage}
			\begin{minipage}{0.188\linewidth}
				\centering
				\fbox{\includegraphics[width=\linewidth]{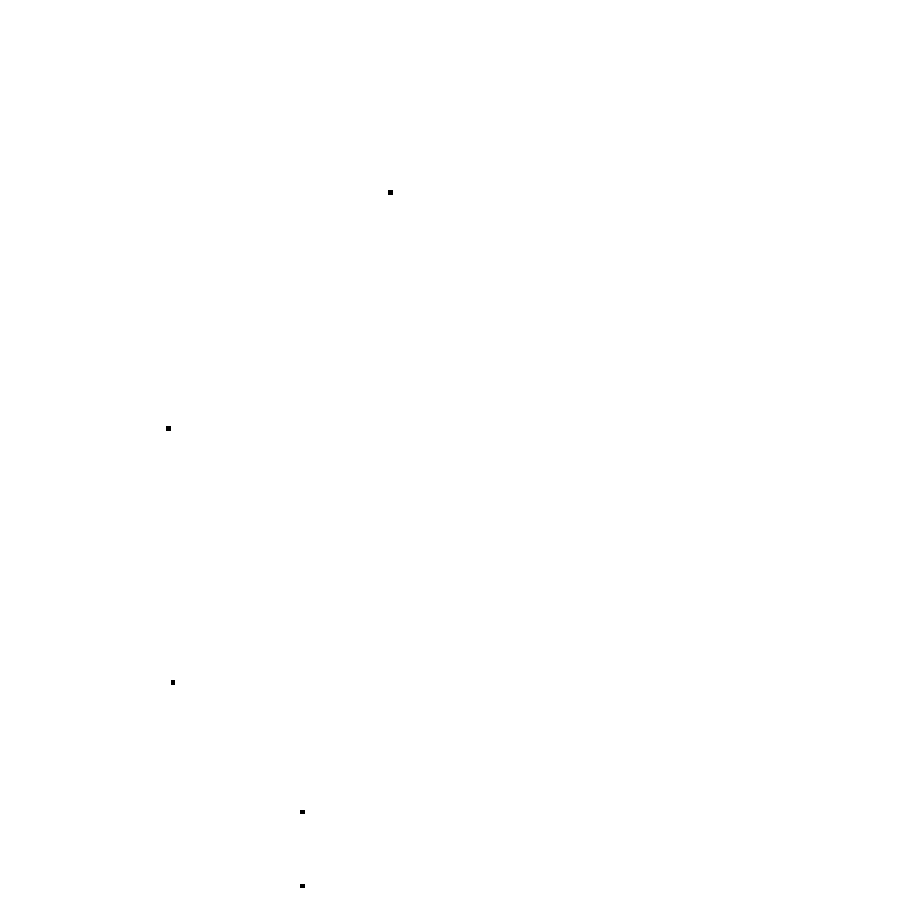}}\\
				\vspace{-2mm}
				{\footnotesize t=100}
			\end{minipage}
			\begin{minipage}{0.188\linewidth}
				\centering
				\fbox{\includegraphics[width=\linewidth]{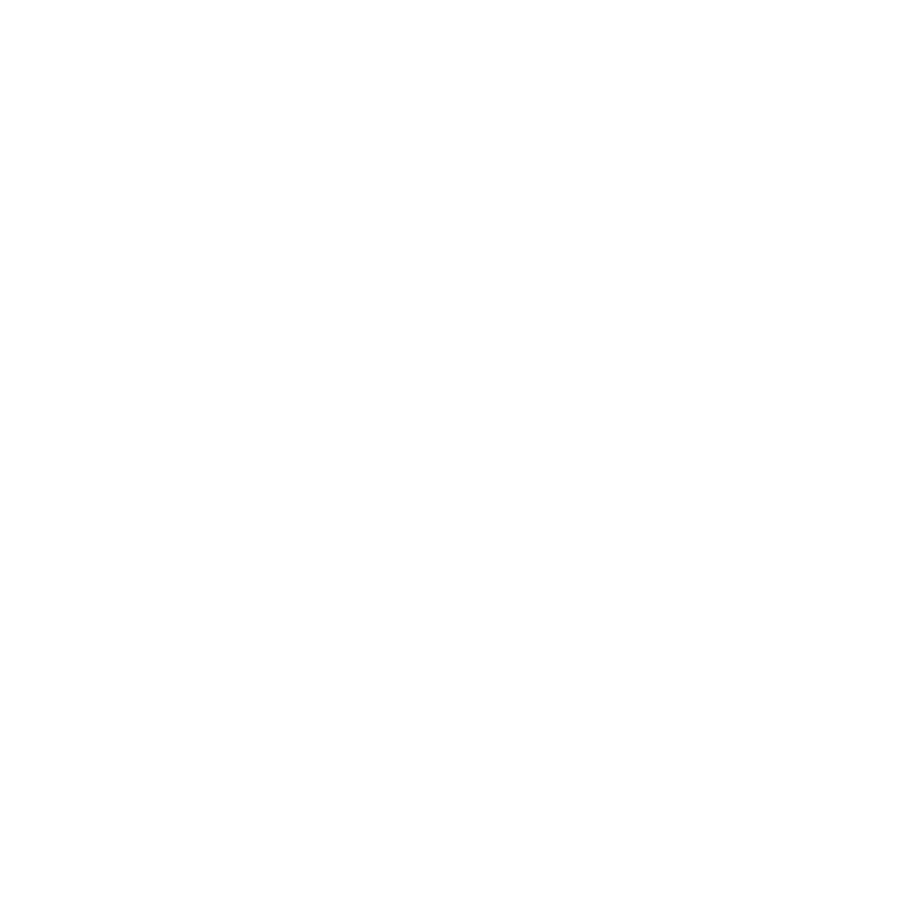}}\\
				\vspace{-2mm}
				{\footnotesize t=1000}
			\end{minipage}
			\vspace{-2mm}
			\caption*{\footnotesize (b) r=4.7}
		\end{minipage}
		\\
		[2mm]
		\begin{minipage}{\linewidth}
				\begin{minipage}{0.188\linewidth}
					\centering
					\includegraphics[width=\linewidth]{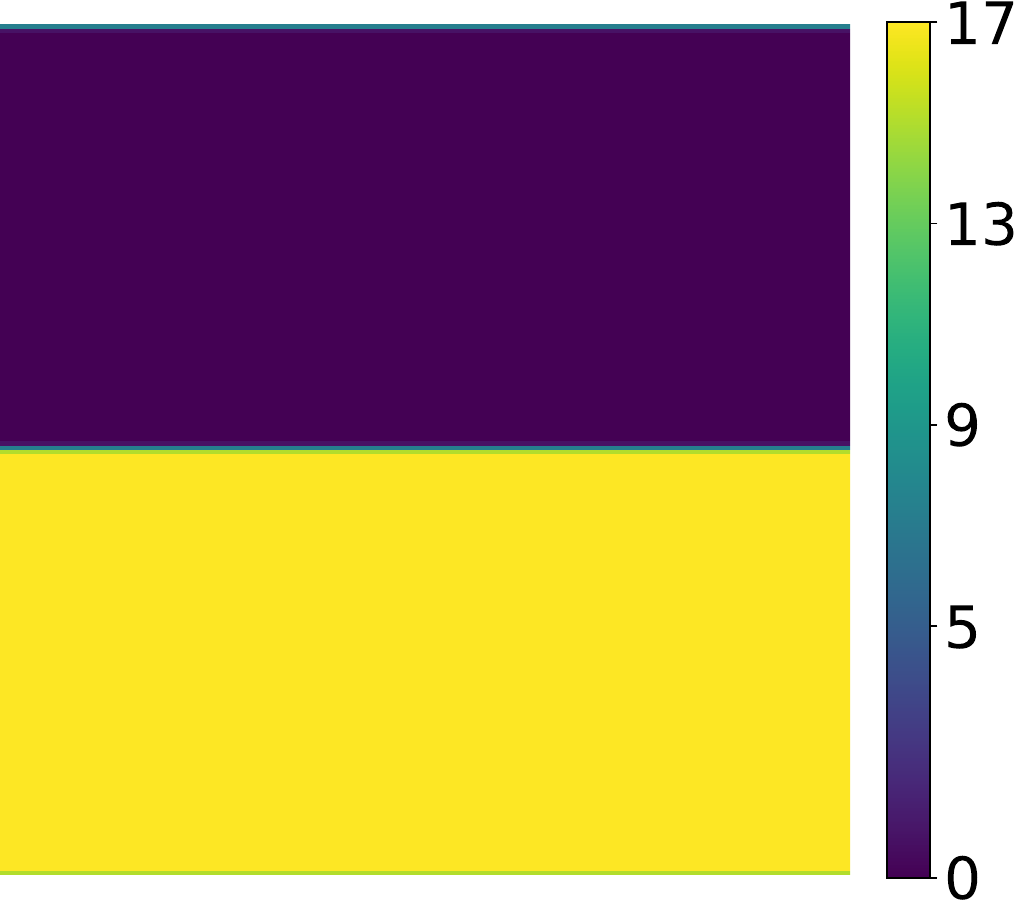}\\
					\vspace{-2mm}
					{\footnotesize t=0}
				\end{minipage}
				\hfill
				\begin{minipage}{0.188\linewidth}
					\centering
					\includegraphics[width=\linewidth]{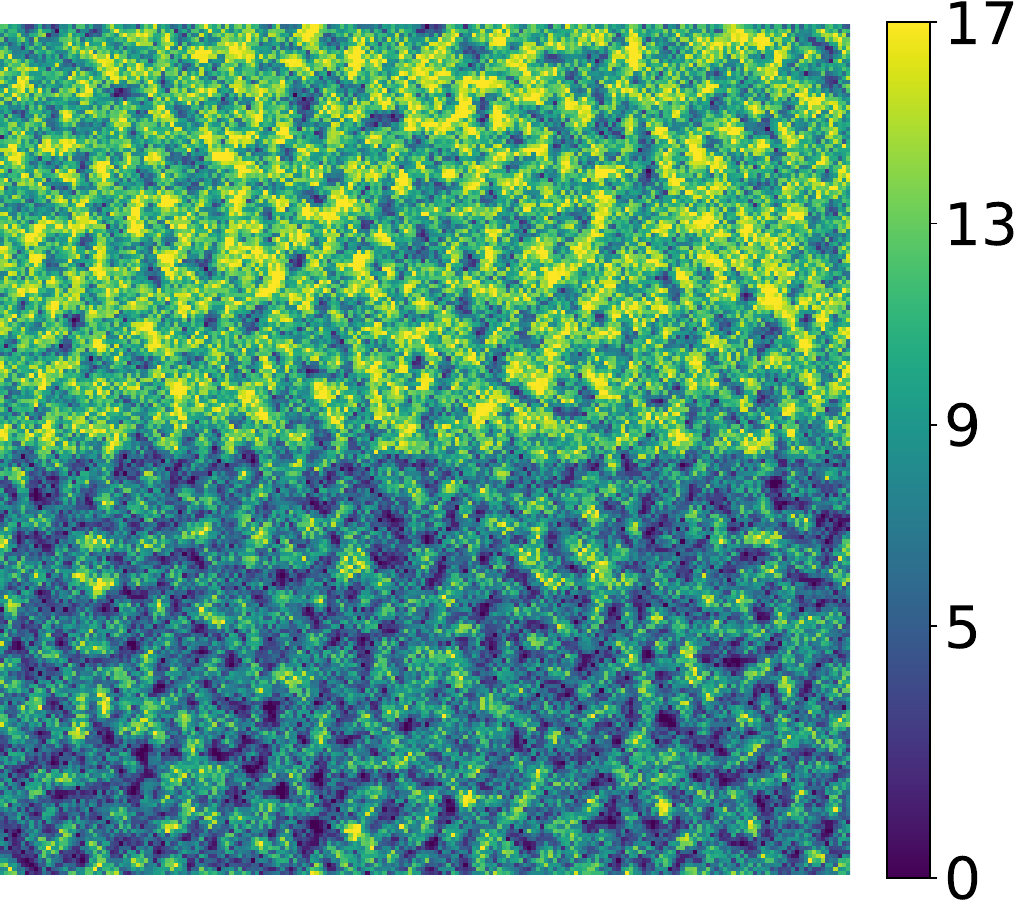}\\
					\vspace{-2mm}
					{\footnotesize t=1}
				\end{minipage}
				\hfill
				\begin{minipage}{0.188\linewidth}
					\centering
					\includegraphics[width=\linewidth]{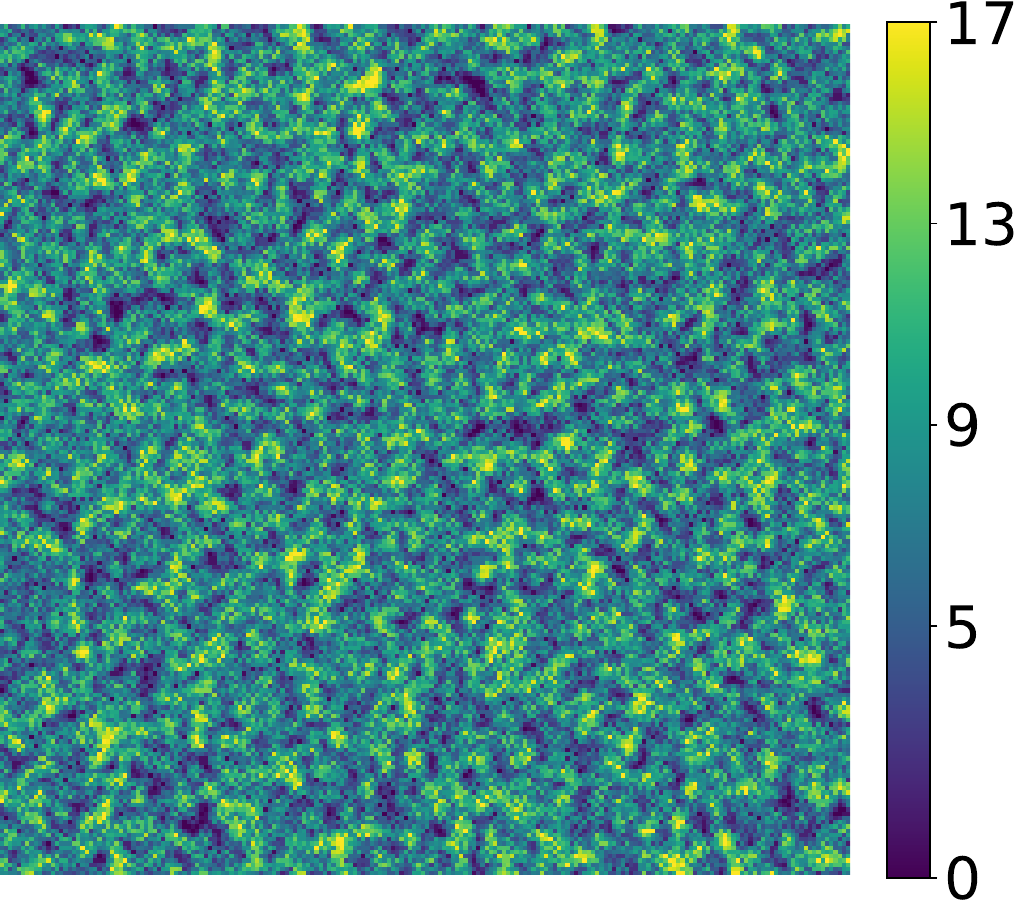}\\
					\vspace{-2mm}
					{\footnotesize t=10}
				\end{minipage}
				\hfill
				\begin{minipage}{0.188\linewidth}
					\centering
					\includegraphics[width=\linewidth]{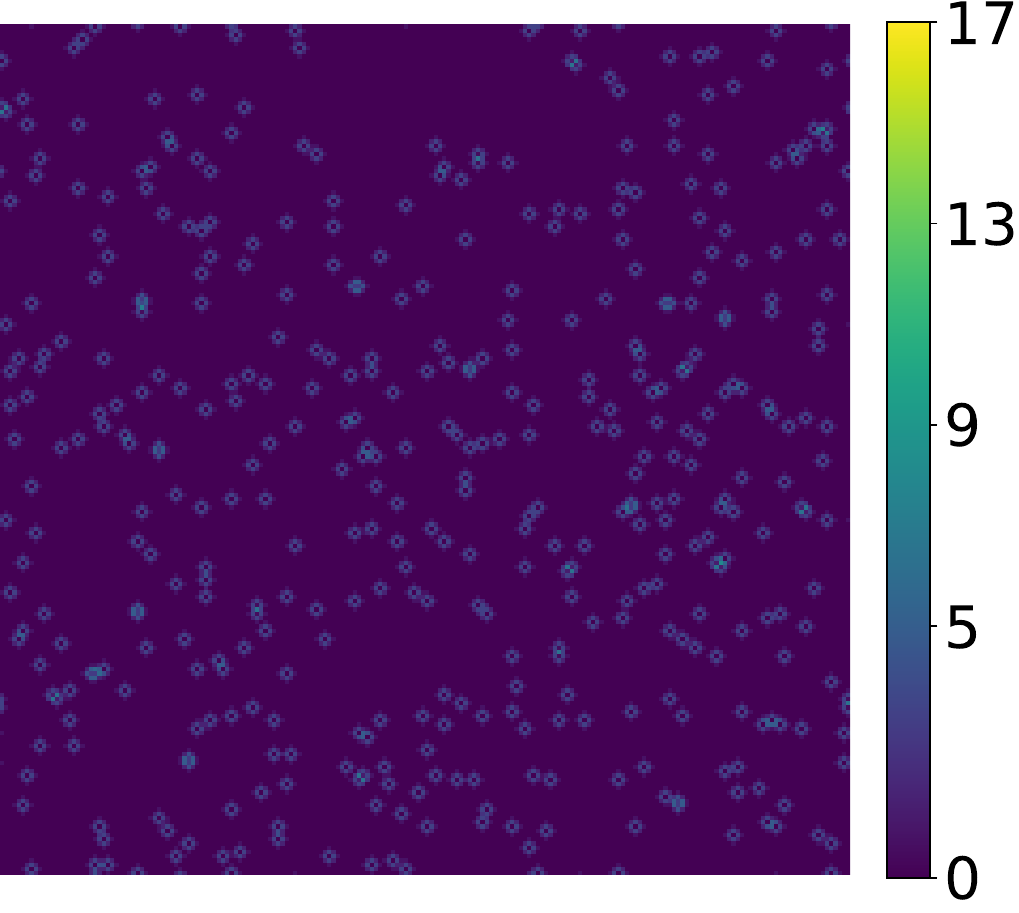}\\
					\vspace{-2mm}
					{\footnotesize t=100}
				\end{minipage}
				\hfill
				\begin{minipage}{0.188\linewidth}
					\centering
					\includegraphics[width=\linewidth]{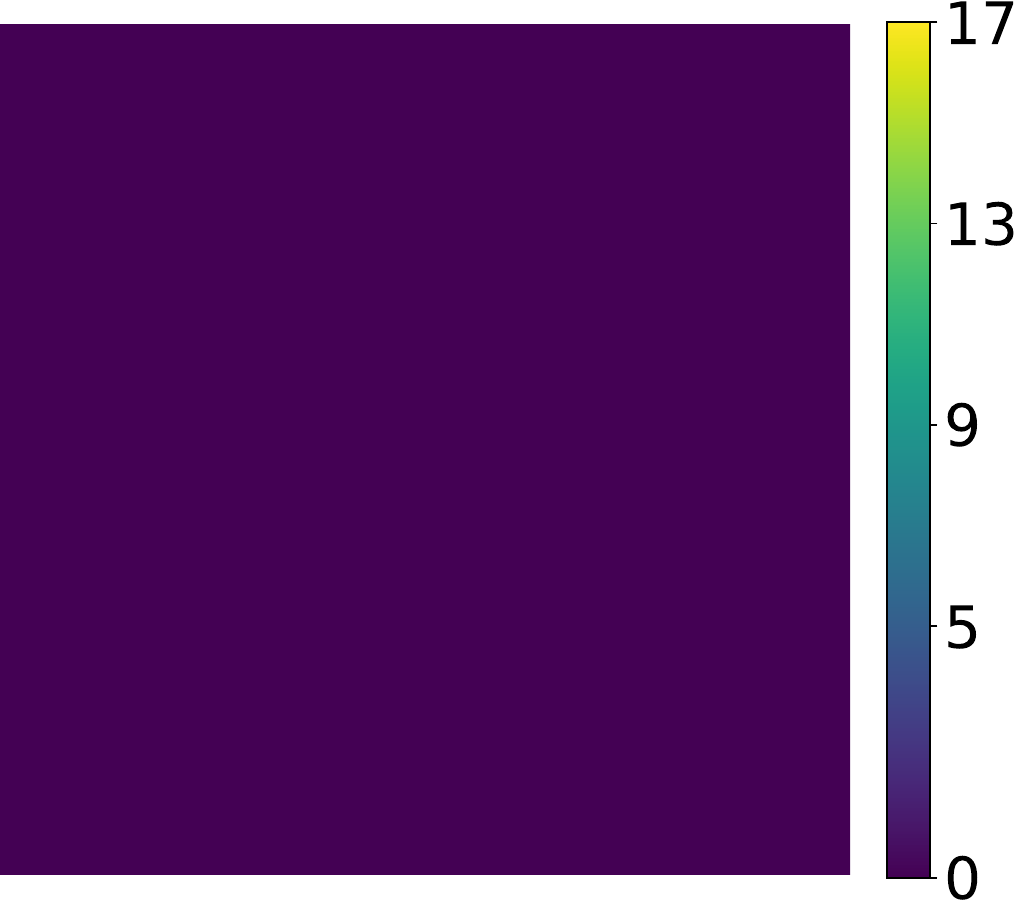}\\
					\vspace{-2mm}
					{\footnotesize t=1000}
				\end{minipage}
				\vspace{-2mm}
				\caption*{\footnotesize (c) r=4.1 (Payoff heatmaps)}
			\end{minipage}
				\\
			[2mm]
			\begin{minipage}{\linewidth}
				\begin{minipage}{0.188\linewidth}
					\centering
					\includegraphics[width=\linewidth]{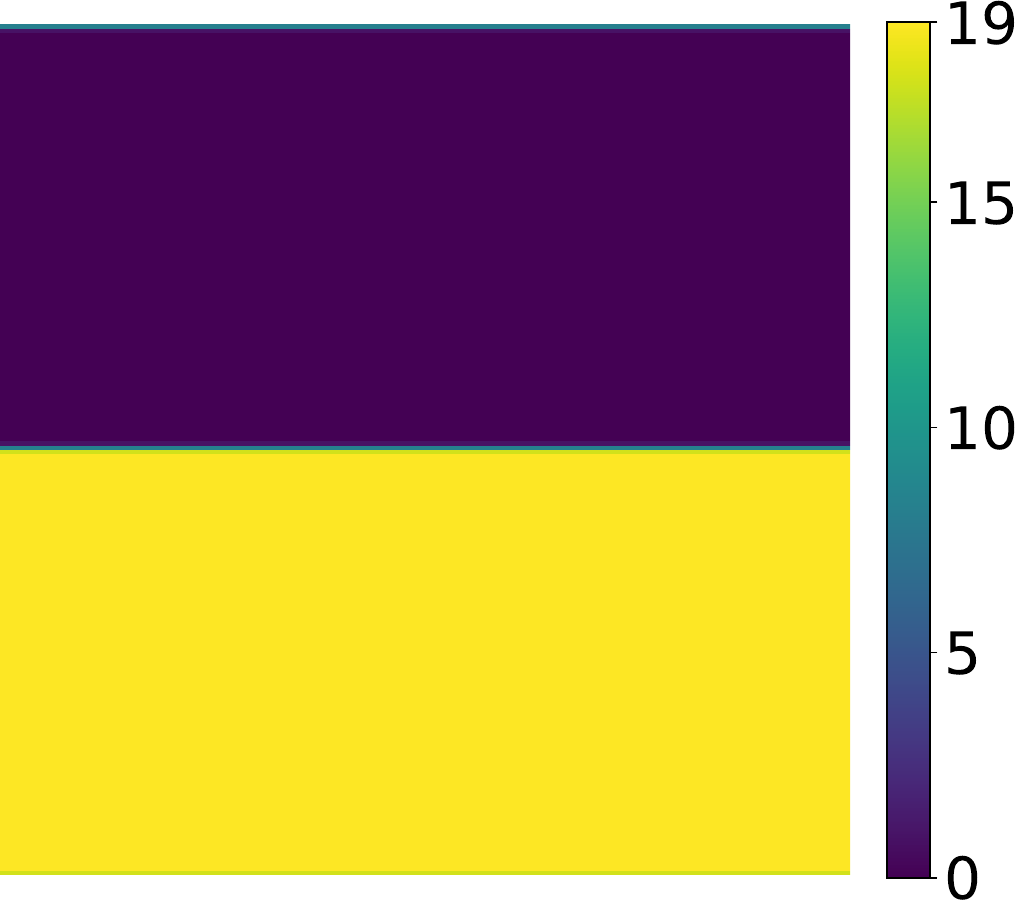}\\
					\vspace{-2mm}
					{\footnotesize t=0}
				\end{minipage}
				\hfill
				\begin{minipage}{0.188\linewidth}
					\centering
					\includegraphics[width=\linewidth]{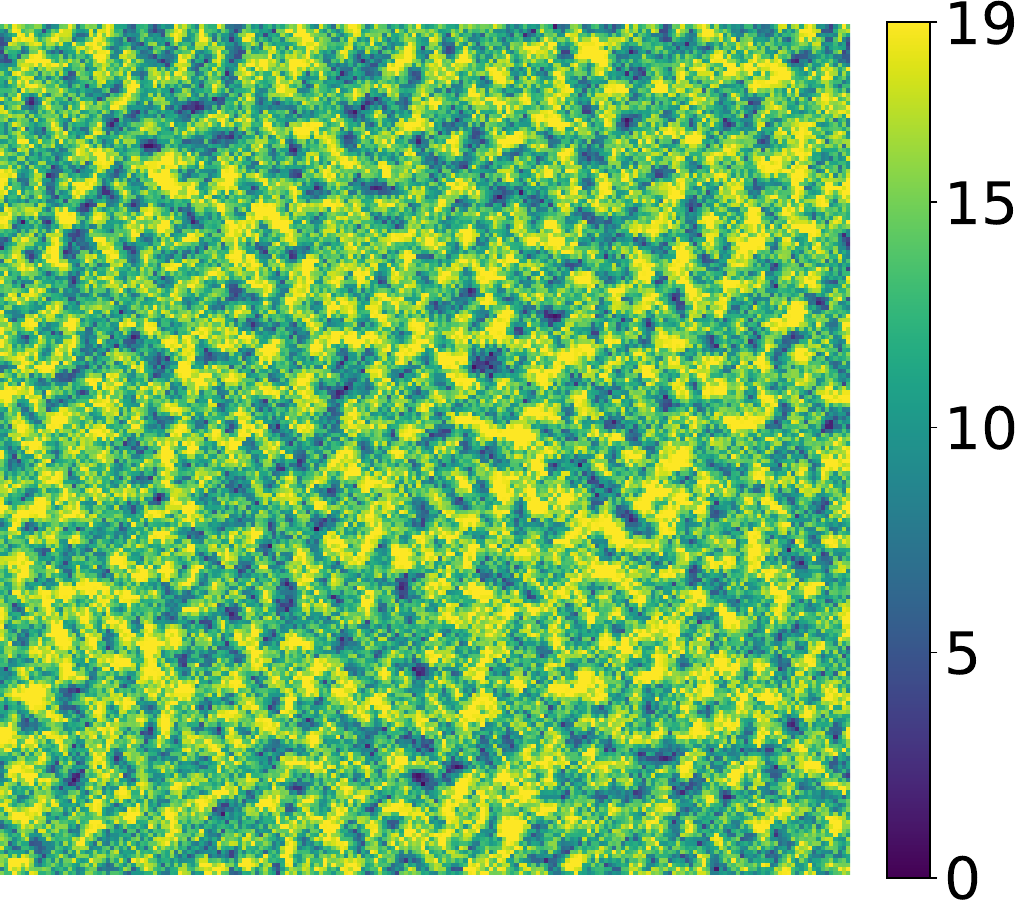}\\
					\vspace{-2mm}
					{\footnotesize t=1}
				\end{minipage}
				\hfill
				\begin{minipage}{0.188\linewidth}
					\centering
					\includegraphics[width=\linewidth]{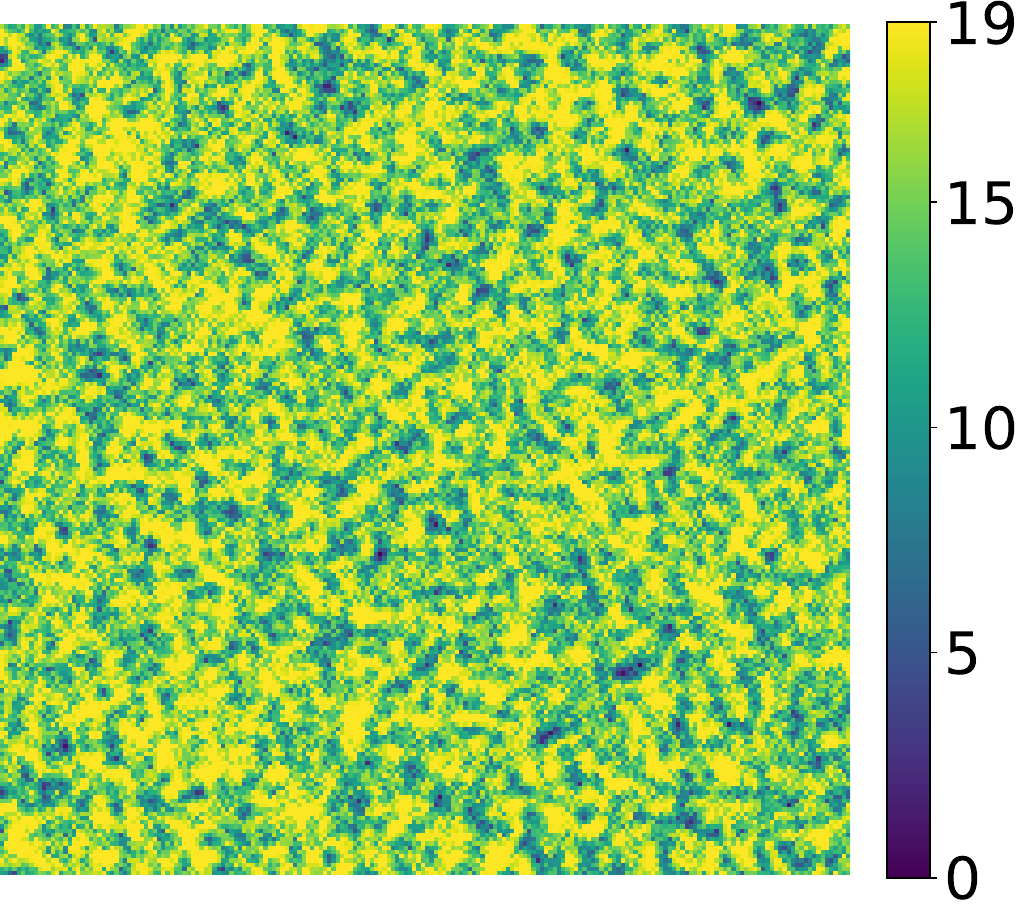}\\
					\vspace{-2mm}
					{\footnotesize t=10}
				\end{minipage}
				\hfill
				\begin{minipage}{0.188\linewidth}
					\centering
					\includegraphics[width=\linewidth]{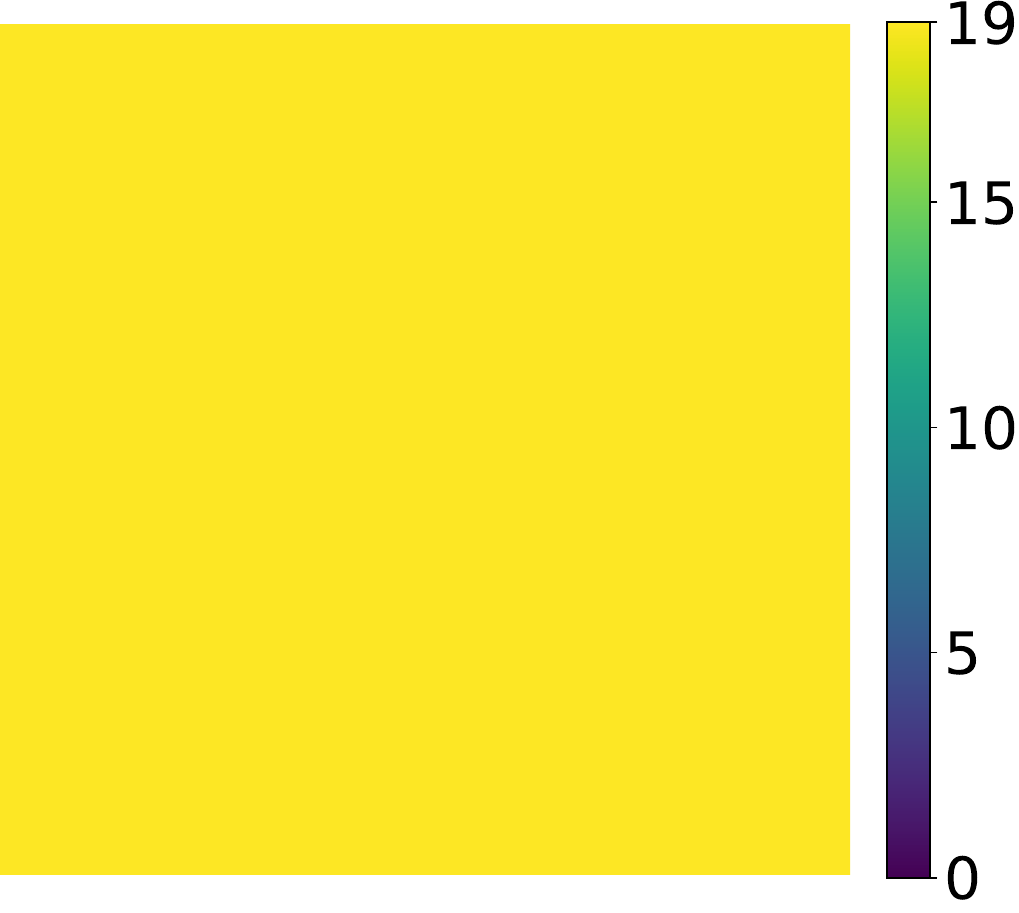}\\
					\vspace{-2mm}
					{\footnotesize t=100}
				\end{minipage}
				\hfill
				\begin{minipage}{0.188\linewidth}
					\centering
					\includegraphics[width=\linewidth]{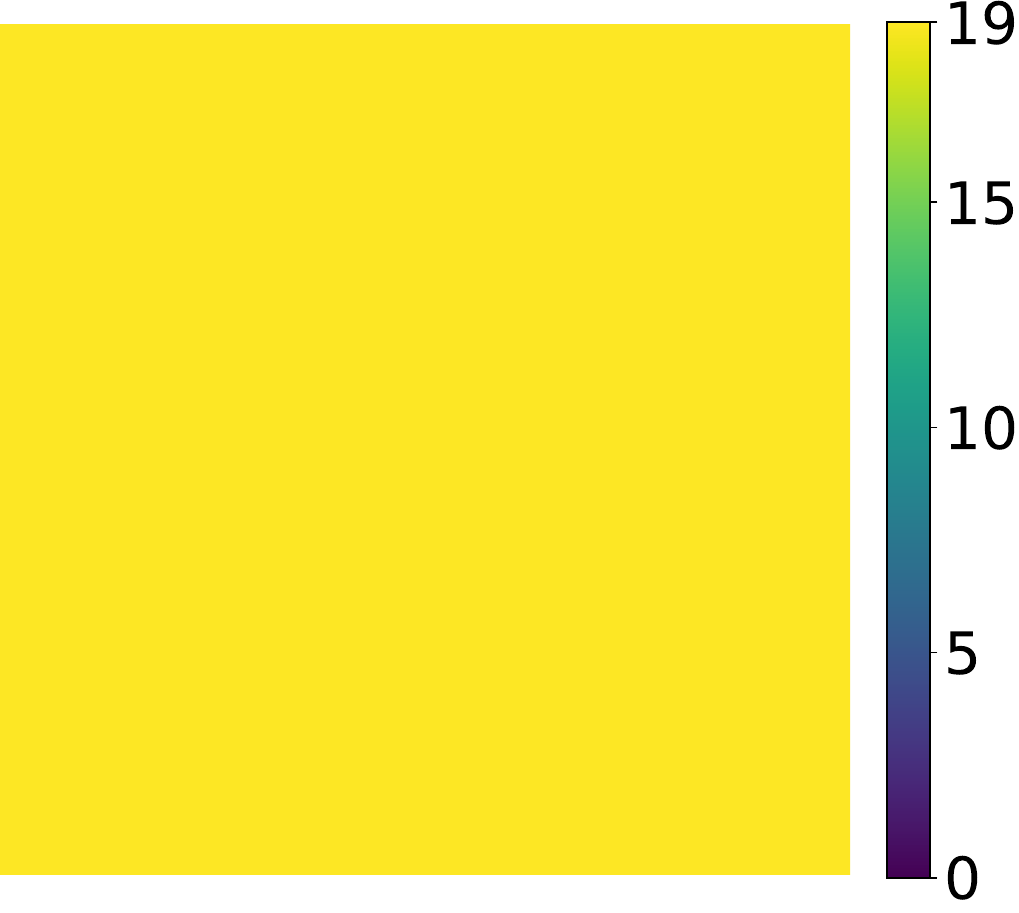}\\
					\vspace{-2mm}
					{\footnotesize t=1000}
				\end{minipage}
				\vspace{-2mm}
				\caption*{\footnotesize (d) r=4.7 (Payoff heatmaps)}
			\end{minipage}
		\caption{MAPPO-LCR under half-and-half initialization.
			Subfigures (a) and (b) show cooperation–defection evolution curves for $r=4.1$ and $r=4.7$ with time on the x-axis and fractions on the y-axis.
			Subfigures (c) and (d) display state snapshots at $t=0,1,10,100,1000$ and their reward maps.
			White denotes cooperation and black denotes defection.
			Reward maps progress from yellow (high) to purple (low).}
		\label{fig:half_combined}
	\end{figure*}

	We evaluate MAPPO-LCR under a half-and-half initialization where the upper region begins with defectors and the lower region begins with cooperators.
	Figure~\ref{fig:half_combined} reports results for $r=4.1$ and $r=4.7$.
	Each experiment includes one evolution curve and five state snapshots at $t=0,1,10,100,1000$, together with the corresponding reward maps. 
	White marks cooperators and black marks defectors. 
	Yellow indicates high reward and purple indicates low reward.
	The state snapshots show no spatial clusters during training. 
	Agents mix uniformly through the grid, and no persistent spatial structure survives.
	This behavior follows from the MAPPO-LCR structure. 
	The centralized critic propagates global value signals, and the LCR term spreads neighborhood feedback smoothly. Policy updates therefore do not promote spatial grouping.
	For $r=4.1$, the first snapshot still reflects the initial split. 
	The upper region contains more cooperators and the lower region contains more defectors.
	This structure fades quickly. 
	Cooperation collapses before $t=200$. 
	The reward maps confirm this decline. 
	Low-payoff regions expand as defection becomes dominant.
	For $r=4.7$, the system mixes rapidly from the beginning. 
	The spatial boundary disappears almost immediately because the critic and the LCR term couple both regions strongly. 
	Defection disappears before $t=80$, and all agents become cooperators. 
	The reward maps reflect this transition. High-payoff values spread across the grid as cooperation takes over.
	These results show that MAPPO-LCR removes spatial asymmetry quickly and relies on value learning rather than imitation. 
	Cooperation persists only when the enhancement factor exceeds the threshold needed to sustain cooperative dynamics.

	\subsection{MAPPO-LCR with bernoulli random initialization}
	\label{exp_b}
	
		\begin{figure*}[htbp!]
		\begin{minipage}{0.47\linewidth}
			\begin{minipage}{\linewidth}
				\centering
				\includegraphics[width=\linewidth]{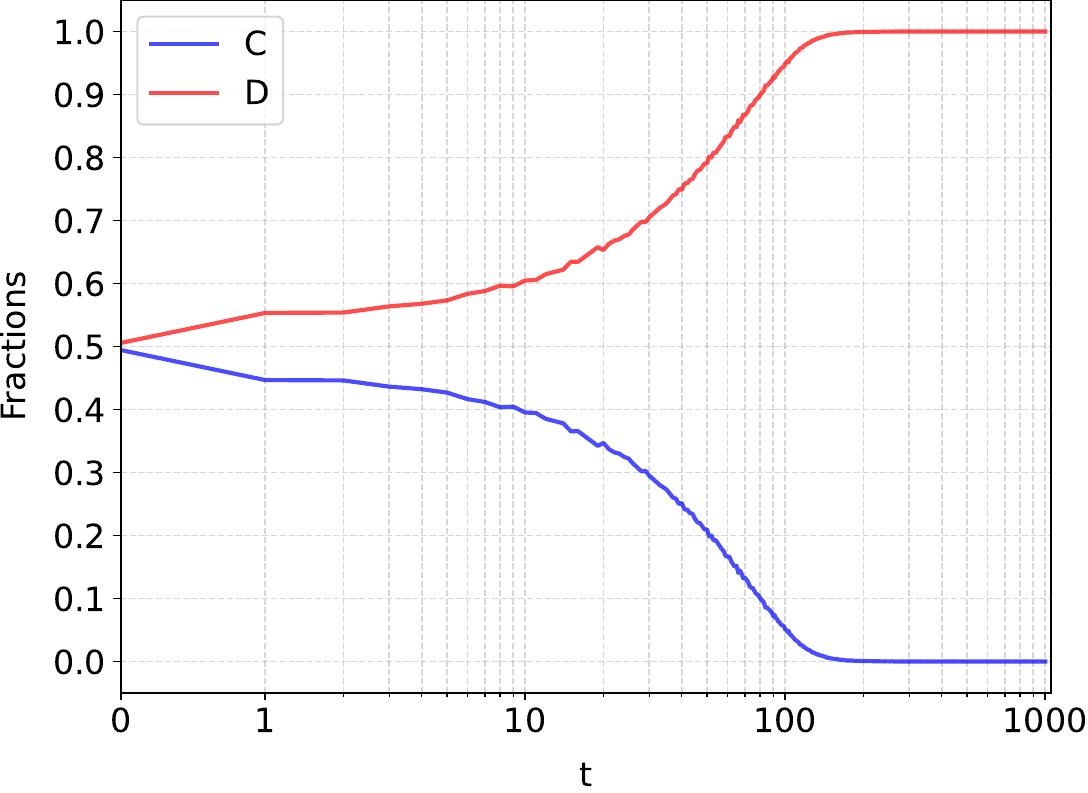}\\
			\end{minipage}
			\vspace{2mm}
			\\
			\begin{minipage}{0.188\linewidth}
				\centering
				\fbox{\includegraphics[width=\linewidth]{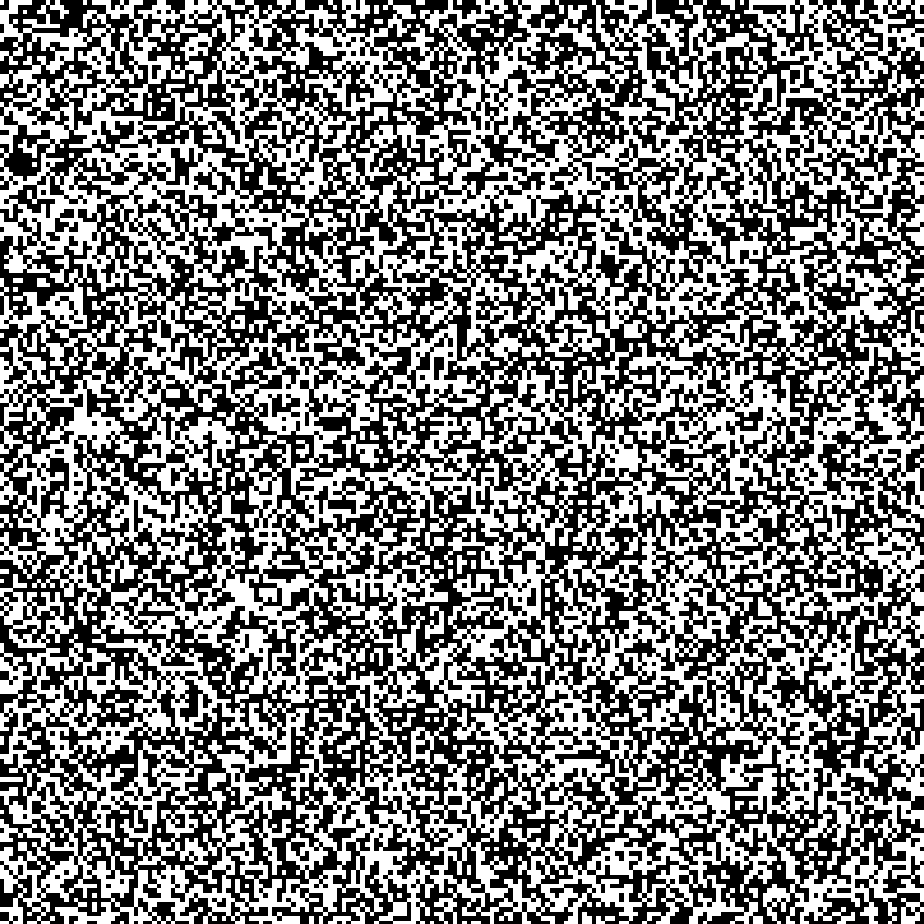}}\\
				\vspace{-2mm}
				{\footnotesize t=0}
			\end{minipage}
			\begin{minipage}{0.188\linewidth}
				\centering
				\fbox{\includegraphics[width=\linewidth]{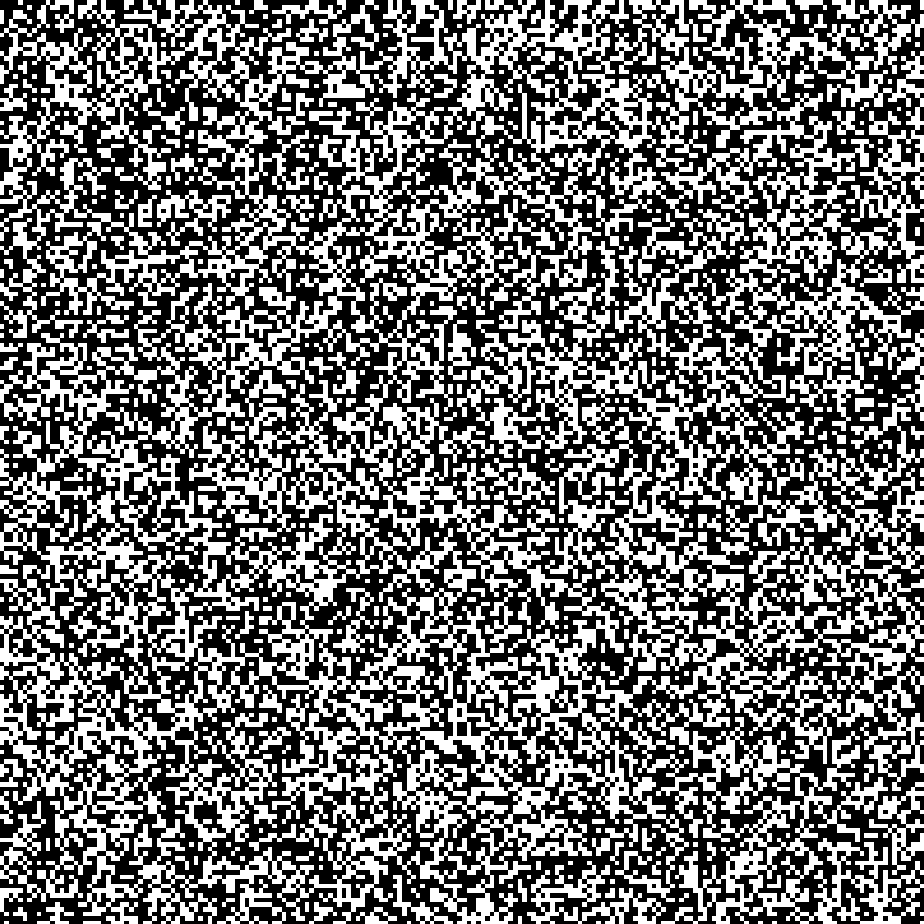}}\\
				\vspace{-2mm}
				{\footnotesize t=1}
			\end{minipage}
			\begin{minipage}{0.188\linewidth}
				\centering
				\fbox{\includegraphics[width=\linewidth]{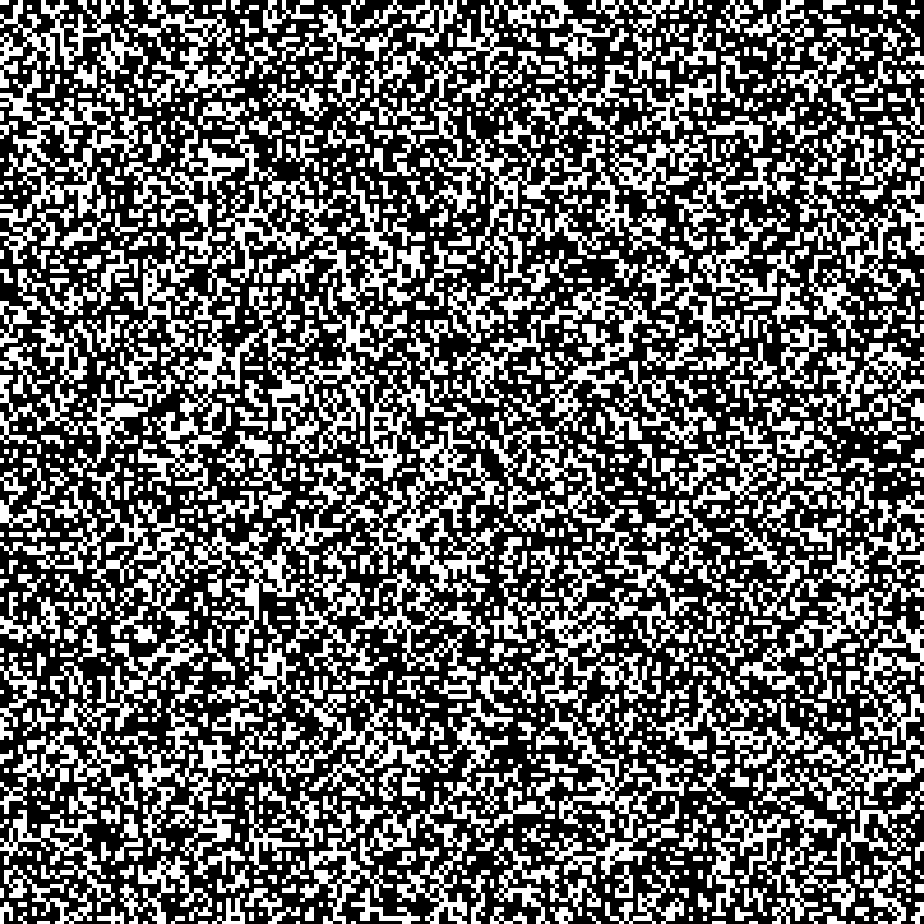}}\\
				\vspace{-2mm}
				{\footnotesize t=10}
			\end{minipage}
			\begin{minipage}{0.188\linewidth}
				\centering
				\fbox{\includegraphics[width=\linewidth]{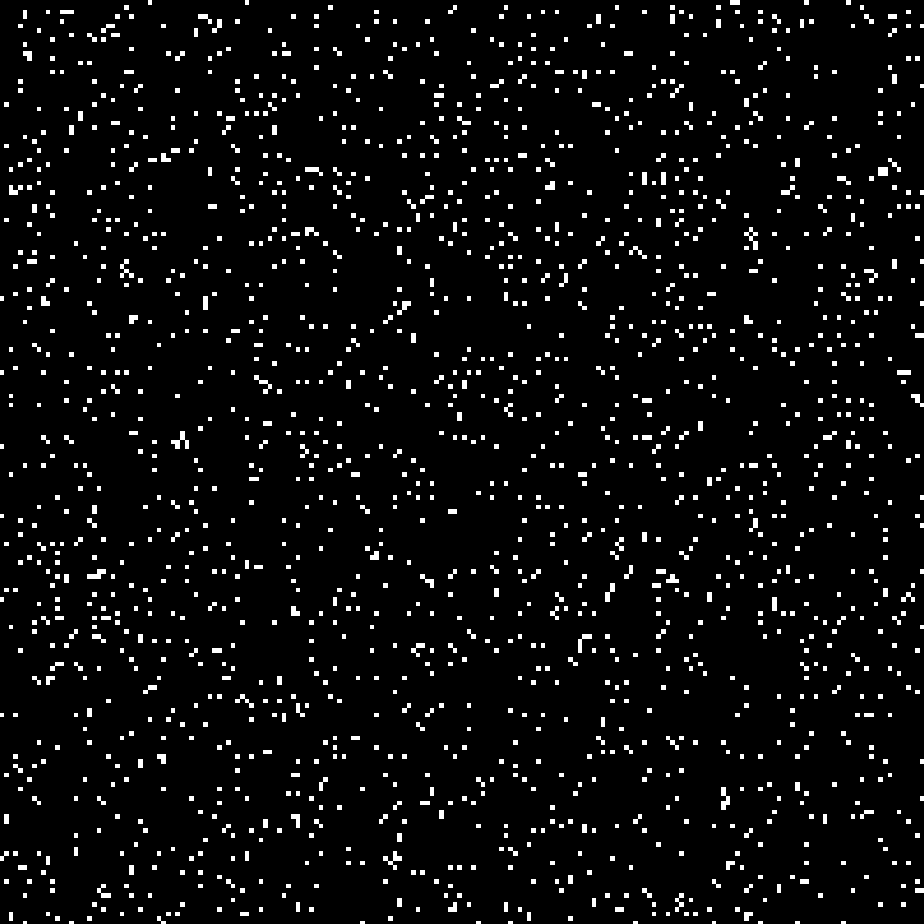}}\\
				\vspace{-2mm}
				{\footnotesize t=100}
			\end{minipage}
			\begin{minipage}{0.188\linewidth}
				\centering
				\fbox{\includegraphics[width=\linewidth]{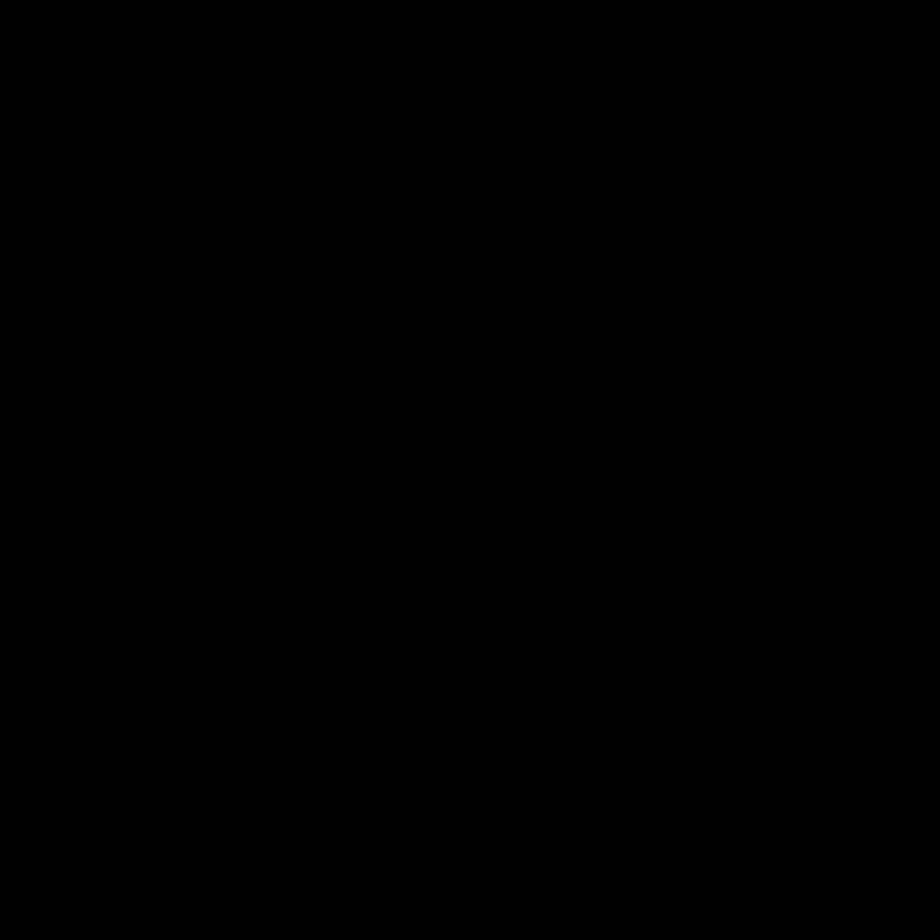}}\\
				\vspace{-2mm}
				{\footnotesize t=1000}
			\end{minipage}
			\vspace{-2mm}
			\caption*{\footnotesize (a) r=4.1}
		\end{minipage}
		\hfill
		\begin{minipage}{0.47\linewidth}
			\begin{minipage}{\linewidth}
				\centering
				\includegraphics[width=\linewidth]{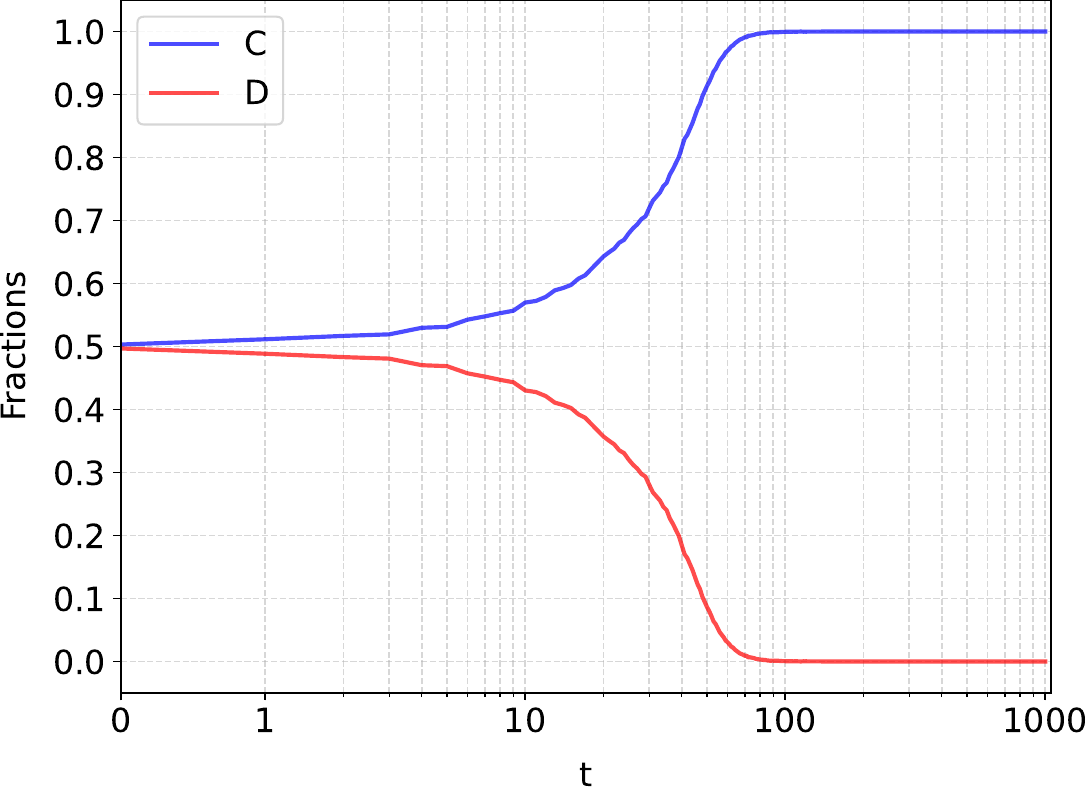}\\
			\end{minipage}
			\vspace{2mm}
			\\
			\begin{minipage}{0.188\linewidth}
				\centering
				\fbox{\includegraphics[width=\linewidth]{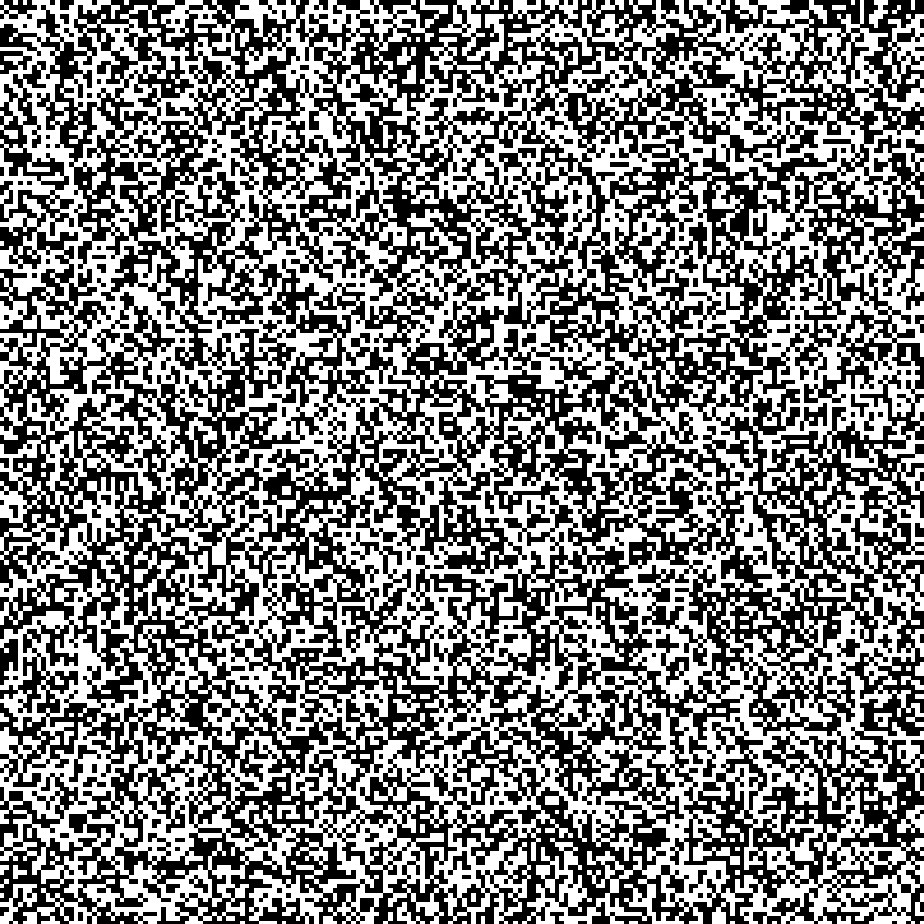}}\\
				\vspace{-2mm}
				{\footnotesize t=0}
			\end{minipage}
			\begin{minipage}{0.188\linewidth}
				\centering
				\fbox{\includegraphics[width=\linewidth]{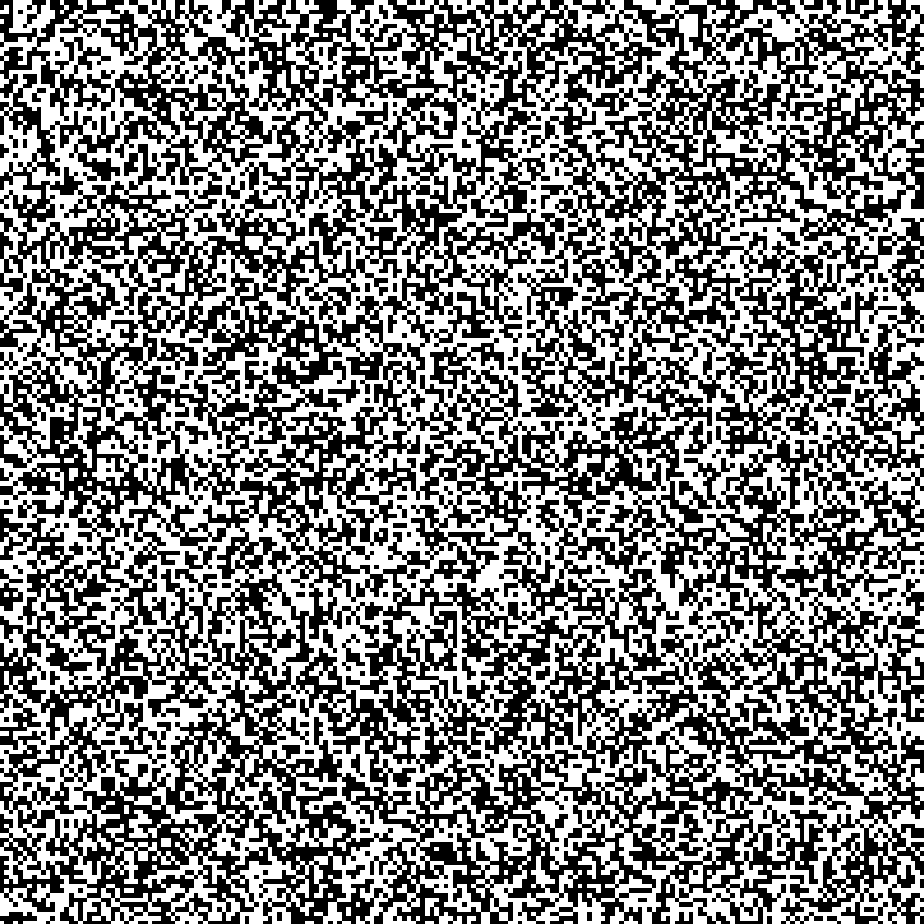}}\\
				\vspace{-2mm}
				{\footnotesize t=1}
			\end{minipage}
			\begin{minipage}{0.188\linewidth}
				\centering
				\fbox{\includegraphics[width=\linewidth]{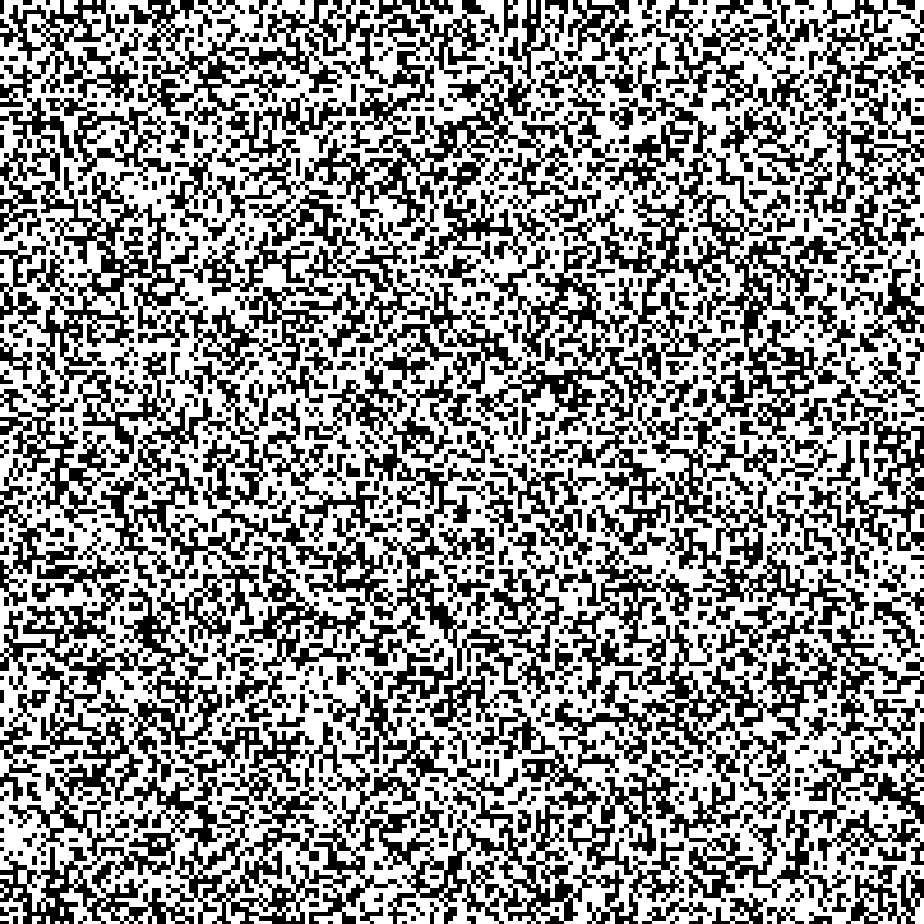}}\\
				\vspace{-2mm}
				{\footnotesize t=10}
			\end{minipage}
			\begin{minipage}{0.188\linewidth}
				\centering
				\fbox{\includegraphics[width=\linewidth]{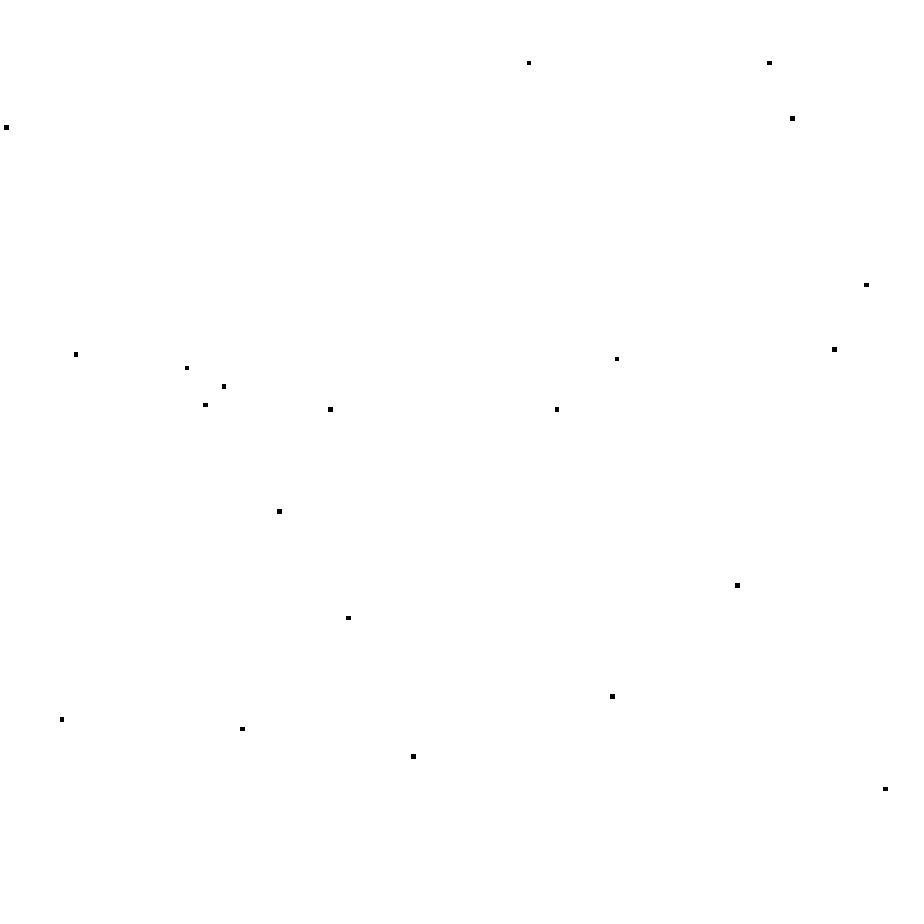}}\\
				\vspace{-2mm}
				{\footnotesize t=100}
			\end{minipage}
			\begin{minipage}{0.188\linewidth}
				\centering
				\fbox{\includegraphics[width=\linewidth]{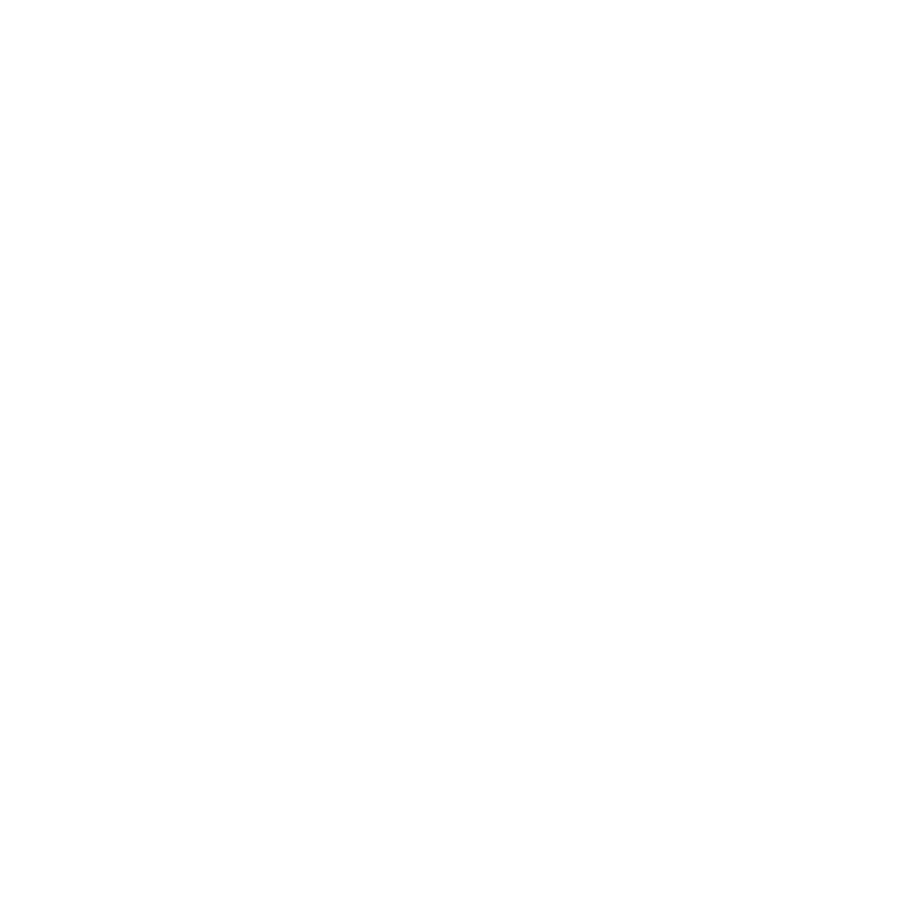}}\\
				\vspace{-2mm}
				{\footnotesize t=1000}
			\end{minipage}
			\vspace{-2mm}
			\caption*{\footnotesize (b) r=4.7}
		\end{minipage}
		\\
		[2mm]
		\begin{minipage}{\linewidth}
			\begin{minipage}{0.188\linewidth}
				\centering
				\includegraphics[width=\linewidth]{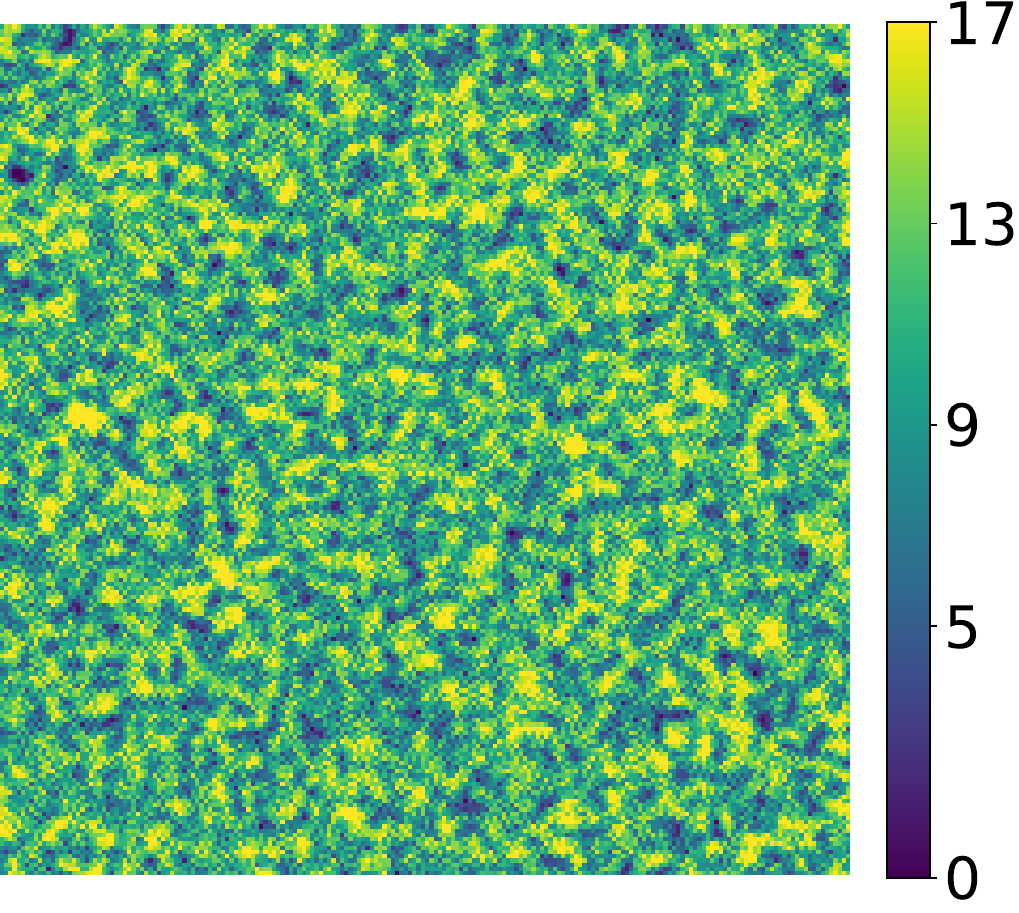}\\
				\vspace{-2mm}
				{\footnotesize t=0}
			\end{minipage}
			\hfill
			\begin{minipage}{0.188\linewidth}
				\centering
				\includegraphics[width=\linewidth]{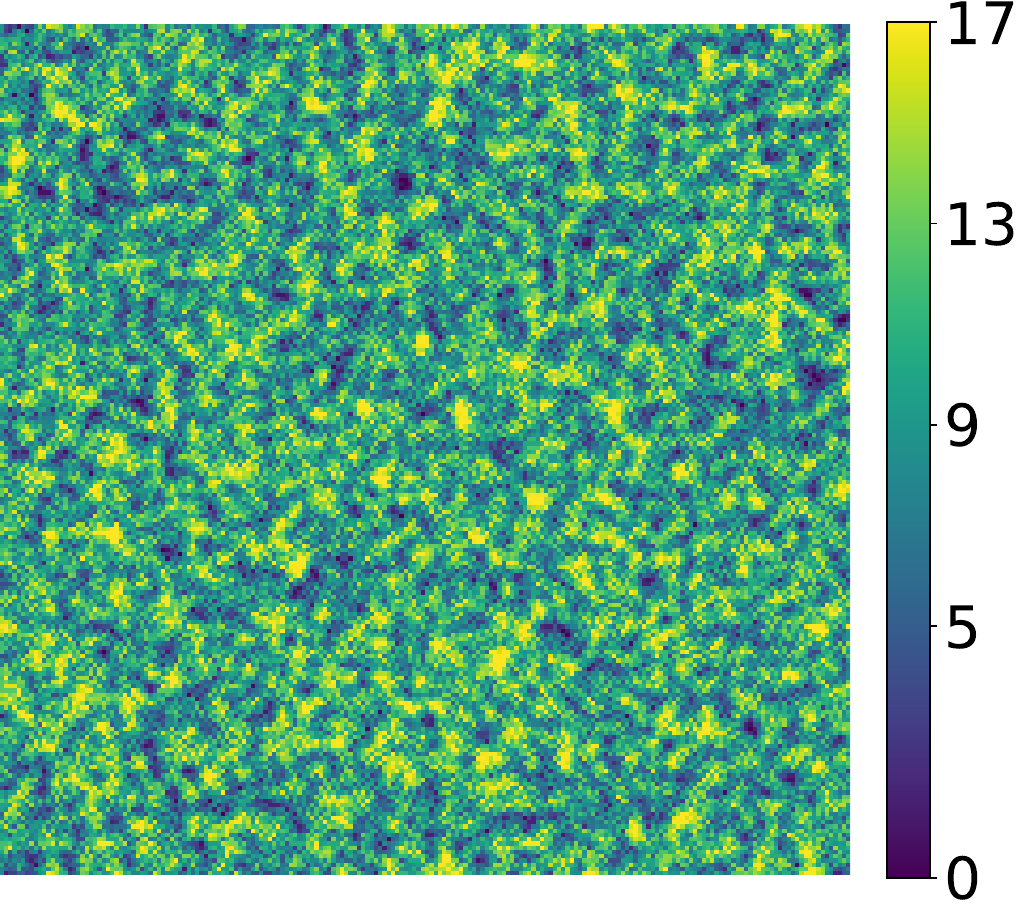}\\
				\vspace{-2mm}
				{\footnotesize t=1}
			\end{minipage}
			\hfill
			\begin{minipage}{0.188\linewidth}
				\centering
				\includegraphics[width=\linewidth]{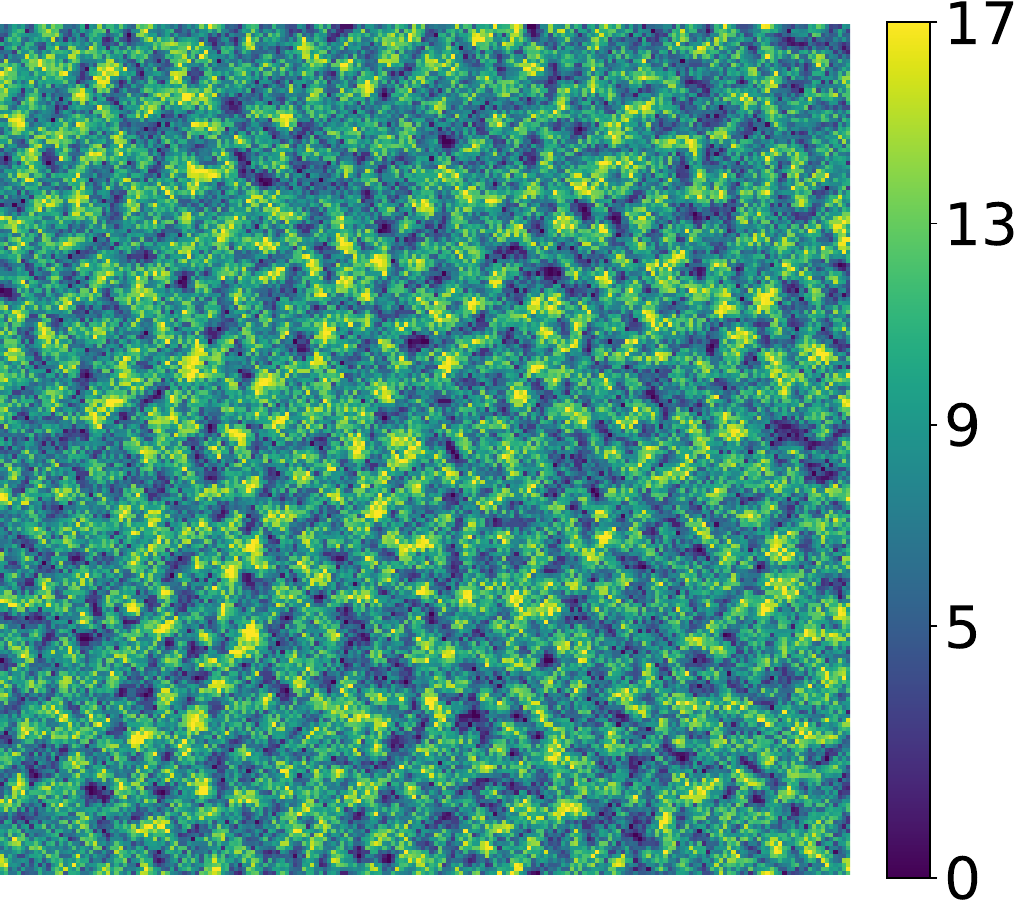}\\
				\vspace{-2mm}
				{\footnotesize t=10}
			\end{minipage}
			\hfill
			\begin{minipage}{0.188\linewidth}
				\centering
				\includegraphics[width=\linewidth]{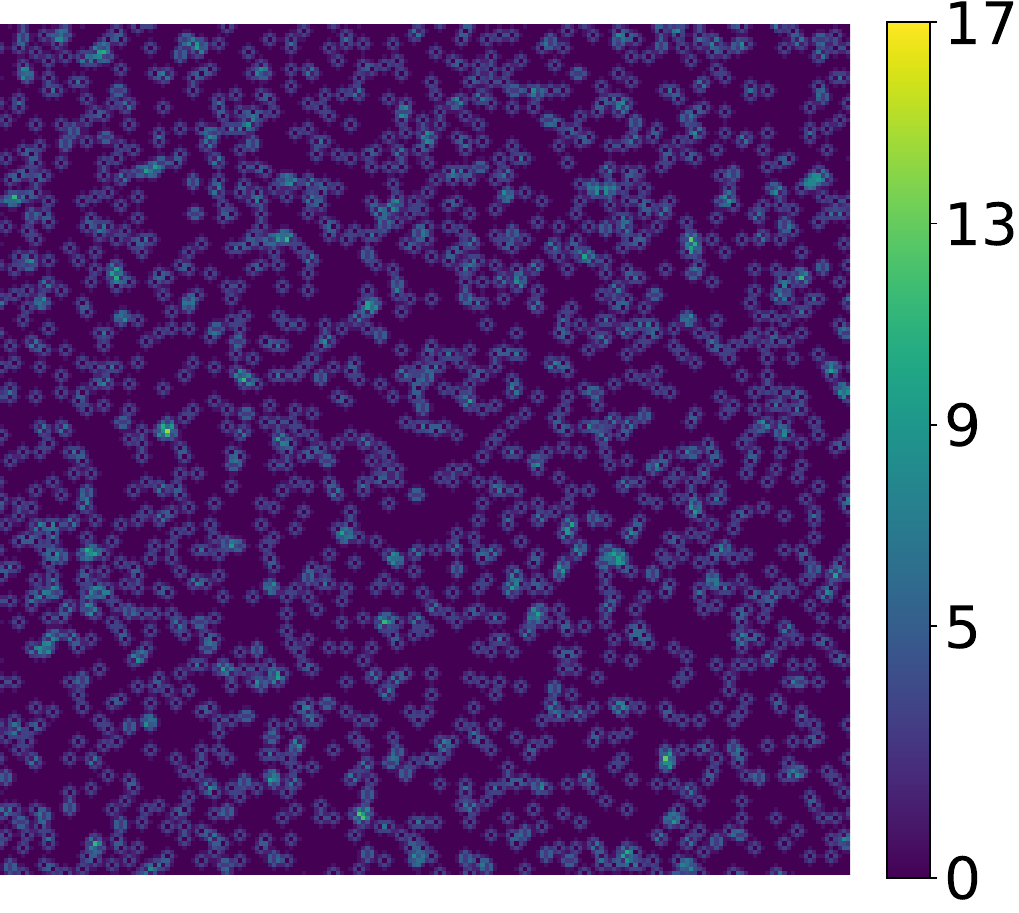}\\
				\vspace{-2mm}
				{\footnotesize t=100}
			\end{minipage}
			\hfill
			\begin{minipage}{0.188\linewidth}
				\centering
				\includegraphics[width=\linewidth]{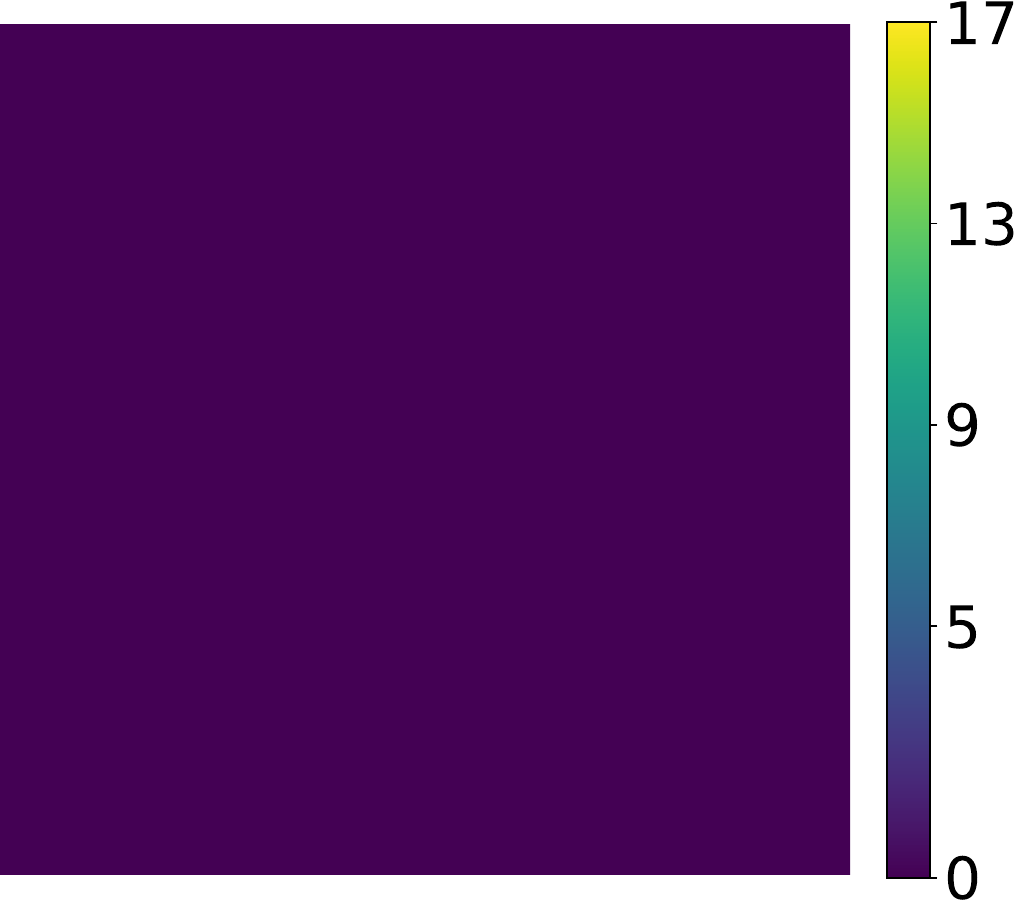}\\
				\vspace{-2mm}
				{\footnotesize t=1000}
			\end{minipage}
			\vspace{-2mm}
			\caption*{\footnotesize (c) r=4.1 (Payoff heatmaps)}
		\end{minipage}
		\\
		[2mm]
		\begin{minipage}{\linewidth}
			\begin{minipage}{0.188\linewidth}
				\centering
				\includegraphics[width=\linewidth]{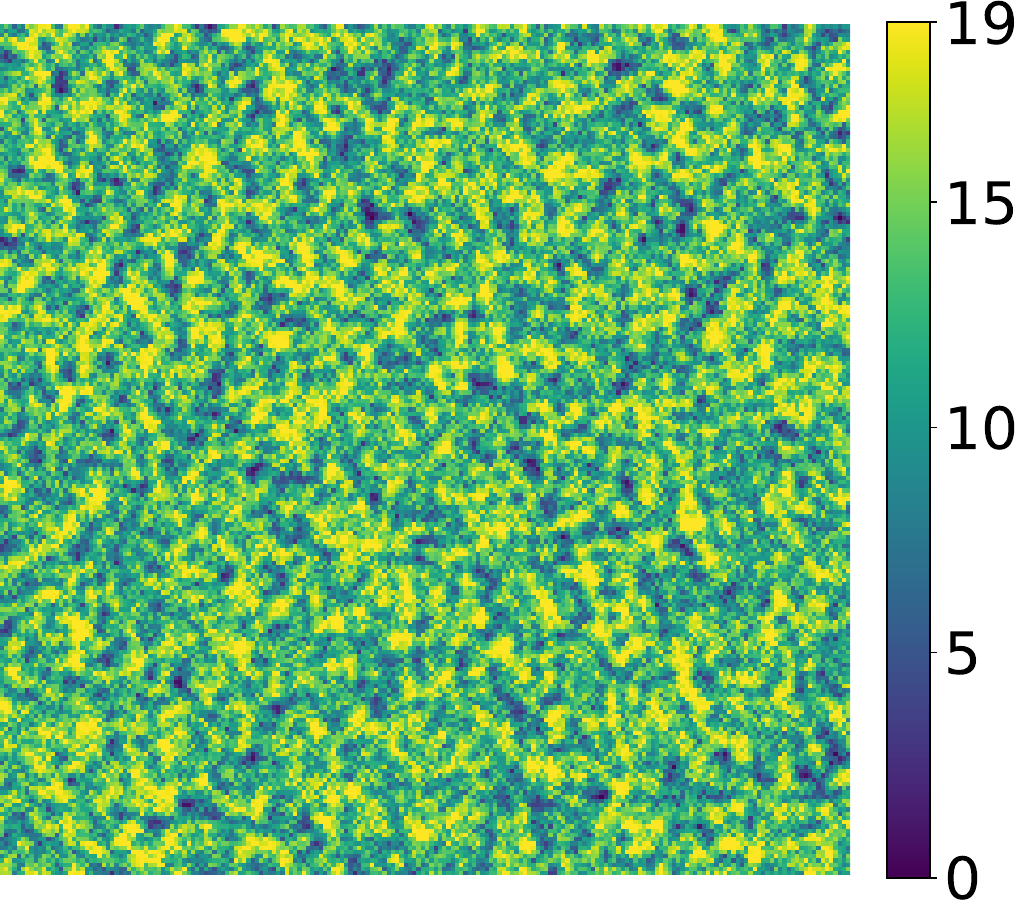}\\
				\vspace{-2mm}
				{\footnotesize t=0}
			\end{minipage}
			\hfill
			\begin{minipage}{0.188\linewidth}
				\centering
				\includegraphics[width=\linewidth]{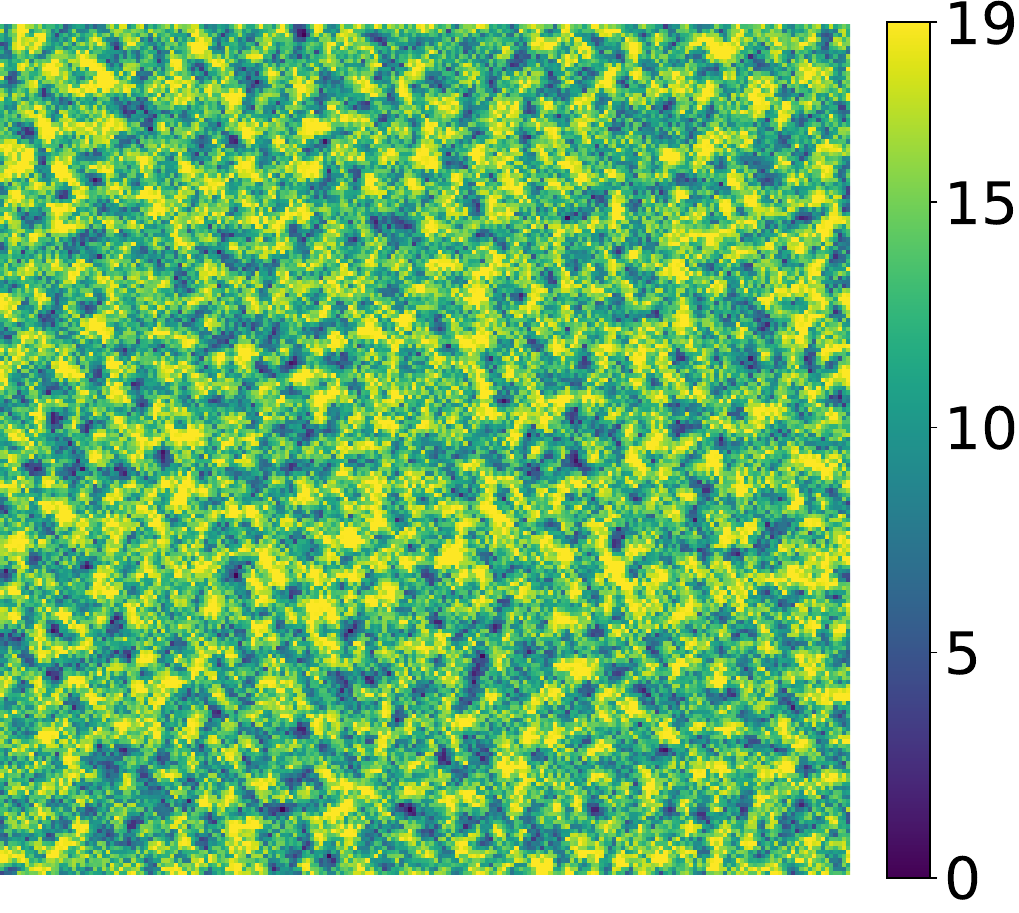}\\
				\vspace{-2mm}
				{\footnotesize t=1}
			\end{minipage}
			\hfill
			\begin{minipage}{0.188\linewidth}
				\centering
				\includegraphics[width=\linewidth]{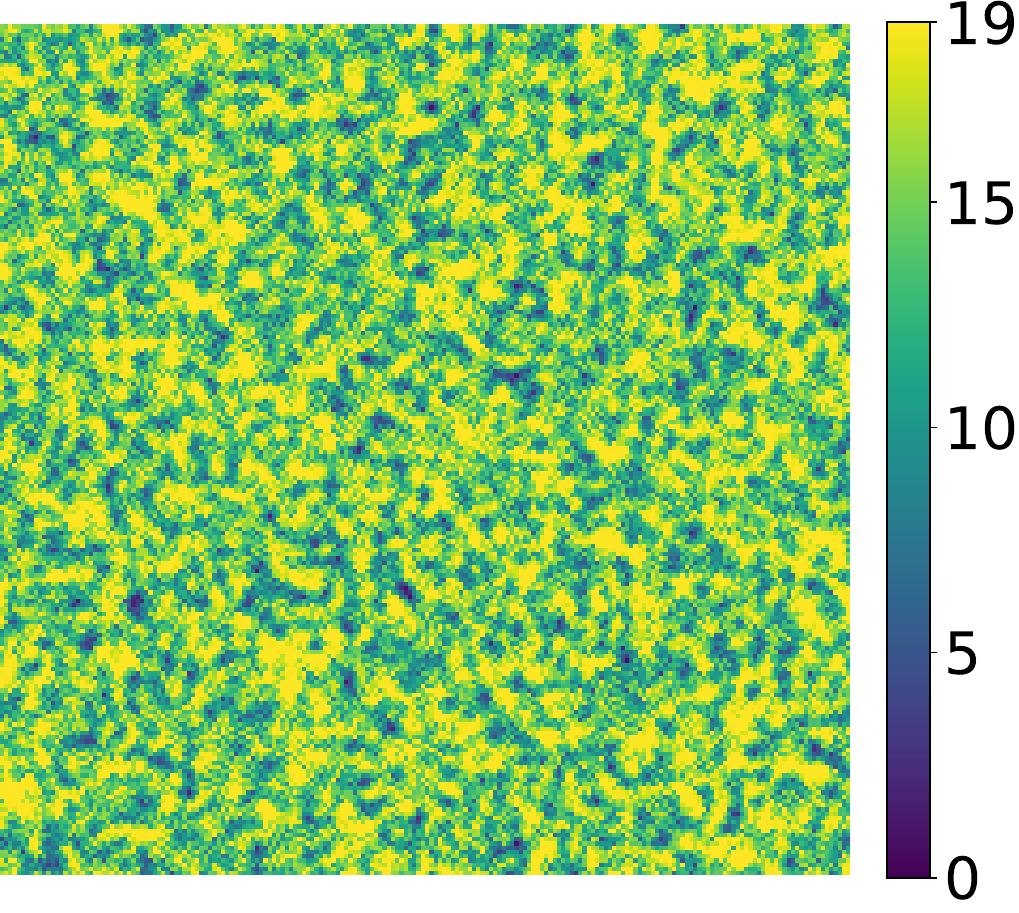}\\
				\vspace{-2mm}
				{\footnotesize t=10}
			\end{minipage}
			\hfill
			\begin{minipage}{0.188\linewidth}
				\centering
				\includegraphics[width=\linewidth]{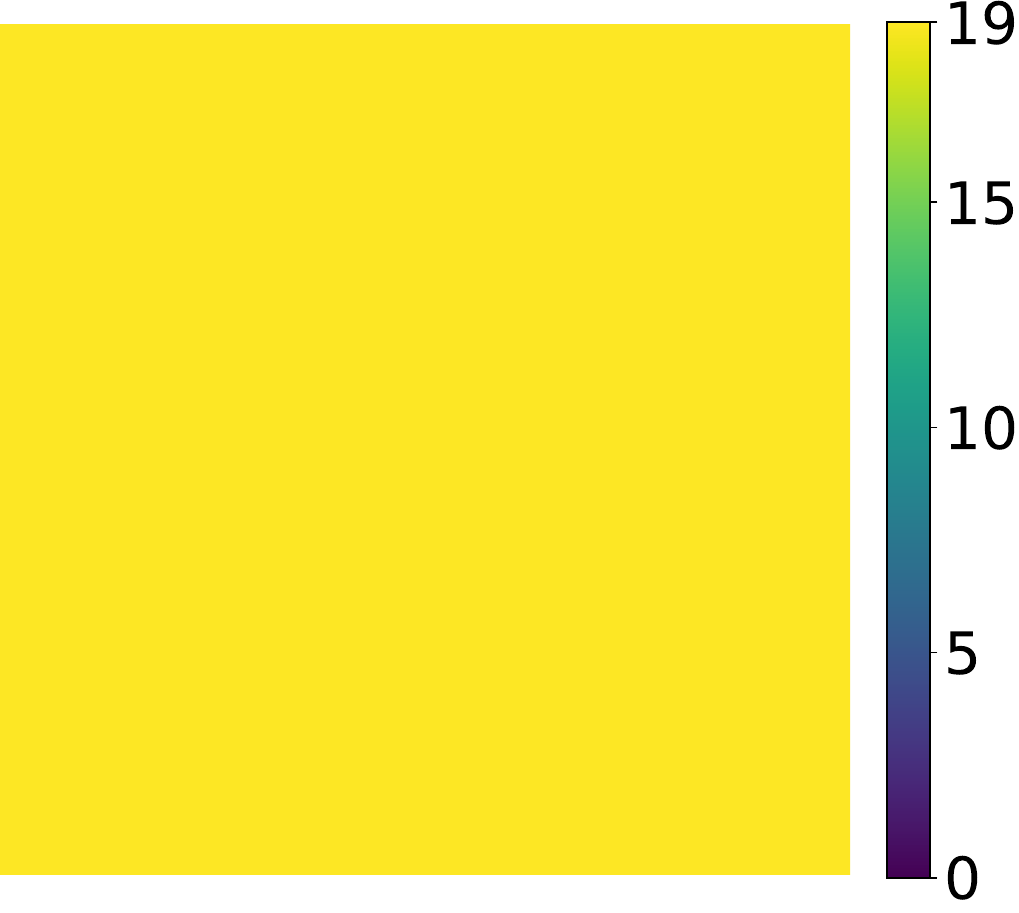}\\
				\vspace{-2mm}
				{\footnotesize t=100}
			\end{minipage}
			\hfill
			\begin{minipage}{0.188\linewidth}
				\centering
				\includegraphics[width=\linewidth]{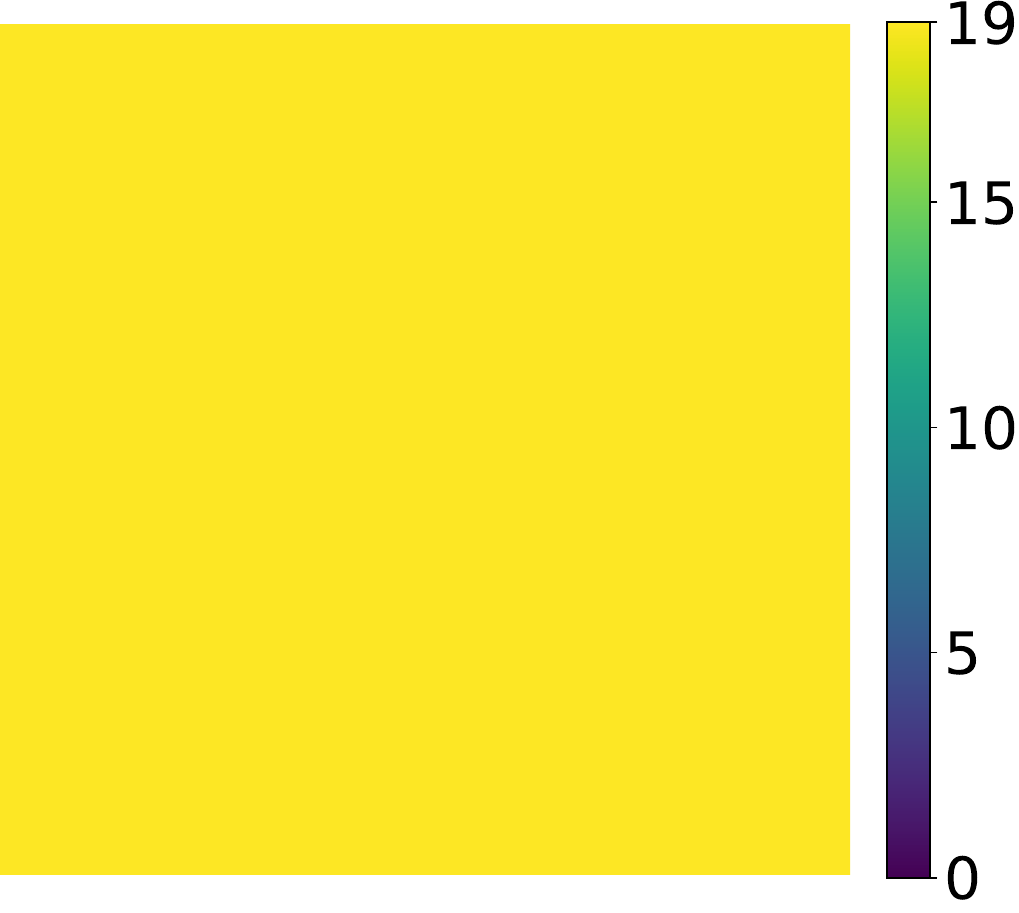}\\
				\vspace{-2mm}
				{\footnotesize t=1000}
			\end{minipage}
			\vspace{-2mm}
			\caption*{\footnotesize (d) r=4.7 (Payoff heatmaps)}
		\end{minipage}
		\caption{Evolution of MAPPO-LCR under Bernoulli initialization.
			Top: cooperation and defection fractions over time.
			Middle: strategy snapshots at $t=0,1,10,100,1000$.
			Bottom: payoff snapshots at the same iterations.
			White indicates cooperators and black indicates defectors.
			Yellow–blue colors correspond to high–low payoffs.}
		\label{fig:mappo_lcr_rand}
	\end{figure*}
	
	We further evaluate MAPPO-LCR under Bernoulli initialization, where each agent starts as a cooperator or defector with equal probability $p=0.5$. 
	Figure~\ref{fig:mappo_lcr_rand} presents the cooperation–defection trajectories and the corresponding state and payoff snapshots for $r=4.1$ and $r=4.7$.
	For $r=4.1$, the proportion of defectors steadily increases and the system converges to full defection before $t=200$. 
	The state snapshots show a persistent mixture of strategies without any spatial structure.
	The payoff snapshots gradually shift toward low-return regions as defection becomes dominant.
	For $r=4.7$, cooperation expands rapidly and defection disappears before $t=90$. 
	State snapshots remain mixed in spatial appearance, but the proportion of cooperators rises sharply. 
	Payoff snapshots become uniformly high once cooperation prevails across the grid.
	The contrast between $r=4.1$ and $r=4.7$ aligns with theoretical expectations. 
	A larger enhancement factor strengthens the return incentive for cooperation.
	The LCR component increases the sensitivity of policy updates to local cooperative density.
	When $r$ is small, the intrinsic payoff of cooperation remains insufficient even with LCR, which drives the system toward full defection. 
	When $r$ is larger, MAPPO-LCR internalizes cooperative feedback more effectively and converges to full cooperation.

	\subsection{MAPPO-LCR with all-defectors initialization}
	\label{exp_ad}
	
\begin{figure*}[htbp!]
			\begin{minipage}{0.47\linewidth}
			\begin{minipage}{\linewidth}
				\centering
				\includegraphics[width=\linewidth]{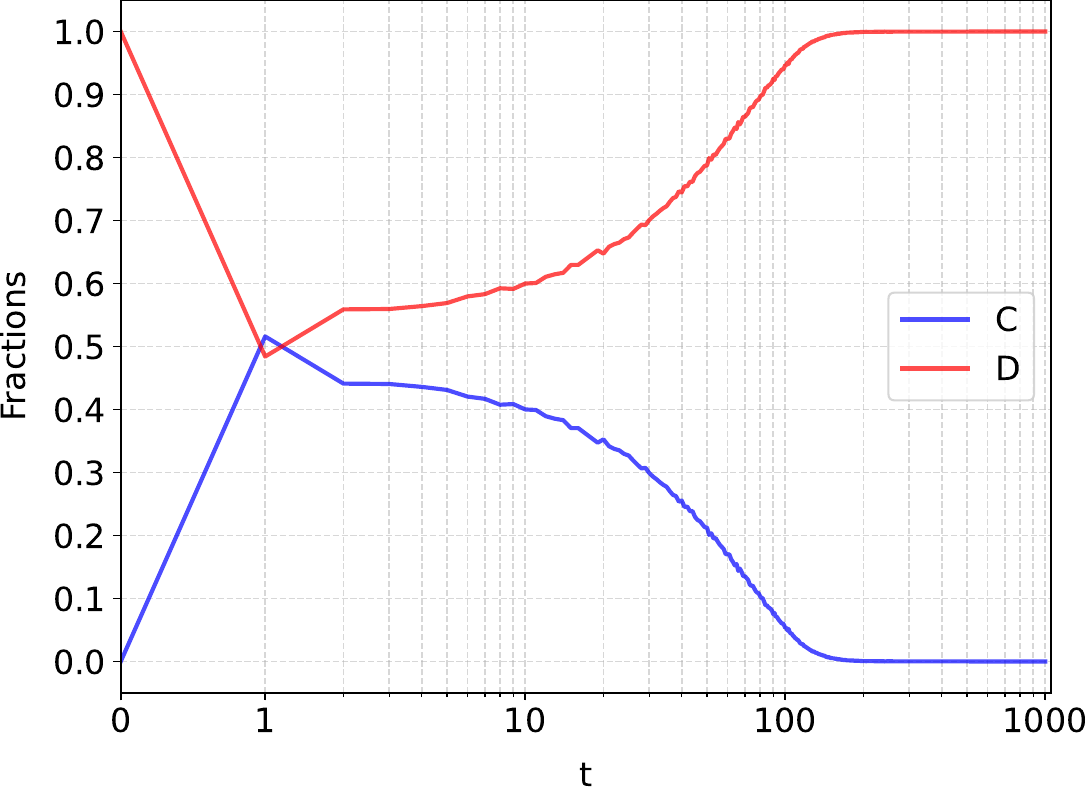}\\
			\end{minipage}
			\vspace{2mm}
			\\
			\begin{minipage}{0.188\linewidth}
				\centering
				\fbox{\includegraphics[width=\linewidth]{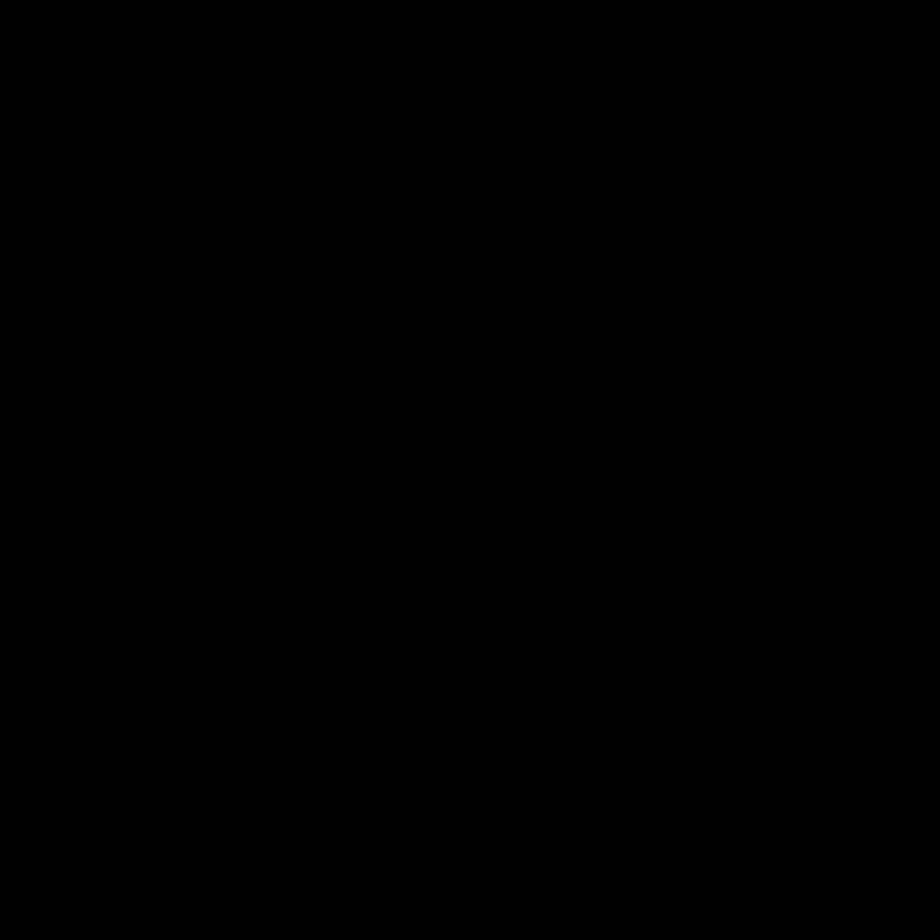}}\\
				\vspace{-2mm}
				{\footnotesize t=0}
			\end{minipage}
			\begin{minipage}{0.188\linewidth}
				\centering
				\fbox{\includegraphics[width=\linewidth]{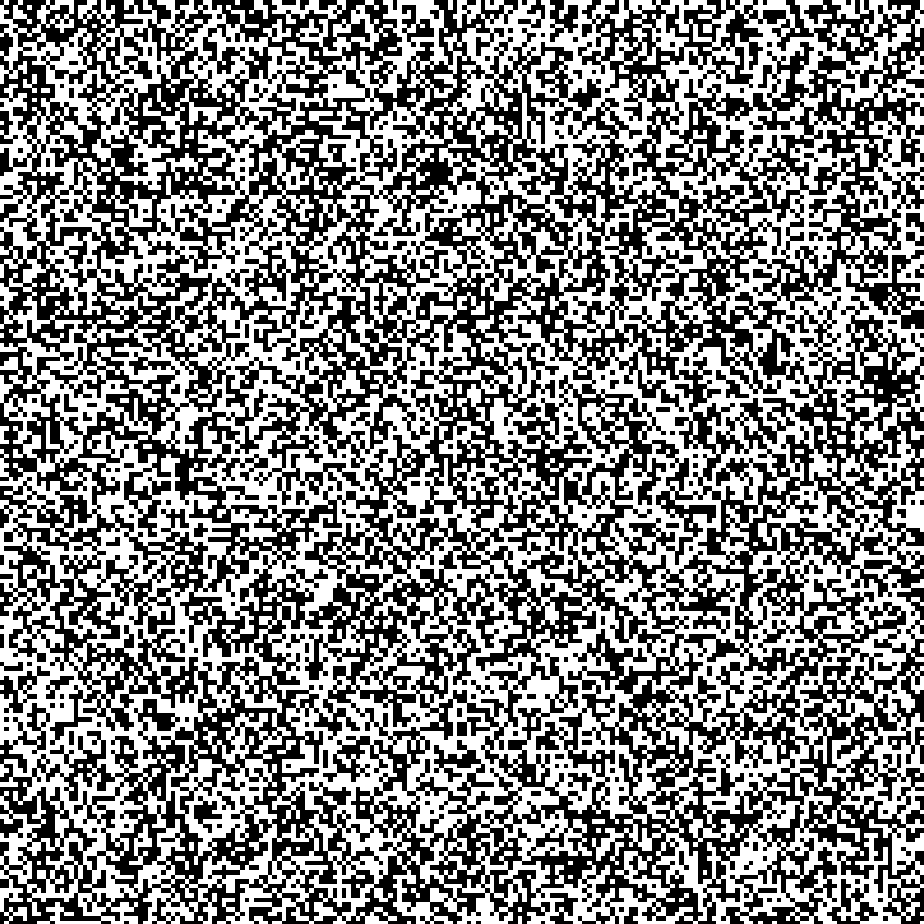}}\\
				\vspace{-2mm}
				{\footnotesize t=1}
			\end{minipage}
			\begin{minipage}{0.188\linewidth}
				\centering
				\fbox{\includegraphics[width=\linewidth]{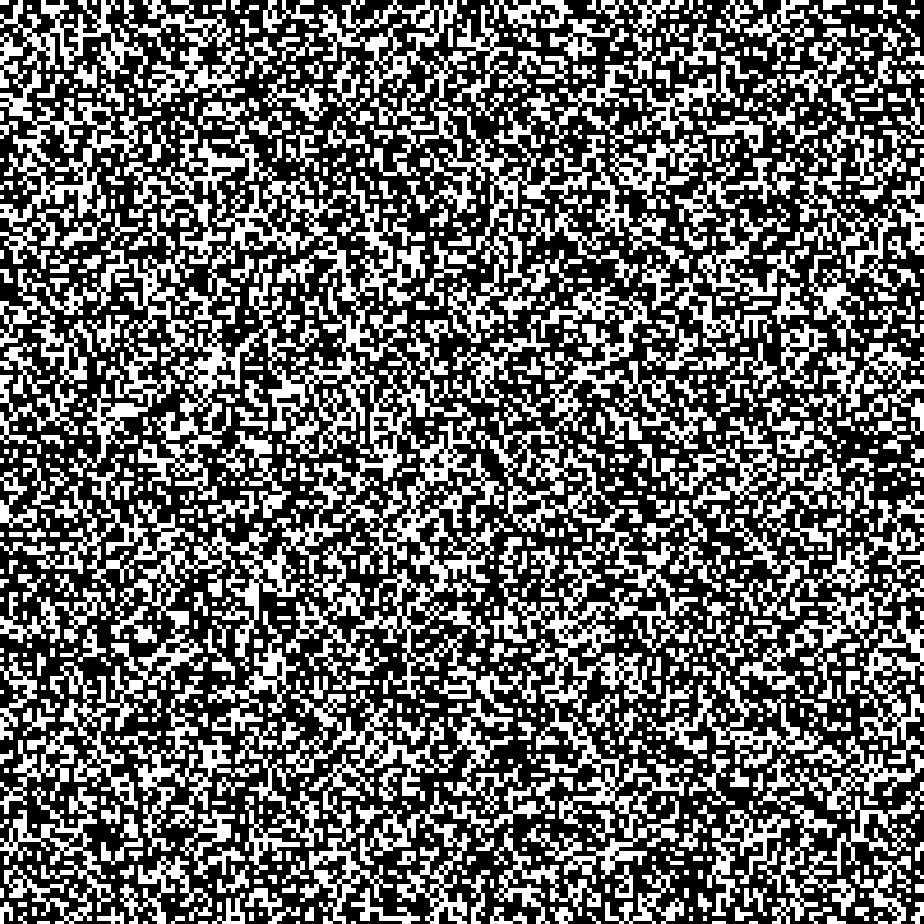}}\\
				\vspace{-2mm}
				{\footnotesize t=10}
			\end{minipage}
			\begin{minipage}{0.188\linewidth}
				\centering
				\fbox{\includegraphics[width=\linewidth]{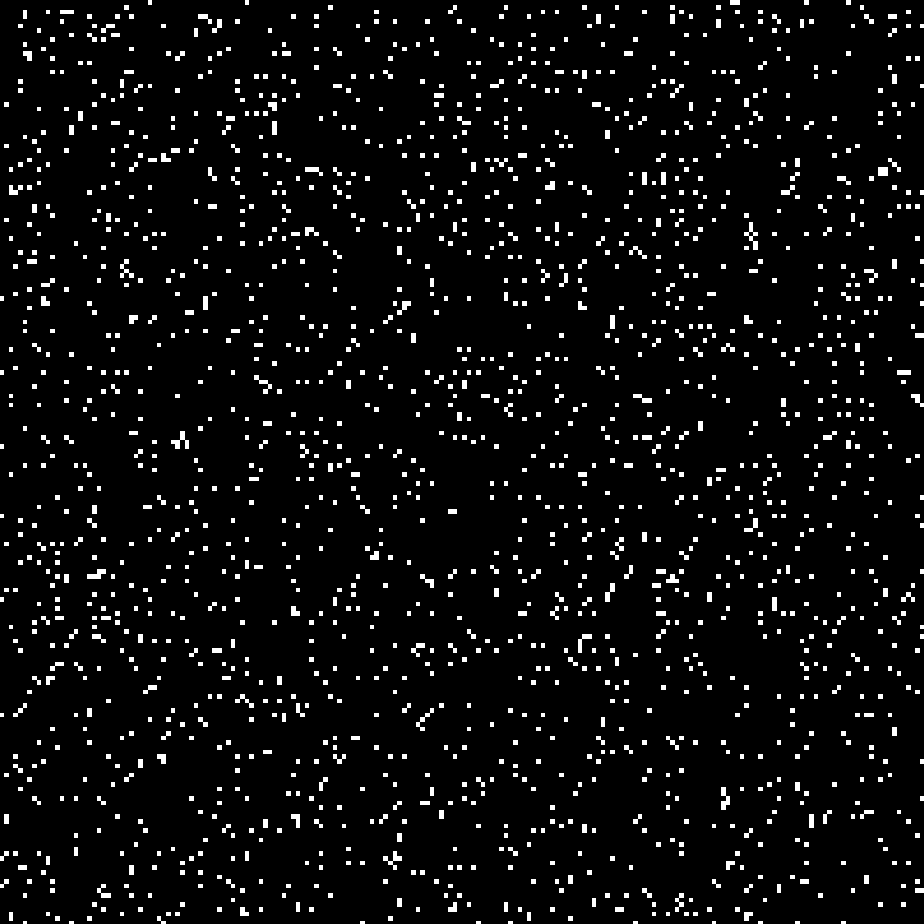}}\\
				\vspace{-2mm}
				{\footnotesize t=100}
			\end{minipage}
			\begin{minipage}{0.188\linewidth}
				\centering
				\fbox{\includegraphics[width=\linewidth]{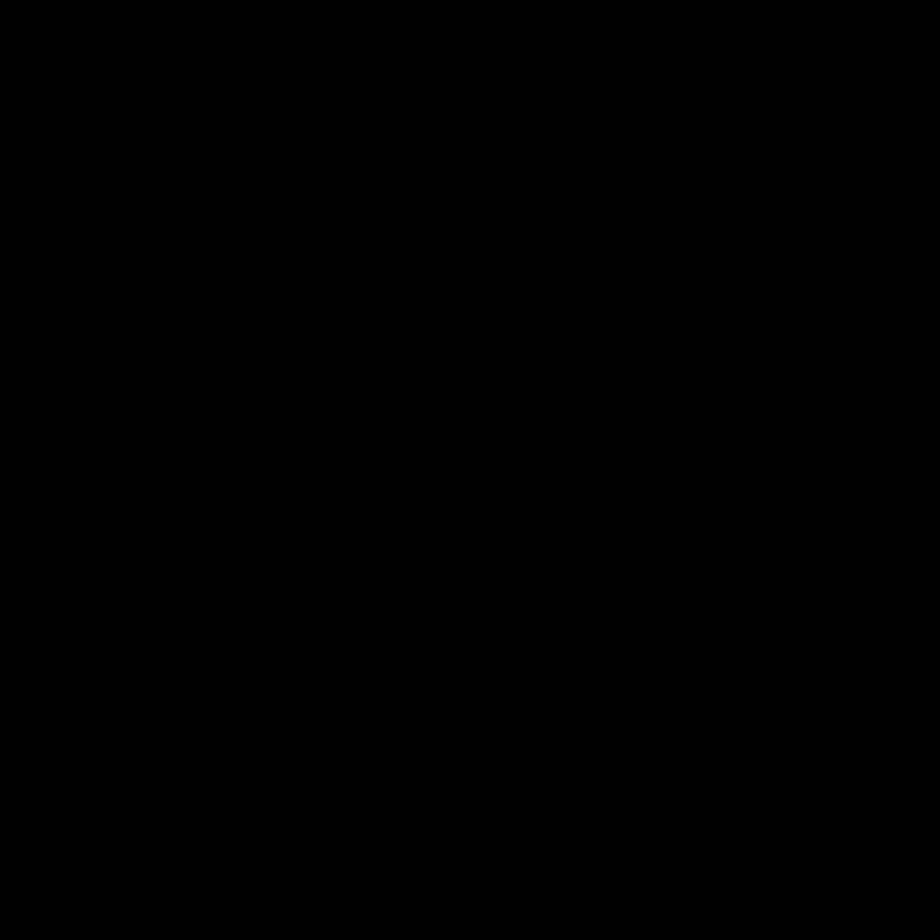}}\\
				\vspace{-2mm}
				{\footnotesize t=1000}
			\end{minipage}
			\vspace{-2mm}
			\caption*{\footnotesize (a) r=4.1}
		\end{minipage}
		\hfill
		\begin{minipage}{0.47\linewidth}
			\begin{minipage}{\linewidth}
				\centering
				\includegraphics[width=\linewidth]{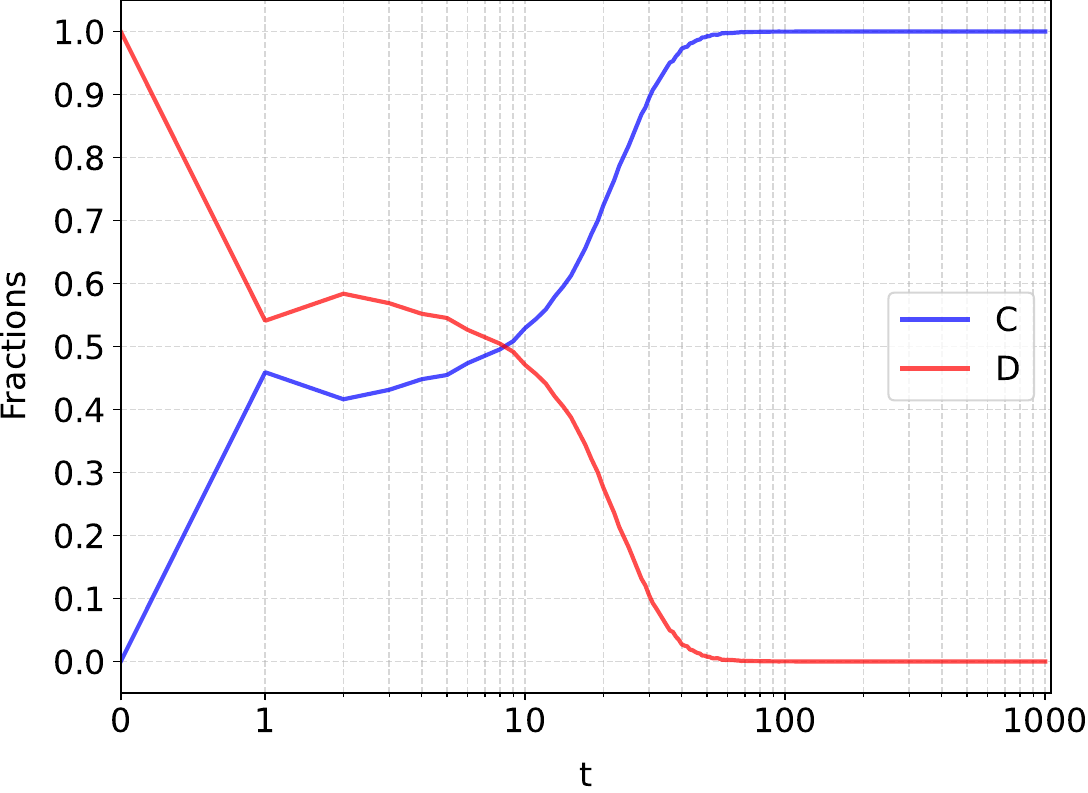}\\
			\end{minipage}
			\vspace{2mm}
			\\
			\begin{minipage}{0.188\linewidth}
				\centering
				\fbox{\includegraphics[width=\linewidth]{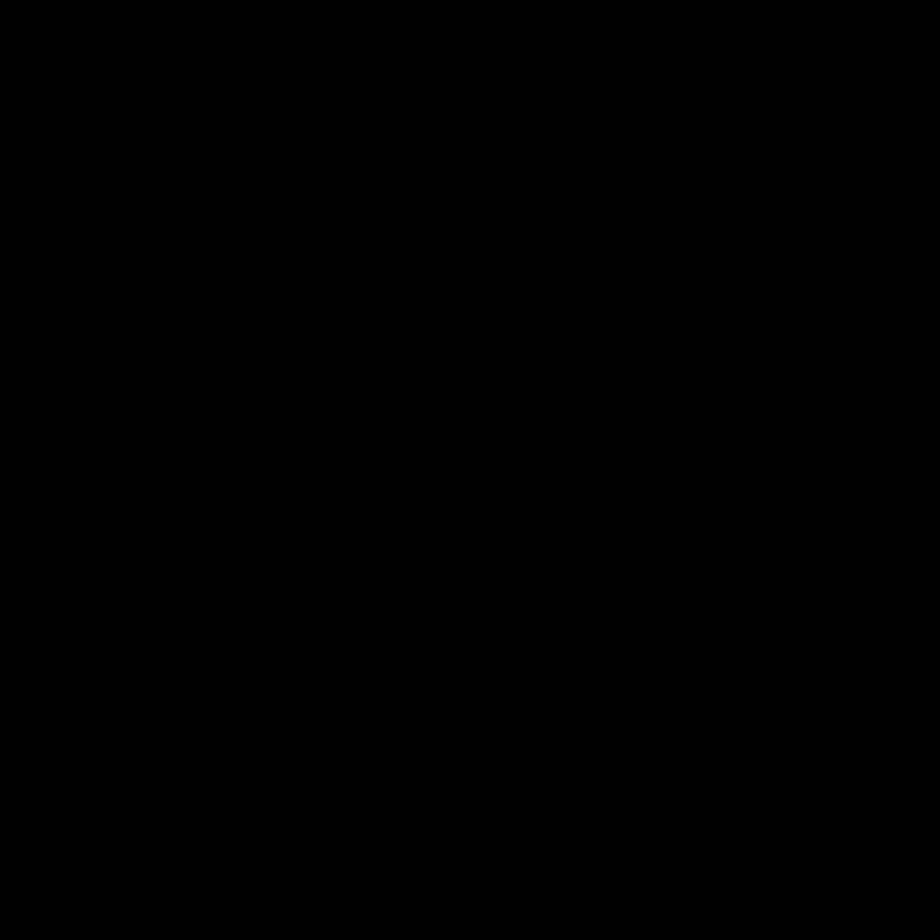}}\\
				\vspace{-2mm}
				{\footnotesize t=0}
			\end{minipage}
			\begin{minipage}{0.188\linewidth}
				\centering
				\fbox{\includegraphics[width=\linewidth]{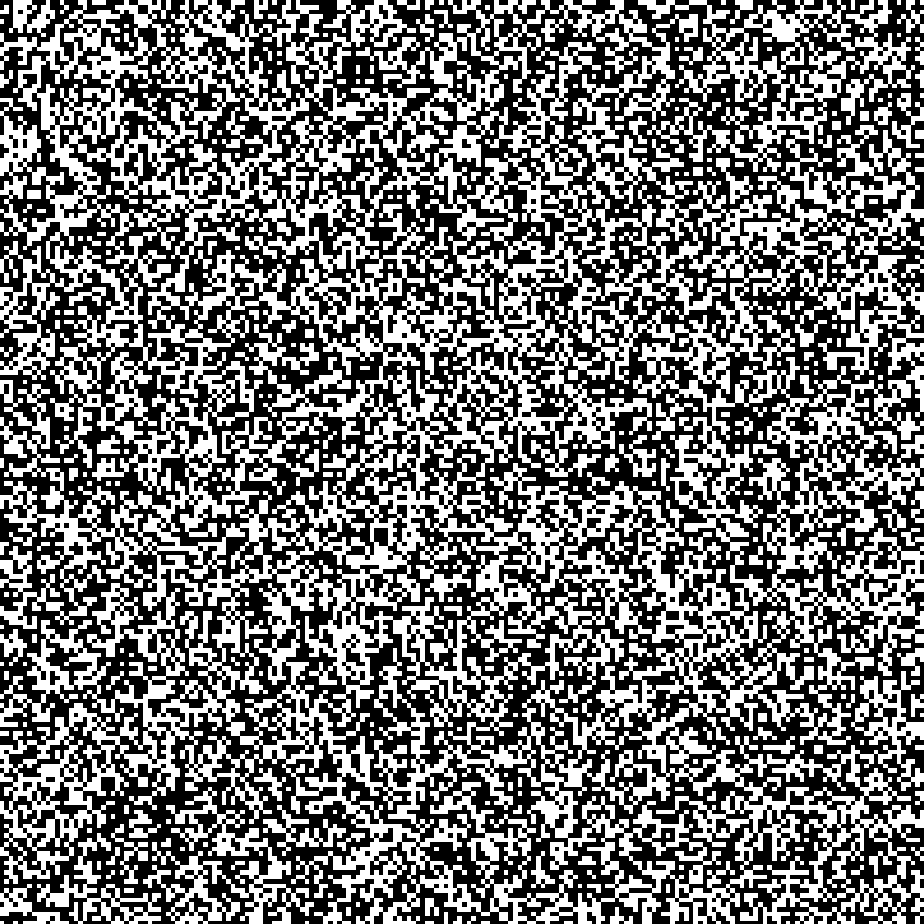}}\\
				\vspace{-2mm}
				{\footnotesize t=1}
			\end{minipage}
			\begin{minipage}{0.188\linewidth}
				\centering
				\fbox{\includegraphics[width=\linewidth]{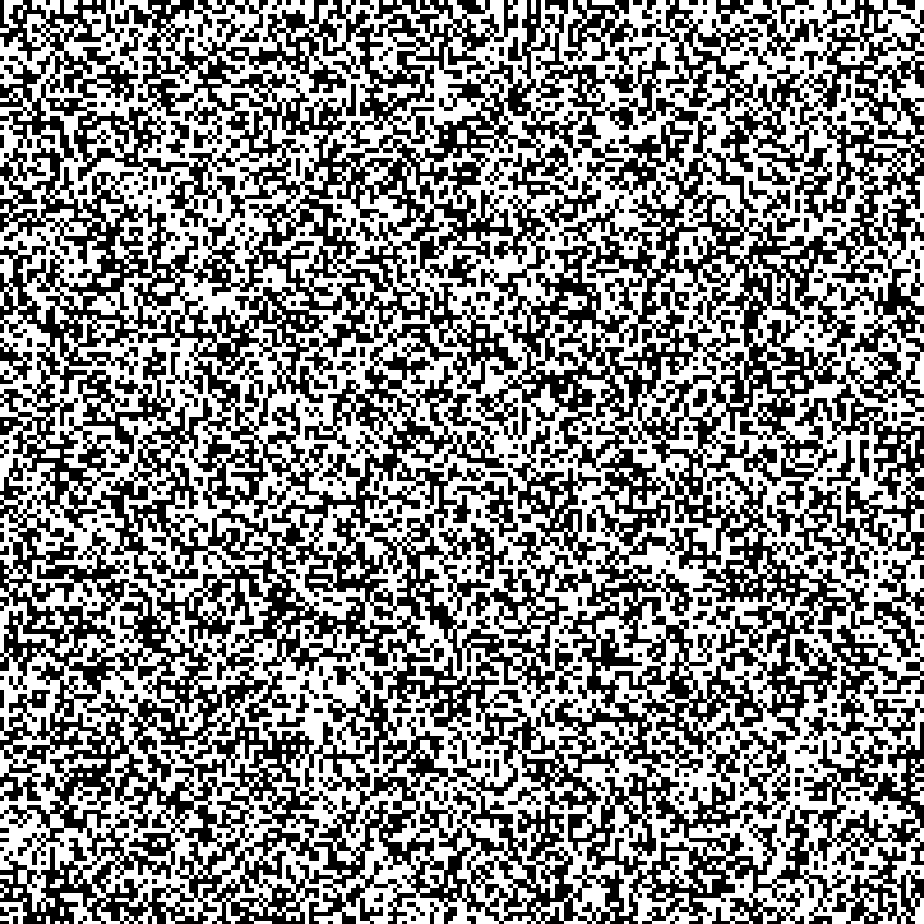}}\\
				\vspace{-2mm}
				{\footnotesize t=10}
			\end{minipage}
			\begin{minipage}{0.188\linewidth}
				\centering
				\fbox{\includegraphics[width=\linewidth]{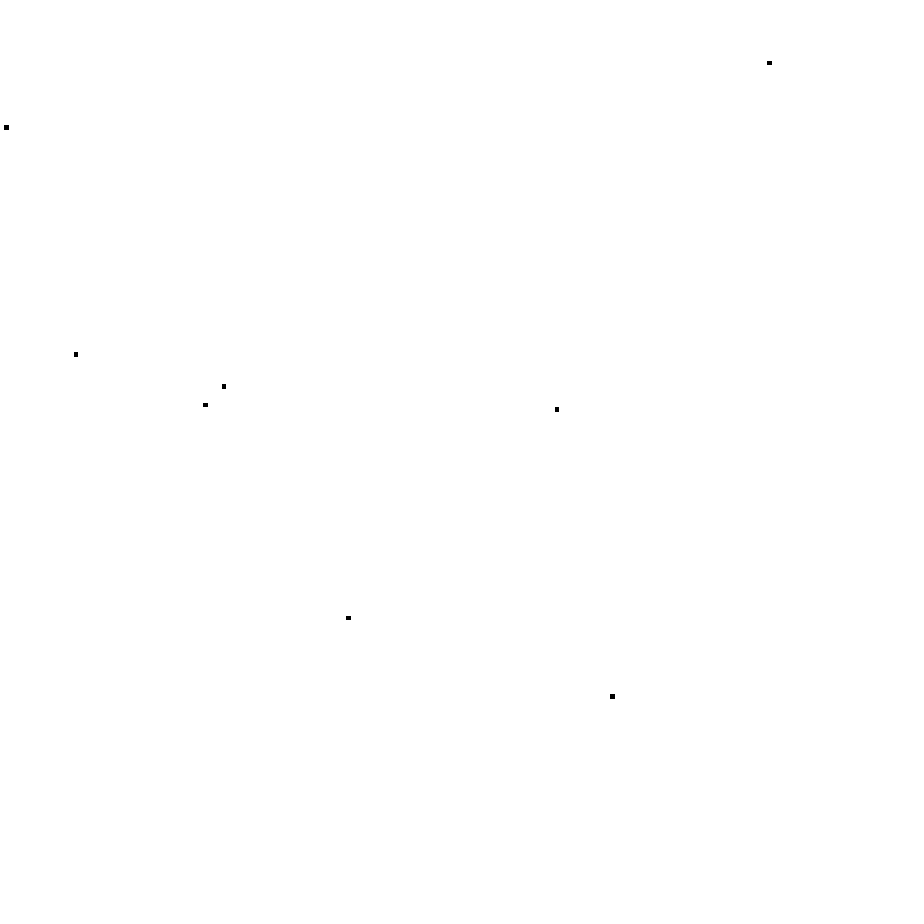}}\\
				\vspace{-2mm}
				{\footnotesize t=100}
			\end{minipage}
			\begin{minipage}{0.188\linewidth}
				\centering
				\fbox{\includegraphics[width=\linewidth]{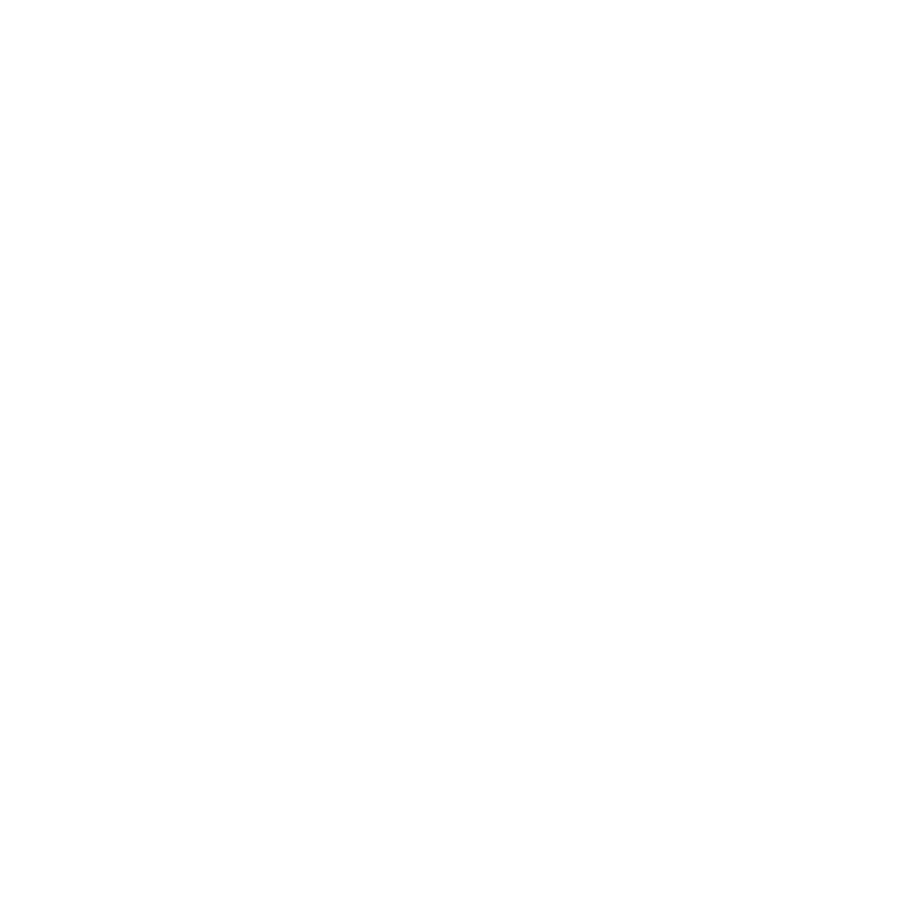}}\\
				\vspace{-2mm}
				{\footnotesize t=1000}
			\end{minipage}
			\vspace{-2mm}
			\caption*{\footnotesize (b) r=4.7}
		\end{minipage}
		\\
		[2mm]
		\begin{minipage}{\linewidth}
			\begin{minipage}{0.188\linewidth}
				\centering
				\includegraphics[width=\linewidth]{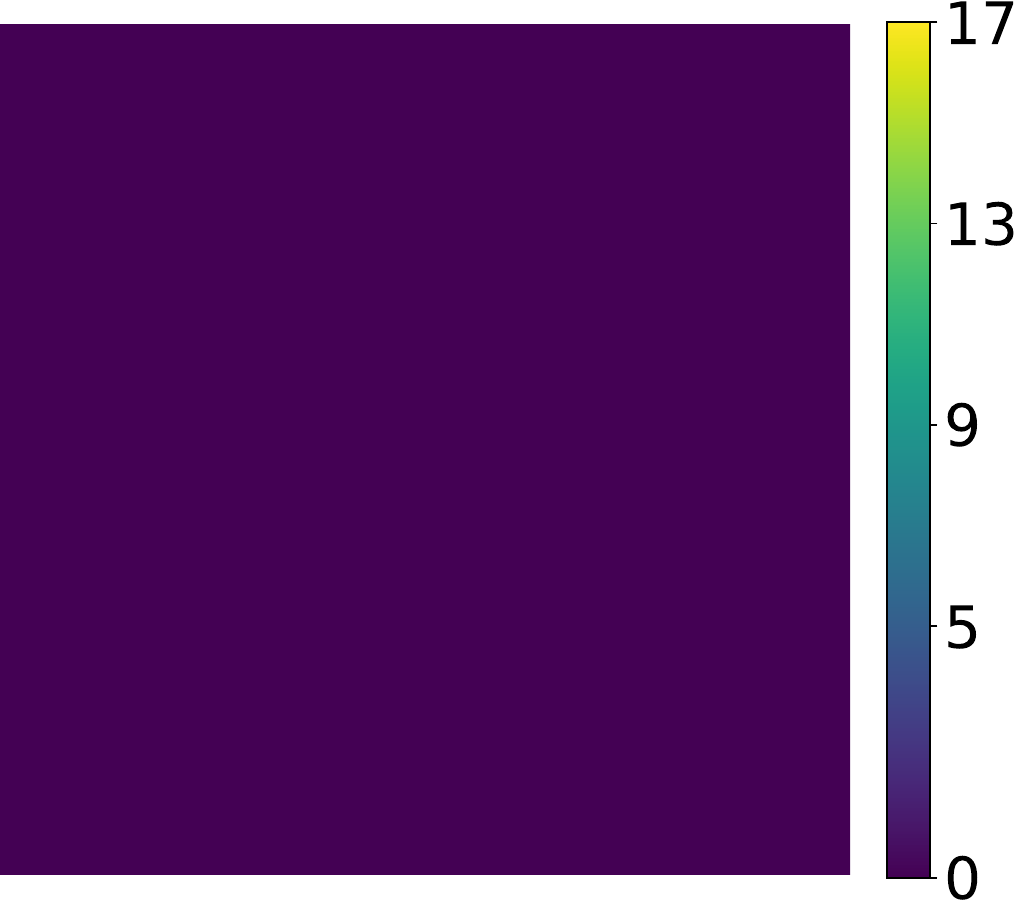}\\
				\vspace{-2mm}
				{\footnotesize t=0}
			\end{minipage}
			\hfill
			\begin{minipage}{0.188\linewidth}
				\centering
				\includegraphics[width=\linewidth]{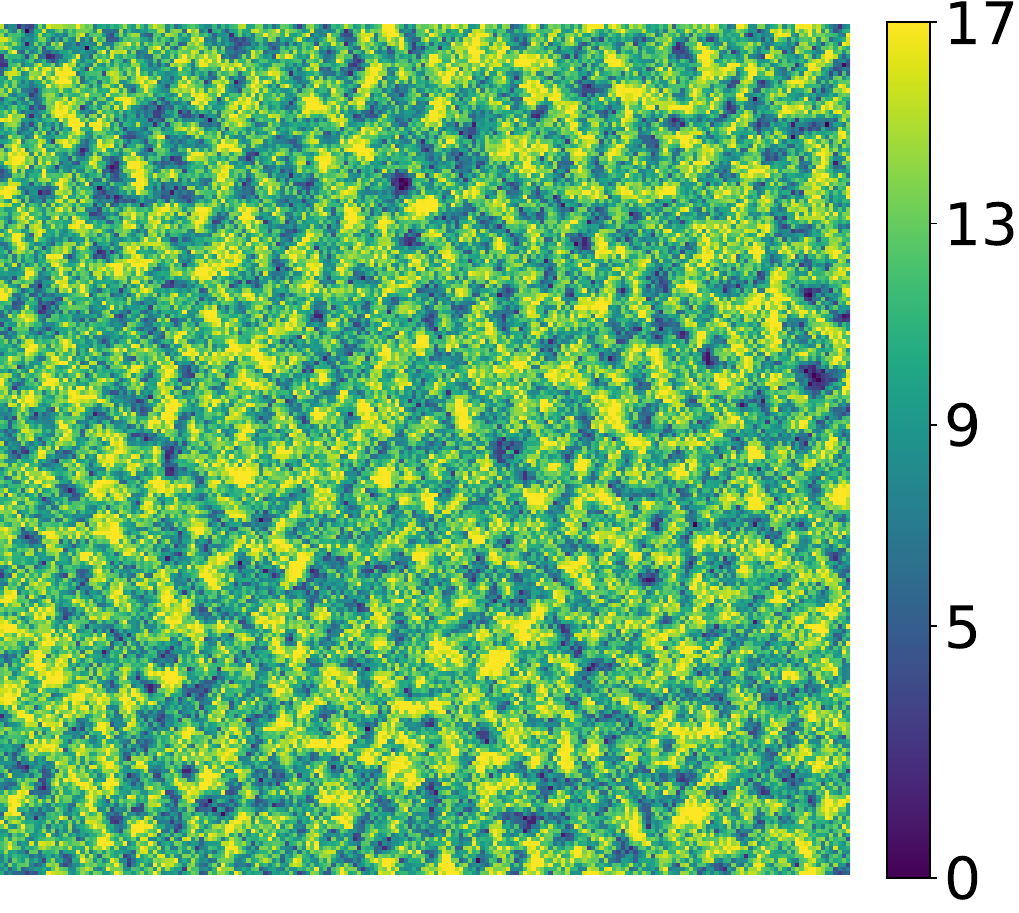}\\
				\vspace{-2mm}
				{\footnotesize t=1}
			\end{minipage}
			\hfill
			\begin{minipage}{0.188\linewidth}
				\centering
				\includegraphics[width=\linewidth]{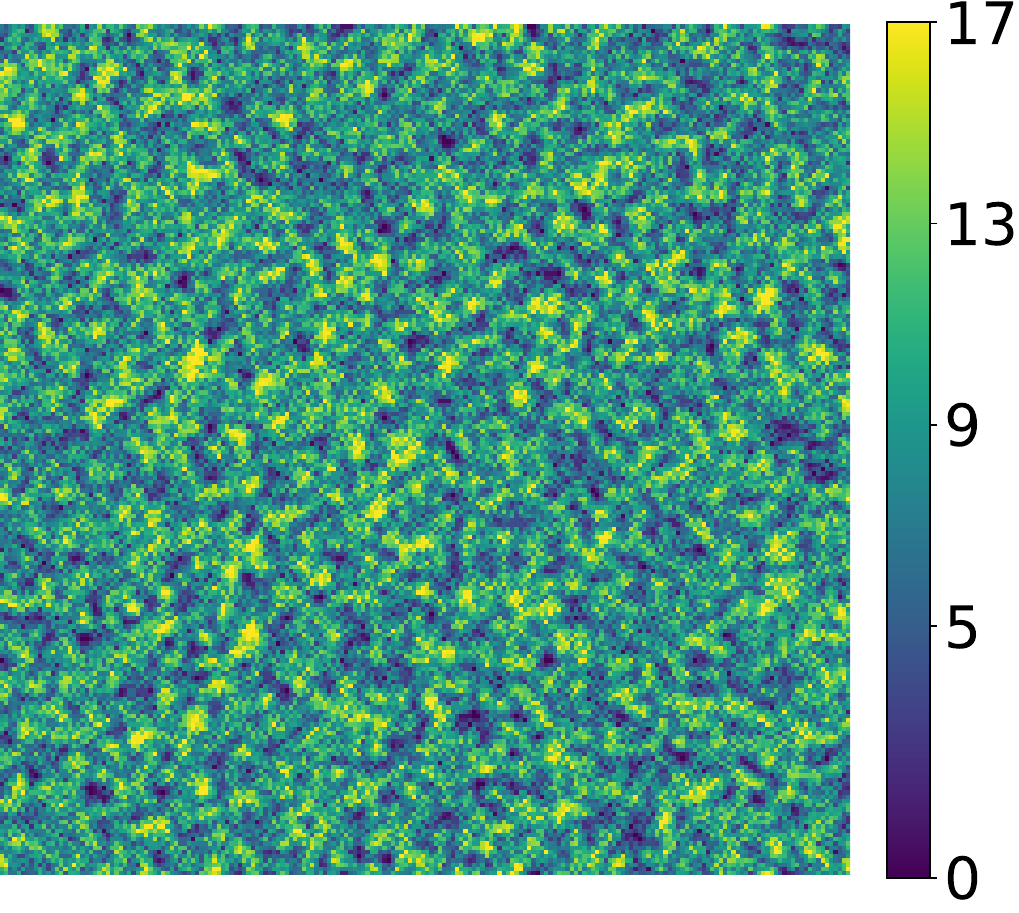}\\
				\vspace{-2mm}
				{\footnotesize t=10}
			\end{minipage}
			\hfill
			\begin{minipage}{0.188\linewidth}
				\centering
				\includegraphics[width=\linewidth]{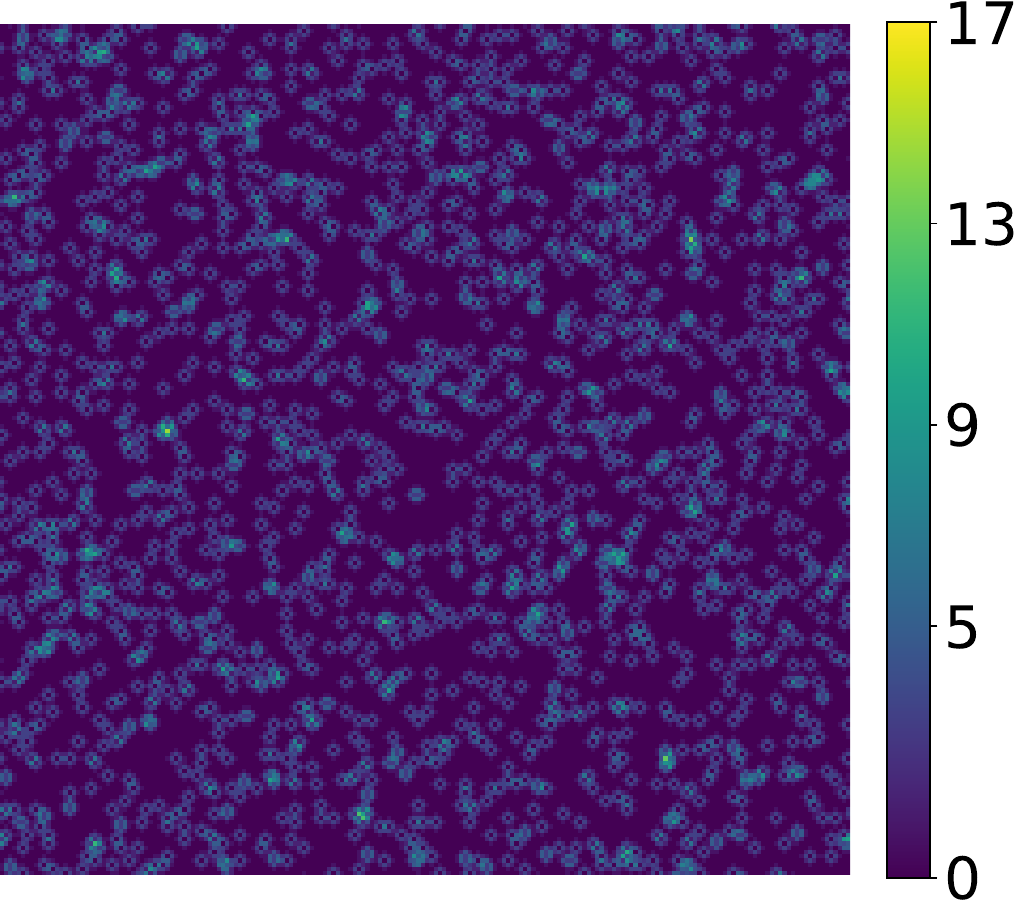}\\
				\vspace{-2mm}
				{\footnotesize t=100}
			\end{minipage}
			\hfill
			\begin{minipage}{0.188\linewidth}
				\centering
				\includegraphics[width=\linewidth]{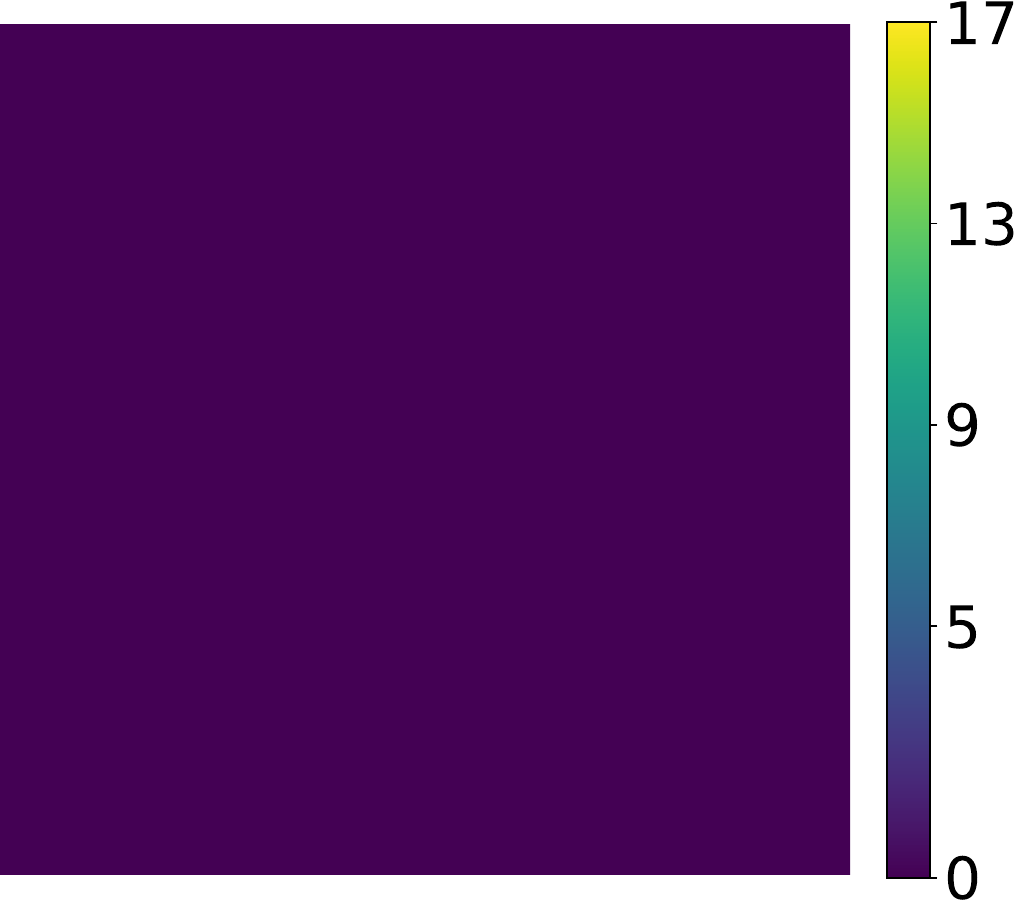}\\
				\vspace{-2mm}
				{\footnotesize t=1000}
			\end{minipage}
			\vspace{-2mm}
			\caption*{\footnotesize (c) r=4.1 (Payoff heatmaps)}
		\end{minipage}
		\\
		[2mm]
		\begin{minipage}{\linewidth}
			\begin{minipage}{0.188\linewidth}
				\centering
				\includegraphics[width=\linewidth]{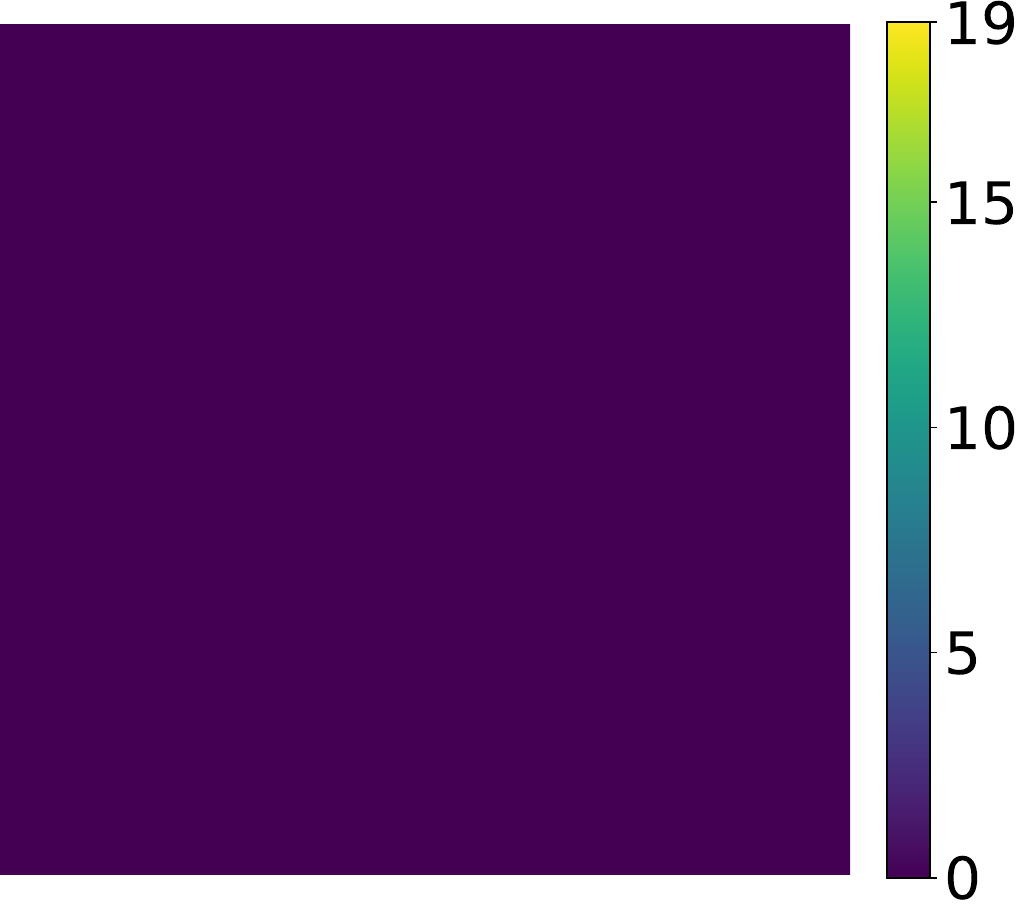}\\
				\vspace{-2mm}
				{\footnotesize t=0}
			\end{minipage}
			\hfill
			\begin{minipage}{0.188\linewidth}
				\centering
				\includegraphics[width=\linewidth]{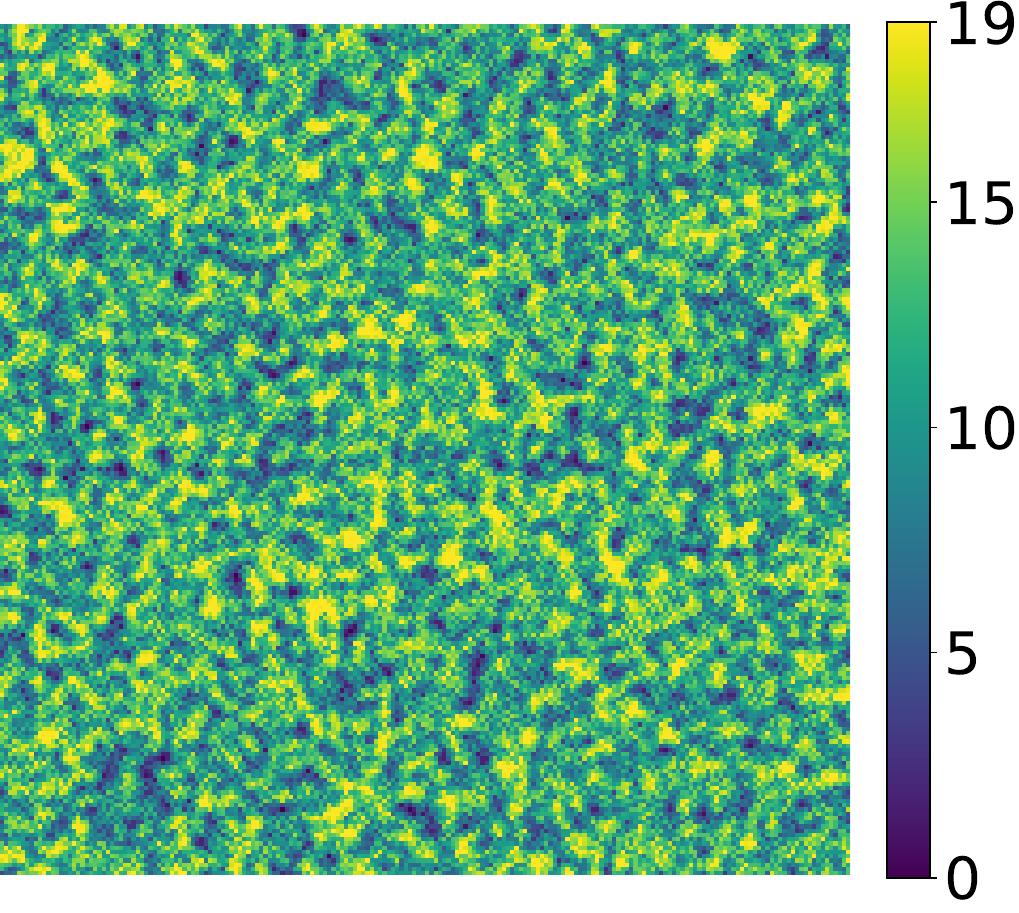}\\
				\vspace{-2mm}
				{\footnotesize t=1}
			\end{minipage}
			\hfill
			\begin{minipage}{0.188\linewidth}
				\centering
				\includegraphics[width=\linewidth]{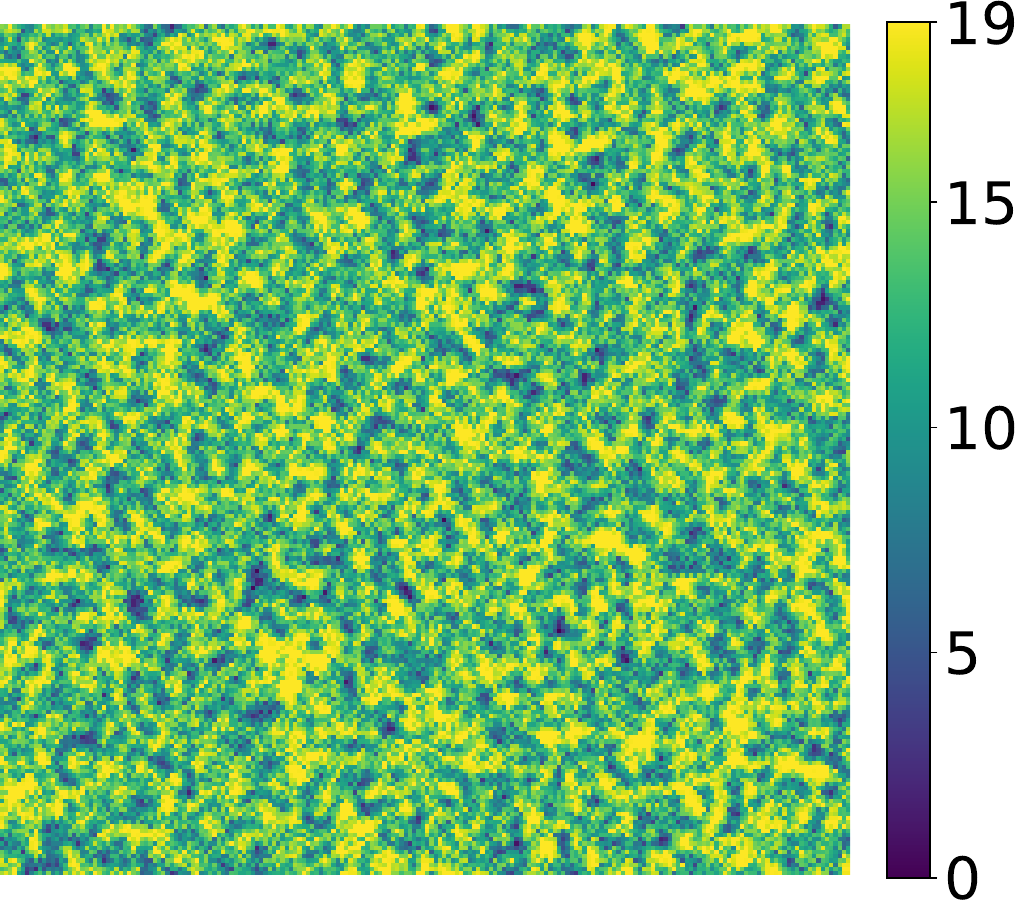}\\
				\vspace{-2mm}
				{\footnotesize t=10}
			\end{minipage}
			\hfill
			\begin{minipage}{0.188\linewidth}
				\centering
				\includegraphics[width=\linewidth]{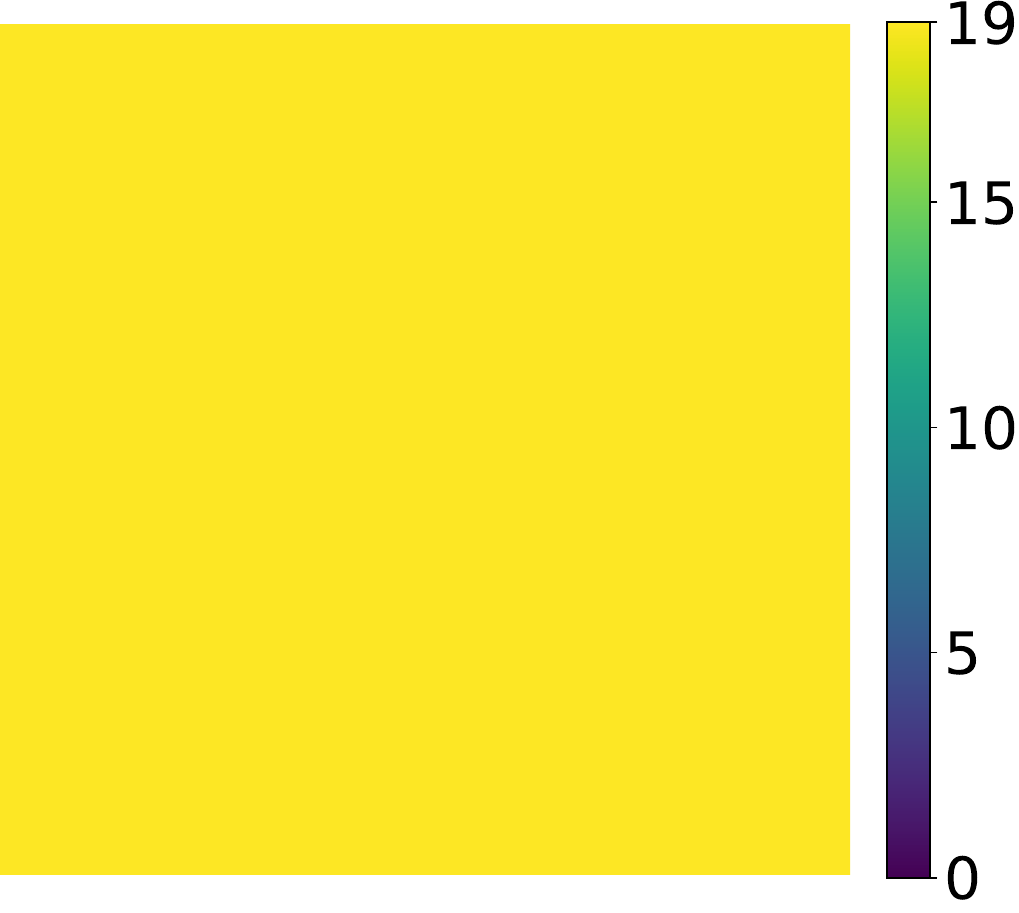}\\
				\vspace{-2mm}
				{\footnotesize t=100}
			\end{minipage}
			\hfill
			\begin{minipage}{0.188\linewidth}
				\centering
				\includegraphics[width=\linewidth]{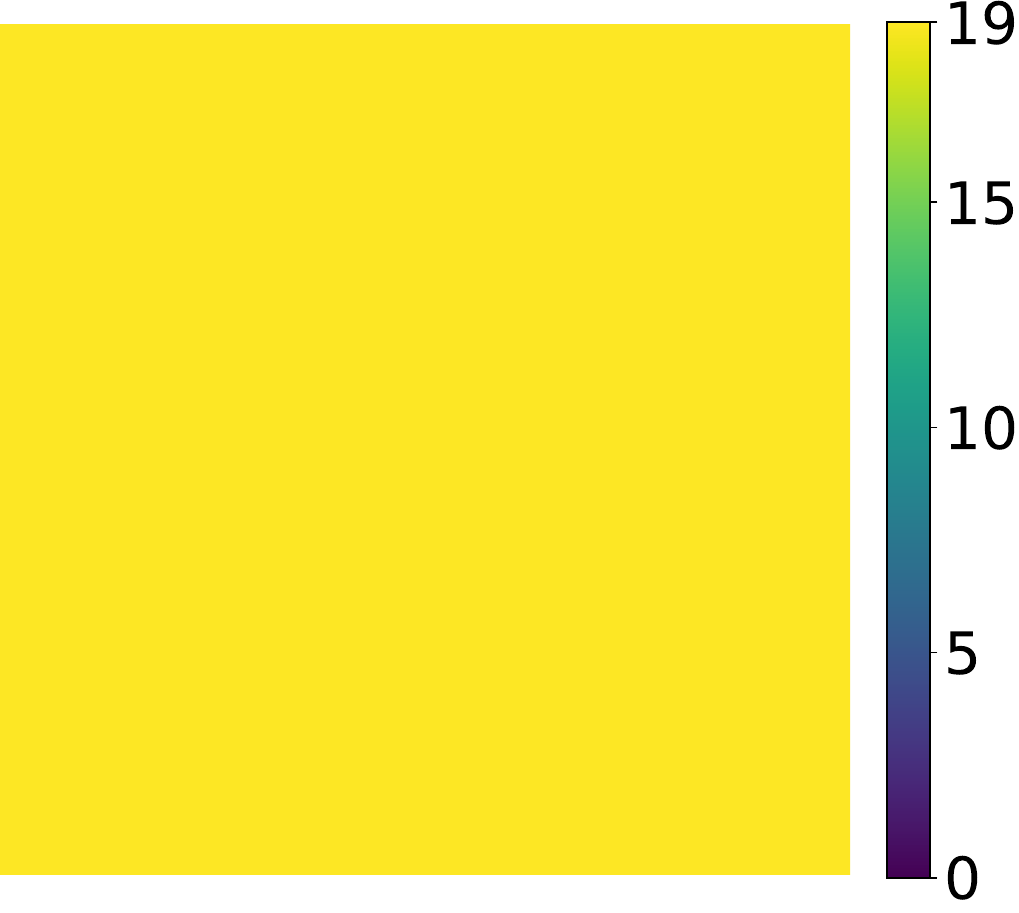}\\
				\vspace{-2mm}
				{\footnotesize t=1000}
			\end{minipage}
			\vspace{-2mm}
			\caption*{\footnotesize (d) r=4.7 (Payoff heatmaps)}
		\end{minipage}
		\caption{MAPPO-LCR with all-defectors initialization.
			Top: evolution of cooperation and defection.
			Middle: state snapshots at $t=0,1,10,100,1000$.
			Bottom: payoff snapshots with high values in yellow and low values in purple.}
		\label{fig:exp_allD}
	\end{figure*}

	We evaluate MAPPO-LCR when all agents begin as defectors.
	Figure~\ref{fig:exp_allD} presents evolution curves, state snapshots, and payoff snapshots.
	When $r=4.1$, cooperation does not appear at any time.
	The evolution curve moves quickly to full defection and remains stable.
	State snapshots stay dark across all iterations, and payoff snapshots show uniform low values.
	This agrees with theory, since cooperation cannot grow below the effective threshold of MAPPO-LCR.
	When $r=4.7$, the system shows a different pattern.
	All agents start as defectors, yet the actor–critic network is randomly initialized.
	The first rollout therefore contains random cooperative actions.
	These early actions generate positive signals for LCR and allow cooperation to expand when $r$ is large enough.
	The evolution curve shows steady growth of cooperation and full cooperation appears before $t=90$.
	State snapshots become bright and well mixed, while payoff snapshots shift from low values to high values as cooperation spreads.
	This behavior demonstrates a structural advantage of MAPPO-LCR.
	Learning-based agents can escape full defection because policy randomness and gradient updates amplify cooperative signals.
	Expected returns guide strategy revision and do not rely on explicit spatial imitation.
	Fermi dynamics cannot escape full defection because no cooperative neighbors exist to start propagation.
	MAPPO-LCR therefore avoids the absorbing trap of full defection and supports cooperation when $r$ is sufficiently large.
	
	\section{Conclusions}
	\label{sec:con}
	
	This paper revisits cooperation dynamics in SPGG from an advanced multi-agent DRL perspective.
	A modern policy-gradient-based multi-agent framework replaces traditional evolutionary rules and tabular learning methods.
	We introduce Multi-Agent Proximal Policy Optimization (MAPPO) to SPGG.
	To the best of our knowledge, this work is the first to apply MAPPO to SPGG.
	By modeling agents as adaptive learners under centralized training and decentralized execution, MAPPO enables richer strategic representations.
	Agents internalize long-term returns and population-level feedback during learning while executing fully decentralized policies.	This formulation reframes SPGG as a learnable multi-agent decision process rather than a rule-driven evolutionary system.
	It provides a flexible foundation for analyzing learning-driven cooperation dynamics in large spatial populations.
	
	Based on this framework, we further examine agent behavior under a LCR mechanism.
	The mechanism is used to analyze how MAPPO responds to neighborhood-level cooperative feedback.
	This analysis reveals how advanced multi-agent learning integrates local information into policy updates.
	It offers insight into reward shaping effects without modifying the underlying game structure.	
	Overall, introducing MAPPO expands the methodological scope of cooperation studies in spatial games.
	The framework bridges evolutionary game theory and modern multi-agent DRL.
	This perspective opens new directions for studying adaptive social behavior and learning dynamics in structured multi-agent environments.
	﻿
	﻿
	
	\section*{CRediT authorship contribution statement}
	
	\textbf{Zhaoqilin Yang}: Writing – original draft, Writing – review and editing, Validation, Methodology, Conceptualization.
	\textbf{Axin Xiang}: Conceptualization, Investigation, Writing – review and editing, Visualization.
	\textbf{Kedi Yang} : Writing – review and editing, Validation, Software, 
	\textbf{Tianjun Liu}: Writing – review and editing, Visualization, Supervision.
	\textbf{Youliang Tian}: Funding acquisition, Resources, Supervision.
	
	\section*{Declaration of competing interest }
	
	The authors declare that they have no known competing financial interests or personal relationships that could have appeared to influence the work reported in this paper.
	
	\section*{Data availability}
	
	No data was used for the research described in the article.
	
	\section*{Acknowledgments}
	This work was supported by the Natural Science Special Project (Special Post) Research Foundation of Guizhou University (No.[2024] 39), Guizhou Provincial Basic Research Program (Natural Science) Youth Guidance Project (No. Qiankehe Foundation QN(2025) 054). 
	Project of Science and Technology Innovation Platform of Guizhou Province under No. CXPTXM[2025]024.
	Foundation of State Key Laboratory of Public Big Data (No.PBD2023-33).
	National Key Research and Development Program of China under Grant 2025YFB3109800; National Natural Science Foundation of China under Grant 62272123. Science and Technology Program of Guizhou Province under Grant [2022]065; Science and Technology Program of Guiyang under Grant [2022]2-4.

	\bibliographystyle{elsarticle-num}
	\bibliography{cas-refs}
\end{document}